\documentclass[a4paper]{article}

\providecommand{\esymbol}{e}
\providecommand{\esymbolv}{\ev}

\usepackage{setspace} 
\usepackage{xr}
\usepackage{wrapfig}
\usepackage{enumitem}
\usepackage{color}
\usepackage{amsmath,amsfonts,amssymb}
\usepackage{epsfig}
\usepackage{graphicx}
\usepackage{colortbl}
\usepackage{url}
\usepackage{lineno}

\definecolor{darkblue}{rgb}{0.05,0.,0.65}
\definecolor{grey}{rgb}{0.9, 0.9, 0.9}

\newcommand{\myparagraph}[1]{\vspace{-3mm}\paragraph{#1}}

\newcommand{\coout}[1]{[{\color{magenta} #1}]}
\newcommand{\co}   [1]{[{\it \color{red} #1}]}
\newcommand{\todo} [1]{[{\color{blue} #1}]}

\usepackage{amsmath,amsfonts,amssymb}
\usepackage{relsize}
\usepackage{ marvosym }

\newcommand{\sign}     {\mbox{sign}}

\newcommand{\inv}{^{-1}}

\newcommand{\md}{{\rm d}}

\newcommand{\e}   {\mbox{\rm e}}

\newcommand{\hv}{{\bf h}}
\newcommand{\cv}{{\bf c}}
\newcommand{\vv}{{\bf v}}
\newcommand{\yv}{{\bf y}}

\newcommand{\kv}{{\bf k}}

\newcommand{\bv}{{\bf b}}

\newcommand{\ev}{{\bf e}}
\newcommand{\xv}{{\bf x}}

\newcommand{\av}{{\bf a}}

\newcommand{\thetav}{{\boldsymbol \theta}}

\newcommand{\nuv}{{\boldsymbol \nu}}

\newcommand{\muv}{{\boldsymbol \mu}}
\newcommand{\kappav}{{\boldsymbol \kappa}}

\newcommand{\Nmat}{\mathbf{N}}

\newcommand{\Gmat}{\mathbf{G}}

\newcommand{\Amat} {\mathbf{A}}

\newcommand{\Kmat}{\mathbf{K}}

\newcommand{\trans}{^{\top}}

\DeclareMathAlphabet{\mathpzc}{OT1}{pzc}{m}{it}
\DeclareMathAlphabet{\mathcalligra}{T1}{calligra}{m}{n}

\providecommand{\esymbol}{e}

\newcommand{\modevector}{\kappav}

\newcommand{\kcat}     {k_{\rm cat}}

\newcommand{\kcatl}{k_{\rm cat,l}}

\newcommand{\Euns}    {E}
\newcommand{\Eun}     {\mathbf{\Euns}}

\newcommand{\Eunint}  {\Eun_{\rm c}} 
\newcommand{\Eunu}    {\Eun_{\rm \esymbol}}

\usepackage{etoolbox}

\newtoggle{bookversion}      \togglefalse{bookversion}
\newtoggle{shortbookversion} \togglefalse{shortbookversion}
\newtoggle{anhang}           \togglefalse{anhang}
\newtoggle{hidecomments}      \togglefalse{hidecomments}

\newcommand{\bookco}[1]{}

\definecolor{brown}{rgb}{0.9,0.69,0.34}
\definecolor{samoabrownlight}{rgb}{0.89,0.69,0.4}
\definecolor{samoabrowndark} {rgb}{0.5,0.3,0.15}
\definecolor{cbasamoabrown1}{rgb}{0.87,0.6,0.23}
\definecolor{cbasamoabrown2}{rgb}{0.87,0.6,0.23}
\definecolor{cbabrown1}{rgb}{0.87,0.6,0.23}
\definecolor{cbabrown2}{rgb}{0.87,0.6,0.23}
\definecolor{cbabrown3}{rgb}{0.87,0.6,0.23}
\definecolor{cbabrown4}{rgb}{0.87,0.6,0.23}
\definecolor{cbaecoblue1}{rgb}{0.8,0.8, 1.0}
\definecolor{cbaecoblue2}{rgb}{0.7,0.7, 1.0}
\definecolor{cbaecoblue3}{rgb}{0.87,0.6,0.23}
\definecolor{cbaecoblue4}{rgb}{0.87,0.6,0.23}
\definecolor{cbablue2}{rgb}{0.87,0.6,0.23}
\definecolor{cbapink}{rgb}{.99,0.92,0.75}
\definecolor{cbabeige1}{rgb}{0.86, 0.797, 0.625} 
\definecolor{cbabeige2}{rgb}{0.93, 0.812, 0.56}  
\definecolor{cbabeige3}{rgb}{1.0, 0.97, 0.88}  
\definecolor{cbahelleslila}{rgb}{1.0, 0.99, 1.0}  
\definecolor{cbalightgrey}{rgb}{0.95,0.95,0.95}
\definecolor{cbatablecolor1}{rgb}{0.86, 0.797, 0.625} 
\definecolor{cbatablecolor2}{rgb}{1,1,1}         

\newcommand{\myvalue}      {value}
\newcommand{\Value}        {Value}
\newcommand{\gain}         {gain}

\newcommand{\investment}   {investment}

\newcommand{\enzymeinvestment}{{enzyme \investment}}

\newcommand{\price}        {price}

\newcommand{\burden}       {burden}

\newcommand{\enzymeprice}  {enzyme \price}

\newcommand{\fluxburden}   {flux \burden}

\newcommand{{\fluxvalue}}  {flux \myvalue}
\newcommand{{\fluxgain}}   {flux \gain}
\newcommand{\metabolicobjective}{metabolic objective}

\newcommand{\valid}  {valid}

\newcommand{{\valueflow}}    {value flow}
\newcommand{{\Valueflow}}    {Value flow}

\newcommand{ {\flow}}        {flux profile}
\newcommand{ {\Flow}}        {Flux profile}

\newcommand{\point}         {point} 

\newcommand{\benefitshade}  {{\point} benefit}

\newcommand{\enzymecost}    {enzyme     {\point} cost}

\newcommand{\enzymebenefit} {enzyme     {\point} load}

\newcommand{\fluxpattern}{flux pattern}
\newcommand{\futile}{futile}
\newcommand{\wasteful}{wasteful}
\newcommand{\nonbeneficial}{non-beneficial}

\newcommand{\loadbalance}           {reaction-metabolite balance}

\newcommand{\summationcondition}    {flux variation rule}

\newcommand{\mfFBA}{FBA with minimal fluxes}

\newcommand{\MCA}{MCT}

\newcommand{\fluxvalueform}{flux value form}

\providecommand{\esymbol}   {e}
\providecommand{\esymbolv}  {\mathbf{e}}

\providecommand{\prodrate}  {r}

\newcommand{\intprod}   {\prodrate^{\rm int}}

\newcommand{\rate}{\nu}
\newcommand{\ratev}{\nuv}
\newcommand{\ratelaw}{k}

\newcommand{\enzdegrate}{\lambda^{\rm deg}}

\newcommand{\prodratev}{\mathbf{\prodrate}}

\newcommand{\ffit}        {{\mathcal F}} 
\newcommand{\fluxbene}    {b}
\newcommand{\fluxcost}    {a}
\newcommand{\metcost}     {g}

\newcommand{\hminus}      {h}
\newcommand{\partialcost} {z}
\newcommand{\wsymbol}     {w}
\newcommand{\loadsymbol}  {y}

\newcommand{\partialder}{^{\centerdot}} 

\newcommand{\Nint}    {\Nmat^{\rm int}}
\newcommand{\Ntot}    {\Nmat^{\rm tot}}

\newcommand{\Next}    {\Nmat^{\rm x}}

\newcommand{\Nobj}    {{\Nmat^{\rm ben}}}

\newcommand{\Deltar}{\square}

\newcommand{\fes}    {\fes}
\newcommand{\fel}    {\fel}
\newcommand{\fevs}   {\fevs}

\newcommand{\husdot}   {\hus\partialder}

\newcommand{\gc}    {\gplusv_{\rm \cint}}

\newcommand{\gvtot}    {\gvsymbolv_{\rm v}}
\newcommand{\gvtotsdot}{\gvsymbol_{\rm v}\partialder}
\newcommand{\gvtotl}   {\gvsymbol_{v_{l}}}
\newcommand{\gvtots}   {\gvsymbol_{\rm v}}

\newcommand{\hminusv}{{\bf \hminus}}

\renewcommand{\u}     {\esymbol}
\newcommand{\hul}     {\hminus_{\esymbol_l}}

\newcommand{\husmin} {{\hminus_{\rm \esymbol}^{\rm min}}}
\newcommand{\hus}     {\hminus_{\rm \esymbol}}
\newcommand{\hu}      {\hminusv_{\rm \esymbol}}

\newcommand{\fluxbenev}     {{\bf \fluxbene}}

\newcommand{\bbenefit}  {\fluxbene}

\newcommand{\fluxbenetot}  {\fluxbene}
\newcommand{\fluxbenetotv} {\fluxbenev}
\newcommand{\bvtot}        {\fluxbenetotv_{\rm v}}

\newcommand{\bvtotn}       {{{\fluxbenetotv}^{(n)}_{{\rm v}}}}

\newcommand{\bvtotl}       {{\fluxbenetot _{v_l}}}

\newcommand{\bvdir}        {{\fluxbenev_{\rm v}^{\rm int}}}

\newcommand{\bvdirs}       {{\fluxbene_{\rm v}^{\rm int}}}
\newcommand{\bvdirl}       {{\fluxbene_{v_l}^{\rm int}}}

\newcommand{\hmetv}  {\mathbf{\metcost}}

\newcommand{\hc}     {\hmetv_{\rm c}}

\newcommand{\bpsi}     {\fluxbenev_{\rm x}}

\newcommand{\bpsij}    {\fluxbene_{\rm x_j}}

\newcommand{\acost}     {\fluxcost}

\newcommand{\acostv}    {{\bf \acost}_{\rm v}}
\newcommand{\acostvl}   {\acost_{v_l}}

\newcommand{\apointcostvs}   {\acost_{v}\partialder}

\newcommand{\hvv}   {\acostv}

\newcommand{\hvs}   {\acost_{\rm v}}
\newcommand{\hvl}   {\acost_{v_{l}}}

\newcommand{\hvsmin}{{\acost_{v}^{\rm min}}}

\newcommand{\zcostgeneric}{\partialcost}
\newcommand{\zcostgenerics}{\zcostgeneric}
\newcommand{\zcostgenericl}{\zcostgeneric_l}

\newcommand{\wvsymbol}{{\bf \wsymbol}}

\newcommand{\loadvsymbol}{{\bf \loadsymbol}}

\newcommand{\wtot} {\wvsymbol_{\rm \prodrate}}

\newcommand{\wtots}{\wsymbol_{\rm \prodrate}}
\newcommand{\wtoti}{\wsymbol_{\prodrate_i}}
\newcommand{\wtotl}{\wsymbol_{\prodrate_l}}

\newcommand{\wint}  {\wvsymbol_{\rm \prodrate}^{\rm int}}
\newcommand{\wints} {\wsymbol_{\rm \prodrate}^{\rm int}}
\newcommand{\wintm} {\wsymbol_{\rm \prodrate_{m}}^{\rm int}}
\newcommand{\wintl} {\wsymbol_{\rm \prodrate_{l}}^{\rm int}}

\newcommand{\wext}   {\wvsymbol_{\rm \prodrate}^{\rm ext}}

\newcommand{\wextj}  {\wsymbol_{\prodrate_j}^{\rm ext}}

\newcommand{\wexts}  {\wsymbol_{\rm \prodrate}^{\rm ext}}

\newcommand{\loadv}  {\loadvsymbol}
\newcommand{\loadint}{\loadv_{\rm c}}

\renewcommand{\modevector}{\kappav}
\newcommand{\modevectorcyc}{\modevector_{\rm cyc}}

\newcommand{\fluxenzymecostl}  {\Delta \wsymbol_{\intprod_l:}}

\newcommand{\us}{\esymbol}
\newcommand{\ul}{\esymbol_{l}}
\newcommand{\huv}{\mathbf{\hminus}_{\rm \esymbol}}

\renewcommand{\gc} {\wvsymbol_{\rm c}}

\renewcommand{\gvtot}  {\wvsymbol_{\rm v}}
\renewcommand{\gvtotl} {\wsymbol_{v_{l}}}
\renewcommand{\gvtots} {\wsymbol_{\rm v}}
\renewcommand{\gvtotsdot} {\wsymbol_{\rm v}\partialder}

\renewcommand{\hv}{{\bf h}}

\renewcommand{\Nmat} {{\bf N}}

\renewcommand{\Kmat} {{\bf K}}

\renewcommand{\coout}[1]{}

\definecolor{brown}{rgb}{0.9,0.69,0.34}
\definecolor{lightyellow}{rgb}{1,0.99,0.85}

\newcommand{\psfilesfluxes}{ps-files}

\renewcommand{\todo}[1]{#1}
\renewcommand{\co}[1]{}
\renewcommand{\myparagraph}[1]{}

\begin{document}

\title{Metabolic fluxes and value production}

\date{}

\author{Wolfram Liebermeister\\[3mm] 
Universit\'e Paris-Saclay, INRAE, MaIAGE, 78350 Jouy-en-Josas, France}

\maketitle

\begin{abstract}
  Metabolic fluxes in cells are governed by physical, physiological,
  and economic principles. Here I assume an optimal allocation of
  enzyme resources and postulate a general principle for metabolism:
  each enzyme must convert less valuable into more valuable
  metabolites to justify its own cost. The ``values'', called economic
  potentials, describe the individual contributions of metabolites to
  cell fitness.  Local value production implies that the cost of an
  enzyme must be balanced by a benefit, given by the economic
  potential difference the catalysed reaction multiplied by the
  flux. Flux profiles that satisfy this principle -- i.e.~for which
  consistent potentials can be found -- are called economical.
  Economical fluxes must lead from lower to higher economic
  potentials, so certain flux cycles are incompatible with any choice
  of economic potentials and can be excluded.  To obtain economical
  flux profiles, non-beneficial local patterns, called futile motifs,
  can be systematically removed from a given flux distribution.  The
  principle of local value production resembles thermodynamic
  principles and complements them in models. Here I describe a
  modelling framework called Value Balance Analysis (VBA) that uses
  the two principles and yields the same solution as enzyme cost
  minimisation (in kinetic models) and flux cost minimisation (in
  FBA).  Given an economical flux distribution, kinetic models in
  enzyme-optimal states and with these fluxes can be constructed
  systematically.  VBA justifies the principle of minimal fluxes and
  the exclusion of futile cycles, predicts enzymes that could be
  plausible targets for regulation, provides criteria for the usage of
  enzymes and pathways, and explains the choice between high-yield and
  low-yield flux modes.  By linking flux analysis to kinetic models,
  it provides a realistic picture of fluxes, kinetics, and enzyme
  investments in cells: fluxes are linked to enzyme efficiencies and
  protein data, assuming a balance between enzyme investments,
  described by a local production of value.
\end{abstract}


\textbf{Abbreviations:} PFK: phosphofructokinase; FBPase:
  fructose bisphosphatase; ATP: adenosine triphosphate; ADP: adenosine
  diphosphate; F6P: fructose 6-phosphate; F16BP: fructose
  1,6-bisphosphate; P: phosphate.

\co{edit section on economic and thermodynamic constraints}
  
  \co{replace all Delta by Deltar? problem mit Delta G, leute sind so daran gewöhnt!}
  \co{JA! put all missing REFs} \co{JA! LESEN! Flux networks in
    metabolic graphs: Article in Physical Biology 6(4):046006, DOI:
    10.1088/1478-3975/6/4/046006 Patrick B. Warren / Silvio M. Duarte
    Queiros / Janette L. Jones} \co{P.B. Warren and
    J.L. Jones. Duality, thermodynamics, and the linear programming
    problem in constraint-based models of metabolism. Phys Rev Lett.,
    99(10):108101, 2007.}  \co{JA! SI: elementary modes = minimal
    supports of stationary flux distributions} \co{cite Nigam + Liang
    2007 paper (in cba/literature)} \co{JA! example for decreasing
    value along flux (bcs of cofactors und ``nebengewinnen''): from
    glucose to CO2!}

  \coout{generally use ``compatible'' to describe agreement between fluxes and potentials}
  \coout{shadow value ist ok; shadow value und shadow price jeweils
    verwenden, wo sinnvoll; relativ viel (anstelle von Lagrange mult)
    verwenden} \coout{``economics of conversion'' and ``economics of
    concentrations''} \coout{introduce ``flux motifs'' as sign
    patterns; wie mode gegenueber profile; dann weiter verwenden}

\coout{Box with formulae (cheat sheet mit symboltabelle? oder nach CBA perspective?)

  Flux burden = $\frac{\mbox{enzyme investment}}{\mbox{rate}} = \frac{\mbox{enzyme price}}{\mbox{efficiency}}$

  $av = \frac{\hus \esymbol}{v}=\frac{\husdot}{v} = \frac{\hus}{\ratelaw}$

  Small efficiency $\ratelaw \rightarrow$ large burden $av \stackrel{balance}\longrightarrow$ large flux benefit!

  e.g. little substrate little force

  Linear chain:

  Economic potentials = embodied value = $\sum$ upstream flux prices

\includegraphics[width=5.5cm]{/home/wolfram/projekte/cba/zeichnungen/cba_flux_dings1.jpg}

  $\rightarrow$ more complicated in nonlinear pathway!

  $w_{i} = \sum_{l=1}^{i} a_{v_{l}}$

Embodiment of enzyme cost 

\includegraphics[width=5.5cm]{/home/wolfram/projekte/cba/zeichnungen/cba_flux_dings2.jpg}

  $w \cdot v$ = value production in metabolite!

  ``point value production'' symbol?}

\coout{\co{WOERTER}  \co{DEF word for thermo and econ feasible .. VBA-feasible? ja uea}
  \coout{doch VBA - Value balance analysis; dann in CBA labour: VFA -
    value flux analysis // klarmachen, dass das erst der anfang der
    value balance analysis ist, die analyse von wertstroemen kommt
    dann in cba labour}
\co{JA! oefter flux mode statt {\flow}
  verwenden?}  \co{CBM (constraint-based modeling)} \co{CBA fluxes (auch CBA kin) erst "flux
  sign pattern" einfuehren, dann kurz "flux pattern"} \co{"Protein
  sector" is a good word} \co{``flux modelling'' statt ``flux
  analysis''} \co{oefters sagen ``because of the value production
  principle''} \co{metabolic optimality problem = optimality principle
  (def! hier und in CBA opt, CBA lag)}\co{USE ``Optimality principle''
  instead of metabolic optimality problem} \co{FN: precise usage of
  ``enzyme-balanced'' (positive benefit, balanced by positive cost) vs
  ``enzyme-economical'' (positive benefit)}}

\co{NEW SIMULATIONS (fuer hier unf CBA labour) \co{ELAD: Although I
    always think that a good theory is better than any example, some
    people think otherwise and maybe you should try to show
    non-trivial examples of how VBA improves predictions (of
    flux or anything else). I know this is an annoyance, but I think
    it might help you get more interest in your work.}  \co{JA! Use
    two protein data sets, determine the economic potentials, and try
    to find differences for key metabolites. ZB Modell mit/ohne
    sauerstoff; reaktion auf sauerstofferhoehung? JA - mit
    variabilitaetsanalyse fuer oekonoimsche potentiale! extraartikel,
    ``naama''-style} RUN FCM auf grossen modell (zB dougies) \co{JA!
    CBA fuer Dougie-Hefemodell (minimal fluxes at fixed biomass!)} +
  plotte die ec pot, fluesse, wertstroeme usw; hier oder cba labour?}
\co{doch selber fluess in grossen netzwerk rechnen?"  (mit
  flussminimierung und biomasseproduktion als ziel)?  Dann weiter mit
  thermo-CBA kram ...}  \co{schoenes klares beispiel; compute the
  values for a real example; (i) EMP vs ED (ii) aerobic / anaerobic E
  coli; show differences for key metabolites; (iii) CBA Large-scale
  analysis of REAL protein costs and fluxes real proteine data;
  compute marginal protein costs and metabolite values}

\co{WENIGER WICHTIG} \co{Opt-knock / opt-slope analysieren mit CBA
  multiobjective? Kurz erwaehnen in CBA fluxes?}  \co{resp + ferm econ
  pot aufschreiben - wo ist das geblieben? SI?}  \co{elad: show and
  discuss larger {\nonbeneficial} cycles (less obvious ones) try to
  show something surprising with futile cycles check the ones with o2?
  // consider {\mfFBA}.  // use rubisco example as an example in
  supplement (for high/low yield?)?}  \co{Elad: show and discuss
  larger futile cycles (less obvious ones) try to show something
  surprising with futile cycles check the ones with o2?}  \co{prepare
  matlab demo files and SBML files} \coout{include opt-slope-like
  rationale: from {\flow}s and flux {{\myvalue}} equation, get to
  inequalities for the {{\enzymeprice}}s to be satisfied for a local
  optimum} \coout{Abildungen, in denen enzymkosten als Groesse gezeigt
  werden}

\co{WEITERE IDEEN} \co{nadh and nadph in glycolysis and
  gluconeogenesis? is the reason for the two systems that the cell can
  assign chem. pot. and prices independently? compute them in the ycm
  model!} \co{ ycm - modell mit NADH, NADP bilanziert, keine
  biomasse-prod, ATP-produktion ist nutzen, mit verschiedenen
  substraten, mit/ohne sauerstoff -> gute flussverteilungen dafuer
  erraten aus princ min flux} \co{futile cycles zählen in hefe-modell
  mit produktion mehrerer biomasserelevanter metabolite (oder biomasse
  selbst)} \co{ ueberlegen: flussbenefit mit *mehreren* guten
  produkten; princ min flux gibt nur eine loesung; bilanzgl. beschreibt
  viele moegcliche loesungen!} \co{ klären: what kind of info about
  enzyme levels can we get from the balance equation (without knowing
  any elasticities or kinetics)} \co{ bsu-beispiel in ordnung
  bringen. ein paar sinnvolle steady states!  Keq identisch setzen
  fuer alle szenarien!}  \co{ueberlegen: enzymkosten groeßer als ein
  epsilon (anstieg bei u=0); normierung auf gesamtprotein
  biomasseprod: gesamtkosten entsprechen gesamtprotein die relativen
  werte der epsilons fuer einen zustand hängen von den kosten pro
  enzymmolekuel (oder summe ueber pathway) ab // experiment: kosten
  manipulieren durch stickstofflimitierung oder ribosomen-inhibitor}
\co{Moeglichst vieles mit laufendem beispiel illustrieren:
  S.cerevisiae-zentralstoffwechsel 1. vergleich fba, eba und vba
  (annahme: maximale atp-produktion), sampling moeglicher fluesse
  2. abschätzung (sampling?) der metabolitwerte, gegeben die äußeren
  metabolite und eine flussverteilung}

\iftoggle{bookversion}
{\section{Metabolic economics in flux analysis}}
{\section{Introduction}}

\coout{Thread (nochmal ueberpruefen):\\
  o metabolic fluxes\\
  o fba and its problems. say: can be justified!\\
  o idea here: positive investments: positive benefits: flux times value difference\\
  o analogy to thermodyn\\
  o usage in economic fba\\
  o article overview}

\coout{WEG? Fluxes need to satisfy metabolic
  demands, while the catalysing enzymes are costly and limited by
  space restrictions.}

\myparagraph{\ \\Metabolic fluxes in cells} \co{in einleitung ueber
  ``modes'' (nicht profiles) reden und spaeter erst den unterschied
  einfuehren?} Cells invest a great fraction of their proteome in
metabolic enzymes. Measured enzyme amounts reflect varying metabolic
fluxes and metabolic demands, and emerge from enzyme kinetics and
cellular resource allocation. \co{LESEN! (naamas gruppe; pdf in
  cba/literature) Metzl-Raz, E. et al. Principles of cellular resource
  allocation revealed by condition-dependent proteome profiling. Elife
  6, doi:10.7554/eLife.28034 (2017).}  Understanding the principles
behind metabolic fluxes and enzyme allocation may help us explain
phenomena such as the Warburg effect \cite{wapn:24}, engineer
metabolic pathways, and understand metabolic evolution (e.g.~the
choice between alternative pathway designs).  While a cell's metabolic
network may carry many possible flux distributions, only some of them
are biologically plausible.  To predict metabolic fluxes,
constraint-based methods such as Flux Balance Analysis (FBA) methods
\cite{vapa:94b} consider all possible flux profiles (defined here as
stationary flux distributions) and require stationarity, thermodynamic
feasibility, or resource economy. Based on such principles,
unrealistic {\flow}s and infeasible flux patterns (describing flux
directions and active reactions) can be discarded. Known flux
directions will restrict the possible flux profiles. In models, flux
directions may be chosen \emph{ad hoc}, from empirical knowledge, or
based on general principles: thermodynamics, for example, requires
fluxes to run from higher to lower chemical potentials, i.e.~in the
direction of thermodynamic forces
\cite{belq:02,bbcq:04,faqb:05,hjbh:06,kuph:06,hebh:07,hohh:07,fmsy:11}. \coout{these
  references come again in thermo section} Even if the chemical
potentials are unknown, this law implies that flux profiles must be
loopless, in the sense that the do not contain submodes without any
net metabolic conversion \cite{prtp:06}. But none of these principles
determines the fluxes precisely: within the physical limits, many
{\flow}s remain possible and which flux profiles are realised depends
on kinetics, available substrates, and enzyme regulation. Since many
of these details are unknown, some models explain fluxes not by their
physical causes, but by their purpose: that is, by metabolic
production at a limited use of enzyme resources!

\myparagraph{Metabolic fluxes and enzyme investments} How can we
understand fluxes and enzyme levels through economic considerations?
Resource allocation models rely on a simple premise: if a cell
converts nutrients into valuable products (e.g.~biomass), this
provides a benefit (e.g.~sustaining life), and it is this benefit that
justifies costly, enzyme-catalysed fluxes: to keep the benefit high,
protein resources must be optimally allocated to different cellular
subsystems to maximise the metabolic benefit while minimising protein
cost. In the late 19$^{\rm th}$ century, a similar competition for
resources within organisms \cite{roux:81} was invoked to explain the
optimal shapes and structures of bones \cite{wolf:92}.  In metabolism,
we can assume similar resource allocation principles: metabolic fluxes
are coupled not only dynamically (through mass balance and kinetics)
but also through enzyme demands, and the ``cost budgets'' for
different pathways are traded against each other\footnote{Enzyme costs
  can be defined empirically or theoretically. Empirically, costs have
  been defined by the growth deficits after a forced expression of
  idle enzyme. For a theoretical definition, one may assume that the
  total protein amount in a cell is limited and that increasing the
  amount of enzymes reduces the available amount of other proteins,
  which then reduces the cell's benefit. In pathway models, we can
  summarise these indirect costs outside the pathway modelled by a
  cost function for enzyme levels, to be subtracted from the metabolic
  benefit.}.  Such trade-offs can be described by assuming that cells
minimise the enzyme amount per metabolic production
\cite{wbgl:11,wnfb:18}. Hence, in flux prediction, protein levels play
multiple roles: first, measured protein levels can be used to estimate
fluxes, assuming that fluxes increase with eth enzyme levels; and
second, the relation between fluxes and protein levels may be directly
used in models, e.g.~to describe how metabolic pathways compete for a
limited protein budget.

\co{JA! SCHREIBEN (kurz!): wie fluesse (zb flussvorhersage) und
  proteinmengen (zb proteomik, oder wissen um platzbegrenzung)
  zusammenbringen? dann (naechster abschnitt): fba hat ein problem
  damit, weil kinetik und metabolitkonz ignoriert werden, dann kommt
  es zu ad hoc-regeln ..  but that also means that model parameters
  are not realistic, but effective and model specific, and that they
  don't generalise between models (eg in model combination)}

\myparagraph{Flux balance analysis and its limitations} Trade-offs
between flux benefits (such as biomass production) and flux costs (a
proxy for enzyme demand) have been implemented in FBA.  \coout{klar
  einfuehren: sagen: moegliche annahme (assumption of known/fixed
  efficiencies, for calculation): v = e kapp; minimise sum e at fixed
  b(v), dh minimise $\sum_{l} k^{\rm app}_{l}|v_{l|}$ (principle of
  minimam sum of fluxes); bzw crowding, analog .. // nur der reihe
  nach. erst dogma mit flux sum dann weighted sum als proxy for
  enzymes; dann behandlung der richtigen enzyme cost! flux benfit,
  compromise moeglich ..} In a comparison of different FBA objectives
\cite{scks:07,szzh:12}, flux cost was found to be an important factor
for flux prediction. Flux Cost
Minimisation (FCM) \cite{lieb:18fcm} assumes that
{\flow}s minimise a flux cost $\acost(\vv)$ while realising a
predefined metabolic benefit $\bvtot \cdot \vv = \fluxbene$.  In FBA
with flux minimisation, this cost function is taken to be the sum of
absolute fluxes \cite{holz:04,holz:06}. Other variants such as FBA
with molecular crowding or CAFBA \cite{mhmm:16}  translate  fluxes into an overall protein
demand, which is then bounded by space constraints or by a limited
protein budget. A penalty or resource constraint on fluxes can
avoid futile flux cycles (where the meaning of ``{\futile}''
depends on one's ideas about cost and benefit functions).  While
classical FBA predicts high-yield {\flow}s, FBA variants with flux
costs can correctly predict the occurrence of low-yield {\flow}s with a
comparably lower enzyme cost. 

\myparagraph{Justification of flux costs or flux minimisation as a
  model assumption} In reality, metabolic fluxes do not cause a burden
themselves: it is their demand for enzyme and metabolite
concentrations that makes them costly. So how can we justify a
principle of minimal flux costs (which ignores metabolite
concentrations)?  \co{And what would be a meaningful notion of flux
  costs based on kinetic models?}  Reaction rates are not simply
proportional to enzyme levels. Instead, they depend on metabolite
concentrations, which in steady states depend on the enzyme levels
indirectly.  This explains why the empirical correlations between
enzyme concentrations and fluxes are low \cite{fbrp:10}, and metabolic
fluxes are hard to reconcile with proteomics data.  FBA ignores this
fact: \co{WD mit paar zeilen vorher} to derive enzyme levels from
fluxes, it assumes a simple proportionality $v= k \,e$ (with a
constant catalytic rate $k$). While reaction rates and enzyme levels
are proportional if metabolite concentrations are fixed, in reality
varying enzyme levels lead to varying metabolite concentrations and
therefore to varying enzyme efficiencies.  Kinetic models can capture
this fact, but optimising the enzyme levels in such models is
difficult if networks are large. So, if FBA employs rules of thumb
such as flux minimisation, can we justify these rules -- i.e.~show
that the same {\flow}s would also be discarded by kinetic models under
a principle of minimal enzyme cost?

\begin{figure*}[t!]
\begin{center}
 \includegraphics[width=16.5cm]{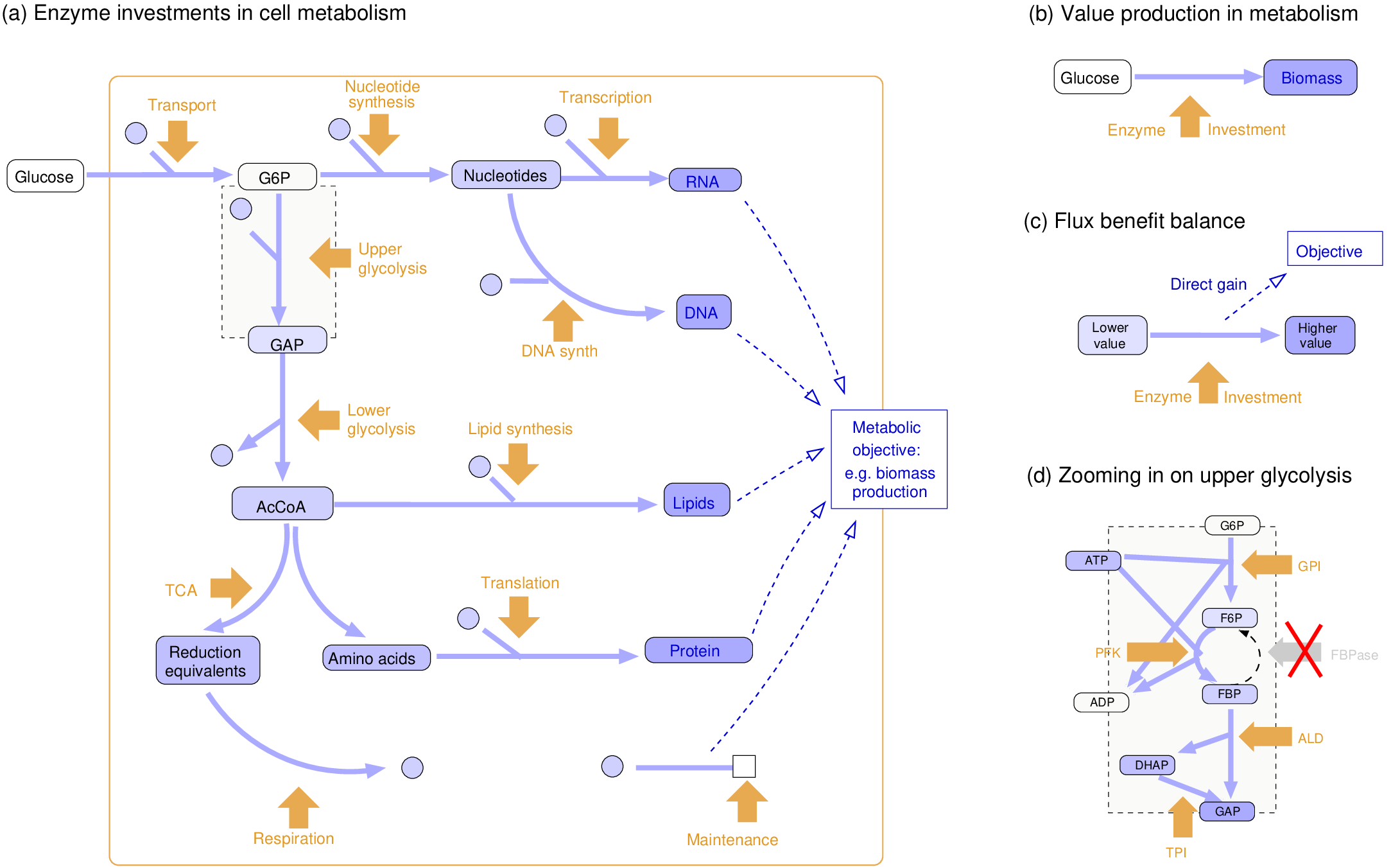}
\end{center}
\caption{\co{JA! move d up, should become b (then fix legend, fix c usw in text). in A, add
    arrow AA production} Balance of enzyme investment and value
  production. The principles shown hold for metabolic networks of
  any resolution and  size. (a) Enzymes (yellow arrows)
  catalyse metabolic reactions. In the network shown, glucose is
  converted into biomass. Each compound has a metabolic value (shades
  of blue): these values increase along the flux, reflecting the
  accumulating {\enzymeinvestment}s. \coout{comment on value flows;
    show them?} (b) The net reaction of (a) describes a conversion of
  glucose into biomass. In optimal states, value production (i.e.~the
  value of biomass, multiplied by its production rate) and
  {\enzymeinvestment} (the sum of price-weighted enzyme levels) must
  be balanced. Mathematically, this is caused by the fact that a
  variation of the enzyme level must leave the fitness (benefit minus
  cost) unchanged.  (c) Economic balance in an enzymatic reaction: a
  positive value production is required in order to balance the positive
  {\enzymeinvestment}. Value production results from a conversion of
  less valuable into more valuable metabolites, plus value generated
  by the flux itself (flux gain).  (d) Metabolic pathway (blow-up of
  upper glycolysis in (a)). \co{While the arrows in (a) refer to the
    large areas in the proteomap, the arrows here refer to individual
    tiles.} To model the pathway in isolation, metabolites on the
  pathway boundary are described as external (they need not be
  mass-balanced in the pathway model).  By choosing their economic
  potentials, we define a pathway objective. In the example, a futile
  cycle is suppressed, e.g.~by repressing the FBPase enzyme (crossed
  out). }
 \label{fig:metabolicEconomics} 
 \end{figure*}

\myparagraph{Producing valuable compounds requires
  {\enzymeinvestment}s} Here I argue that penalising high fluxes in
FBA is indeed a correct way to describe economical enzyme usage, even
if in reality metabolite concentrations are co-optimised (which FBA
ignores). To model this, I describe the proteome as an ``investome'',
assuming that enzyme investments in a reaction must be balanced (and
justified) by their ``usefulness'' for the cell. To link this to
metabolic network models, I further \co{JA! auch in CBA kin:} argue
that this usefulness can be described as a production of value within
the catalysed reaction, i.e.~a conversion of less valuable substrate
into more valuable product. Metaphorical terms like ``investment'' and
``value'' will be defined below.  The value production principle
(Figure \ref{fig:metabolicEconomics}), depicts the cell as a chemical
plant or a planned economy.  The idea that fluxes are costly (and must
provide benefits) is not new: applied to metabolism as a whole, it is
a main premise of FBA with minimal fluxes. Here, I apply the same
logic to every single reaction and relate it to the ``investome'', the
set of all (e.g.~enzyme) investments across the metabolic network,
defined as price-weighted enzyme amounts; ``prices'' refers to
marginal cost, derived from a cost function (e.g.~cell growth rate or
total enzyme mass). To relate enzyme benefits to a local value
production, we need to assign to each metabolite an economic value,
called called economic potential. In a framework called Metabolic
Value Theory \cite{lieb:18theory,lieb:18lagrange,lieb:14a}, the
optimality conditions for metabolic states are written as balance
equations interlinking the economic values of individual metabolites,
enzymes, or fluxes.

\coout{umbenennen FBP: FDP; FBPase: FBP??}

 \myparagraph{Economic potentials and the constraints on flux
   directions} The value production principle \co{MERGE IN : can be
   expressed using economic potentials. Economic potentials, defined
   as shadow values of mass balance constraints, describe the benefit
   that a cell would obtain from extra supplies of metabolites.} has
 further consequences.  \co{WO? In each reaction, fluxes must lead
   from lower to higher economic potentials (unless the fluxes provide
   some direct benefit) \cite{lieb:18lagrange,lieb:14a}.}
 \todo{\co{WD?}A metabolite's economic potential describes a
   metabolite's contribution to the metabolic objective. The economic
   potentials are not constant, but vary between metabolic states,
   like control coefficients in Metabolic Control Theory ({\MCA}), and
   resemble potentials in thermodynamics: to dissipate Gibbs free
   energy as required by the second law of thermodynamics, fluxes run
   from higher to lower chemical potentials. Thus, chemical potential
   differences guide metabolic fluxes by predefining their
   directions.} Similarly, in models with optimal enzyme usage,
 positive value production requires that fluxes lead from lower to
 higher economic potentials
 \cite{lieb:18theory,lieb:18lagrange,lieb:14a}. \todo{\co{sort rest of
     paragraph} Mathematically, this is expressed by a balance
   equation: the local value production in a reaction (given by the
   economic potential difference, multiplied by the flux) must be
   equal to the enzyme investment (representing the cost of the
   catalysing enzyme), see
   Fig.~\ref{fig:metabolicEconomics}\todo{(b)}. The economic
   potentials play a double role: on the one hand, being defined as a
   use value it describes the metabolite's effective contribution to
   the metabolic objective (e.g.~biomass production).  On the other
   hand, in optimal states this use value must be equal to the
   metabolite's ``embodied value'', describing all substrate and
   enzyme investments that are needed to produce this metabolite, at
   its present production rate, divided by this rate.  \co{FN?
     \co{sort! distinguish ``value'' and ``benefit'' balances} The
     balance between use value and embodied value follows directly
     from the value production principle: a potential difference in a
     reaction reflects a ``value generated'' (per flux), which encodes
     the usefulness of an enzyme! In optimal states, this potential
     difference must be exactly balanced by the ``flux burden'' in the
     same reaction, which quantifies the (enzyme) cost per unit of
     flux. Moreover, value production in a reaction is related to the
     enzyme investment in the reaction (as measured by the partial
     enzyme cost).  Thus, pricy enzymes must generate a high benefit
     per flux. If we assume, for simplicity, that enzyme investments
     are proportional to enzyme mass abundance, this yields a direct
     link between metabolic value, fluxes, and proteomics data.}}
  
 \myparagraph{Value balance analysis} Flux profiles that are
 compatible with the value production principle \co{schon eraklert?}
 are called economical. In practice, such {\flow}s can be determined 
 in two ways: by choosing fluxes together with economic potentials
 (such that all fluxes follow the potential differences) or by
 starting from a given {\flow} and removing all {\nonbeneficial}
 {\fluxpattern}s.  Ideally, we should choose potentials and fluxes
 that can also be realised by plausible kinetic models.  To make the
 economic potentials more realistic, heuristic assumptions and
 constraints can be employed, e.g.~taking into account $\kcat$ values
 or measured enzyme investments. In analogy to thermodynamically
 feasible {\flow}s, I define economical {\flow}s. Here I explore the
 consequences of this principle for predicting metabolic fluxes.  By
 employing the value production balance as a constraint, we obtain a
 variant of FBA called VBA that considers local value production along
 with the existing principles of stationarity and energy
 dissipation. As shown below, the notion of economical fluxes matters
 not only for VBA, but  also for existing FBA methods such as
 FBA with flux cost minimisation. \co{then econ FBA + heuristics} \todo{The
 theory behind it links} flux modelling to the underlying kinetic
 models. \co{consistent
   scheme by which kinetic models are ``aufgehoben'' (see hegel) in
   flux analysis} \todo{It also applies to cell} models, in which resources allocation
 and economic cycles do not only concern metabolic reactions, but also
 macromolecule synthesis and other cell processes. \co{explain!}  
 
 \myparagraph{Overview of the article} Here I postulate a new
 principle for flux modelling: a balance between enzyme investment and
 value production.  By \todo{considering} enzyme investments, enzyme
 efficiencies and thermodynamics, plausible fluxes and enzyme levels
 can be predicted. The predictions rely on the ``value structure'' of
 a metabolic state, that is, a \todo{system of economic variables}
 that describe what we mean by ``enzyme investment'' and ``value
 production''. Here I use laws for these variables, called  economic
 balance equations, in a modelling framework that resembles Energy
 Balance Analysis \cite{belq:02} and describes metabolic fluxes,
 chemical potentials, and economic potentials and that constrains flux
 directions by thermodynamic and economic constraints. The framework
 is called Value Balance Analysis (VBA). As a basic assumption, VBA
 requires flux profiles to be economical, that is, free of futile
 submodes.  I show that VBA is closely related to FCM and to the
 principle of minimal fluxes.  Based on the value production
 principle, I address more general questions. First, how can we define
 and detect futile cycles and remove them from given flux
 distributions?  And second, which pathways should cells use under
 different conditions and how do these choices depend on values
 embodied in the metabolites?  \coout{I then consider how usage of
   enzymes and choice between metabolic strategies depend on the
   interplay between {\enzymeinvestment}s and metabolites' economic
   potentials.}  VBA provides tools to answer these questions.
 Combining flux modelling and kinetic models, it relates the principle
 of minimal fluxes to an underlying a principle of minimal enzyme cost
 (or minimal enzyme and metabolite cost)
 \cite{sche:91,tnah:13,dole:17}, gives a specific meaning to economic
 variables in FBA (e.g.~of cost weights for fluxes, which are usually
 defined ad hoc), and relates them to {\enzymeinvestment}s in
 underlying kinetic models.  Mathematical details can be found in the
 Supplementary Information (SI) and at www.metabolic-economics.de.
 Matlab code for metabolic value theory and VBA is available on github
 \cite{github:meteco}.

\section{Economic potentials and economical fluxes}
\label{sec:EcPotEcFlux}

 \co{wo kommt  die erklenntnis, dass $\wext=\bpsij$?}
 
 \myparagraph{\ \\Economic potentials, {\enzymeinvestment}s, and the
   reaction balance} VBA employs a principle of local value
 production, which relies on an economic usage of enzyme.  Like in
 FBA, {\flow}s are scored by a linear benefit
 $\fluxbene(\vv) = \bvtot\cdot \vv$ describing overall value
 production. Typical examples are production of ATP or biomass.
 \coout{die folgenden details woanders?  lieber verbal die groessen
   einfuehren und dann sagen, wie sie definiert werden?}  Depending on
 their benefit, flux profiles are either classified as beneficial
 (benefit $\bvtot\cdot \vv>0$), {\wasteful} ($\bvtot\cdot \vv<0$), or
 {\futile} (zero benefit\footnote{By combining this condition with the
   stationarity condition, we obtain the criterion
   $\Nobj \, \vv = {\bvtot\trans \choose \Nint}\, \vv = 0$ for futile
   flux modes. Formally, the condition $\bvtot \cdot \vv=0$ can be
   seen as the mass balance for a hypothetical ``flux benefit
   compound'', so mathematical tools for stationary fluxes
   (e.g.~elementary modes) can be used to analyse futile stationary
   fluxes.}  $\bvtot\cdot \vv=0$). Futile or wasteful profiles are
 called {\nonbeneficial}.  In VBA, {\flow}s must be not only
 beneficial (providing a positive overall benefit), but also
 economical (providing a positive benefit in every reaction); this
 reflects a condition for enzyme-optimal states in kinetic models
 \cite{lieb:14a}.  To highlight the benefit from metabolic net
 production, we split the flux {{\gain}} vector into a
 sum\footnote{The way $\bvtot$ is split is non-unique and can be
   chosen by the modeller.  By setting $\bvtot = \bvdir$ and
   $\bpsi=0$, all flux {{\gain}}s are directly attributed to the
   reactions.  We may attribute some or all flux {{\gain}}s to the
   production of external metabolites (making the second term as
   sparse or small as possible).  Finally, by introducing virtual
   metabolites (with potentials $\wextj$ as proxies for flux
   {{\gain}}s), all flux gains can be formally attributed to external
   production. \co{kommt auch in cba kin}} \co{JA! bei b, dir statt
   int?}  $\bvtot = {\Next}\trans \bpsi + \bvdir$. The first term
 scores the consumption and production of external metabolites
 ($\Next$ is the part of the stoichiometric matrix referring to
 external metabolites), while the second term scores fluxes
 directly. The coefficients $\bpsij$  and $\bvdirl$, respectively,  are called production 
and  flux {{\gain}}s\footnote{In metabolic
   economics, the term ``gain'' generally means ``direct value''.}.
 In FBA, the flux benefit function is part of an optimality problem:
 it would either be maximised under constraints (e.g.~in classical FBA
 or FBA with molecular crowding) or be constrained to a given value
 while some flux cost is minimised (e.g.~in {\mfFBA}).  In VBA, the
 benefit function is used differently: we require that all economical
 {\flow}s $\vv$ are economical, which means that all active
 enzyme-catalysed (or ``enzymatic'') reactions must satisfy the value
 production balance \co{JA! FN: in diesem artikel ``reaction balance
   equation'' lieber ``value production balance'' nennen} \co{JA! auch
   in cba kin die reaction balance in point form so nennen!}
\begin{eqnarray}
 \label{eq:reactionbalanceeq}
\underbrace{[\Deltar \wtots + \bvdirs]}_{\gvtots}\, v = \zcostgeneric,
\end{eqnarray}
with economic potentials $\wtoti$, flux value $\gvtots$, and a
positive $\zcostgeneric >0$, typically representing enzyme
investments.  In other words: value production, representing an
enzyme's contribution to metabolic benefit, must be equal to the
enzyme investment and must therefore be positive. While the economic
potentials of external metabolites are predefined by the production
{\gain}s $\bpsij$, the economic potentials of internal metabolites
remain to be found.  If Eq.~(\ref{eq:reactionbalanceeq}) cannot be
satisfied, the {\flow} is uneconomical.  In contrast, if a flux
profile is economical, then in reactions without direct {\fluxgain}s
\co{FN? or more precisely, with negative flux gain in flux direction;
  oder oben sagem dass wnn nichts anderes erwaehnt wird, sich
  vorzeichen immer auf die flussrichtung beziehen!}
(i.e.~$\bvdirs=0$) the flux must lead from lower to higher
potentials. This means that economic potentials increase along
metabolic fluxes, reflecting the accumulating enzyme investments
embodied in the metabolites.   The value production balance
Eq.~(\ref{eq:reactionbalanceeq}) can also be written differently: after dividing it
by $v$ and defining the flux burden
$\hvs=z/v$, we obtain (again for optimal states) the balance
equation in  ``\fluxvalueform''
\begin{eqnarray}
 \label{eq:reactionbalanceeqFlux}
  \underbrace{\Deltar \wtots + \bvdirs}_{\gvtots} =  \hvs.
\end{eqnarray}
The
{\fluxburden} $\hvs$ describes a cost per flux (typically the cost of
a catalysing enzyme), and
 reaction flux $v$ and flux burden $\hvs$ must have strictly the
same signs (including zeros). \co{explain this ``strictly'' early on?} 
\co{JA! FN: The 
  flux value \co{wv} can also be expressed in terms of internal economic potentials \co{wint}:
  \begin{eqnarray}
    \gvtot = \Deltar \wtot + \bvdir = \Deltar \wint + \Deltar \wext  + \bvdir = \Deltar \wint + \bvtot.
      \end{eqnarray}
    }

    \co{Flux value balance: $\gvtot = \hvs$ (flux value = flux burden)

      Flux benefit balance
$\underbrace{\gvtotl \,v_{l}}_{\gvtotl\partialder} = \underbrace{\hvl\, v_{l}}_{\hvl\partialder > 0}$

(flux benefit = flux cost)
}

\begin{figure*}[t!]
\begin{center}
  \includegraphics[width=16.5cm]{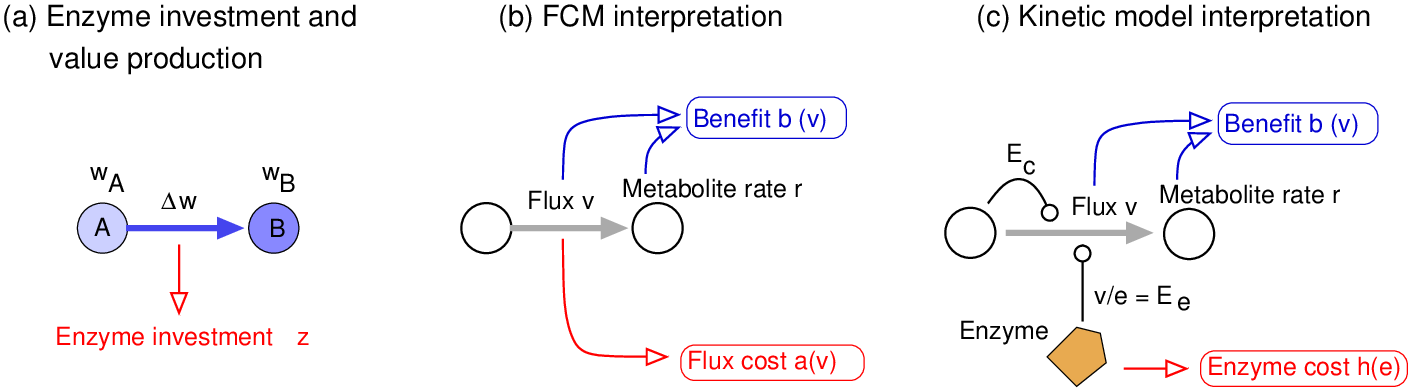}
\end{center}
\caption{\co{a: abstand vor z!} Two explanations of the value
  production balance Eq.~(\ref{eq:reactionbalanceeq}). (a) In each
  enzymatic reaction, the {\enzymeinvestment} $\zcostgenerics$ must be
  balanced with the enzymes' benefit contribution\co{uea?}, which can
  be expressed as a local value production $\gvtotsdot = \gvtots\,v$.
  This balance equation holds for active enzymatic reactions in
  optimal states and can have different interpretations depending on
  the underlying optimality problems considered.  (b) In an underlying
  Flux Cost Minimisation model, the {\enzymeinvestment} is given by
  $\zcostgeneric = \apointcostvs = \frac{\partial \fluxcost}{\partial
    v}\,v$ derived from a given flux cost function
  $\fluxcost(\vv)$. (c) In an underlying kinetic model, it is given by
  $\zcostgeneric = \husdot = \frac{\partial \hminus}{\partial
    \esymbol}\esymbol$ with an enzyme cost function
  $\hminus(\esymbol)$.}
 \label{fig:cbaFluxesInterpretations} 
 \end{figure*}

 \myparagraph{Economical metabolic {\flow}s and {\nonbeneficial}
   submodes} A key concept for describing enzyme-optimal states is the
 notion of economical {\flow}s. A {\flow} $\vv$ is called economical
 if there exists a {\fluxburden} vector $\hvv$ (with the same signs as
 in $\vv$, including all zeros) such that
 $(\hvv-\bvtot) \cdot \delta \vv =0$ holds for any stationary flux
 variation $\delta \vv$ (see SI
 \ref{sec:potentaislfromenzymepar}). \co{FN: explain the concept of a  ``test variation'',
   and that it must be valid?} This criterion, called
 \emph{{\summationcondition}} \cite{lieb:14a}, is a necessary
 condition for kinetic models in enzyme-optimal states
 \cite{lieb:18theory}. Economical {\flow}s satisfy two equivalent
 testable criteria.  First, economical {\flow}s (and no others) can
 satisfy the value production balance Eq.~(\ref{eq:reactionbalanceeq})
 with positive enzyme investments.  For positive {\enzymeinvestment}s
 $\zcostgeneric$ to exist, {\fluxburden}s $\hvs$ and fluxes $v$ must
 have strictly the same signs: from the equality
 $\mbox{sign}(\Deltar \wtots + \bvdirs) = \mbox{sign}(v)$, we obtain
 the principle of value production,
\begin{eqnarray}
 \label{eq:reactionbalanceeq2}
  [\Deltar \wtots + \bvdirs]\, v &>& 0.
\end{eqnarray}
The principle states that active enzymatic reactions produce value at
a positive rate.  \coout{einfaches bild dazu: enzyme labour flows in
  and adds to the value flow!}  Second, economical flux modes can be
equivalently characterised by flux motifs, which are defined as
follows.  If some active reactions in a flux profile $\vv$ would also
be able to carry, by themselves, a (stationary!) flux profile $\modevector$ with the
same flux directions, then $\modevector$ is called a submode of $\vv$,
and $\sign(\modevector)$ is called a flux motif. \co{JA! oder lieber
  buchstaben kappa fuer das motif selbst?  nochmal stefan schuster
  fragen, wie das bei elementarmoden mit vorfaktoren ist} If
$\modevector$ is futile or wasteful with respect to $\bvtot$, the flux
motif is also called futile or wasteful.  The submode criterion
states: a {\flow} is economical if (and only if) it is free of
{\nonbeneficial} motifs!  In fact, this is easy to see.  A flux
variation $\delta \vv$, given by \co{wenn submode dann flux motif;
  (nur signs!)  dann delta v not DEFINED BY, but FOLLOWS; oder lieber:
  signs of submode = motif; wie signs of a mode =pattern} a submode
with a positive prefactor, will increase some fluxes but cannot decrease them.
The higher enzyme demand\footnote{If metabolite concentrations
  are constant, higher fluxes require more enzyme. If metabolite
  concentrations can be adjusted, these adjustments are second-order
  effects and can be ignored.}  makes this variation costly. Since
{\valid} (i.e.~constraint-respecting) variations in an optimal state
must be fitness-neutral, the additional cost must be justified by a
positive benefit. This means: in optimal states, all submodes (and
therefore flux motifs) of our flux distributions must be beneficial.
Algorithms for checking this are described below.

\coout{NOETIG?: konkreter:
    .. actually equivalent to another optimisation approach, ..; flux
    min, weighted flux min, or flucx cost min. they all imply such a
    condition!} 

\begin{figure*}[t!]
{\small \begin{tabular}{lll}
(a) Beneficial, economical fluxes & (b) Futile fluxes& (c) Beneficial,  uneconomical fluxes\\[2mm]
    \includegraphics[width=4.8cm]{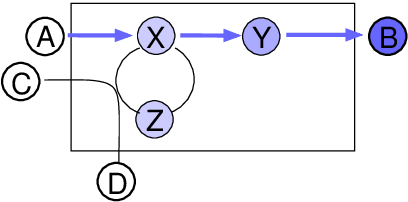}&
    \includegraphics[width=4.8cm]{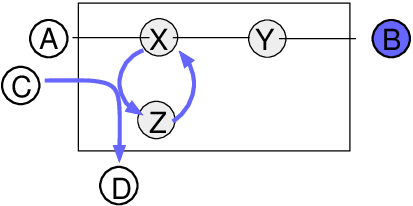}&
    \includegraphics[width=4.8cm]{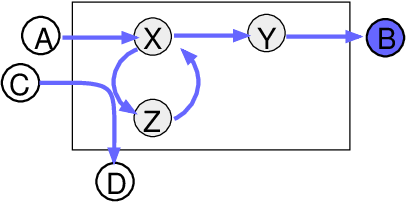}
  \end{tabular}
}
\caption{Economical flux modes, economic potentials, and futile
  cycles.  (a) Example pathway with external metabolite B as the
  valuable product, and its production rate as metabolic objective. In
  steady state, production and consumption of each internal metabolite
  (inside the box) must be balanced. If we describe the pathway by a
  kinetic model and optimise the enzyme levels for a minimal enzyme
  cost per flux, then the flux cycle between X and Z is suppressed in
  optimal states: we obtain a steady flux from A to B (unless the
  enzymes are too costly and all fluxes vanish).  Such a {\flow} is
  both beneficial (providing a positive overall benefit) and
  economical (i.e.~each active enzyme contributes to this benefit),
  and economic potentials can increase along the flux (shades of blue,
  where brighter colours represents lower values). All this holds for
  any kinetic models, and for any choice of kinetics.  (b) The flux
  cycle between X and Z provides no benefit. It is futile even if it
  may be thermodynamically feasible.  The flux profile in (c), the sum
  of the {\flow}s in (a) and (b), is beneficial but uneconomical
  because enzyme is wasted for driving the cycle. The fluxes in (b)
  and (c) are not compatible with any pattern of economic potentials,
  i.e.~no choice of potentials would allow all fluxes to go from lower
  to higher potentials.  This means that the flux mode is
  uneconomical.  The {\flow} in (c) contains a futile submode (the one
  shown in (b)) and so it  cannot be economical.}
    \label{fig:examplefutilecycle}
\end{figure*}

  \myparagraph{The two criteria for economical fluxes are equivalent}
  Our criteria for economical fluxes -- the consistence with economic
  potentials and the absence of \nonbeneficial\ submodes -- are
  closely related. In Figure \ref{fig:examplefutilecycle}, the {\flow}
  in (a) is economical, as proven by the fact that economic potentials
  can increase along the flux.  In contrast, to make the flux cycle in
  (b) economical, economic potentials would have to increase in a
  circle, which is logically impossible. Any {\flow}s with this motif
  are uneconomical (for example, the one in (c)).  A major advantage
  of the submode criterion is that it can be tested locally: we can
  discard an entire {\flow} based on a local pattern, even without
  knowing the entire network!

\myparagraph{Economic potentials and flux cycle in an example pathway}
Figure \ref{fig:examplefutilecycle} shows how economic potentials (of
metabolites) are related to futile motifs (in flux profiles) and to
economical enzyme usage. As a simple example model, we consider a
kinetic model with a production objective and enzyme levels to be
optimised.  While the network structure allows for a cycle flux, this
cycle is never active in enzyme-optimal states, no matter which rate
laws or enzyme cost functions are assumed.  The reason is simple: in
optimal states, the cost of an active enzyme must be balanced by a
positive benefit (or in the language of {\MCA}): by a positive control
on the {\metabolicobjective} function. In the flux cycle, this would not
be possible.  \co{kam schon oben:} As we shall see below, this is in
line with a general principle: if a {\flow} is economical, it is
possible to find economic potentials such that all fluxes run from
lower to higher potentials.  We can see this in Figure
\ref{fig:examplefutilecycle}: in (a), the economic potentials increase
along the {\flow}, as required for an economical flux mode.  In (b),
potentials would have to increase in a cycle, which is logically
impossible: {\flow}s with this cycle are uneconomical, even the
beneficial profile in (c).  Mathematically, this criterion resembles
an important thermodynamic criterion: fluxes must run from higher to
lower chemical potentials (to dissipate energy in every reaction)
\cite{qibe:05}. Importantly, the two (economic and thermodynamic)
constraints are logically independent. If metabolite C has a high
chemical potential, the cycle in Figure \ref{fig:examplefutilecycle}
(c) will be thermodynamically feasible, but economically
futile. \todo{Other flux profiles may also be beneficial, but
  thermodynamically impossible (under physiological concentrations),
  e.g.~one that produces ATP from ADP and phosphate without any other
  conversions.}  Interestingly, FBA with flux minimisation, too,
suppresses futile cycles. This is no coincidence: in models with a
simple production objective, economic and thermodynamic feasibility
becomes interdependent the two optimality principles -- enzyme
optimisation in kinetic models and flux minimisation in FBA -- lead to
the same flux solutions \cite{lieb:18lagrange}.

\begin{figure*}[t!]
 \begin{center}
 \includegraphics[width=15.5cm]{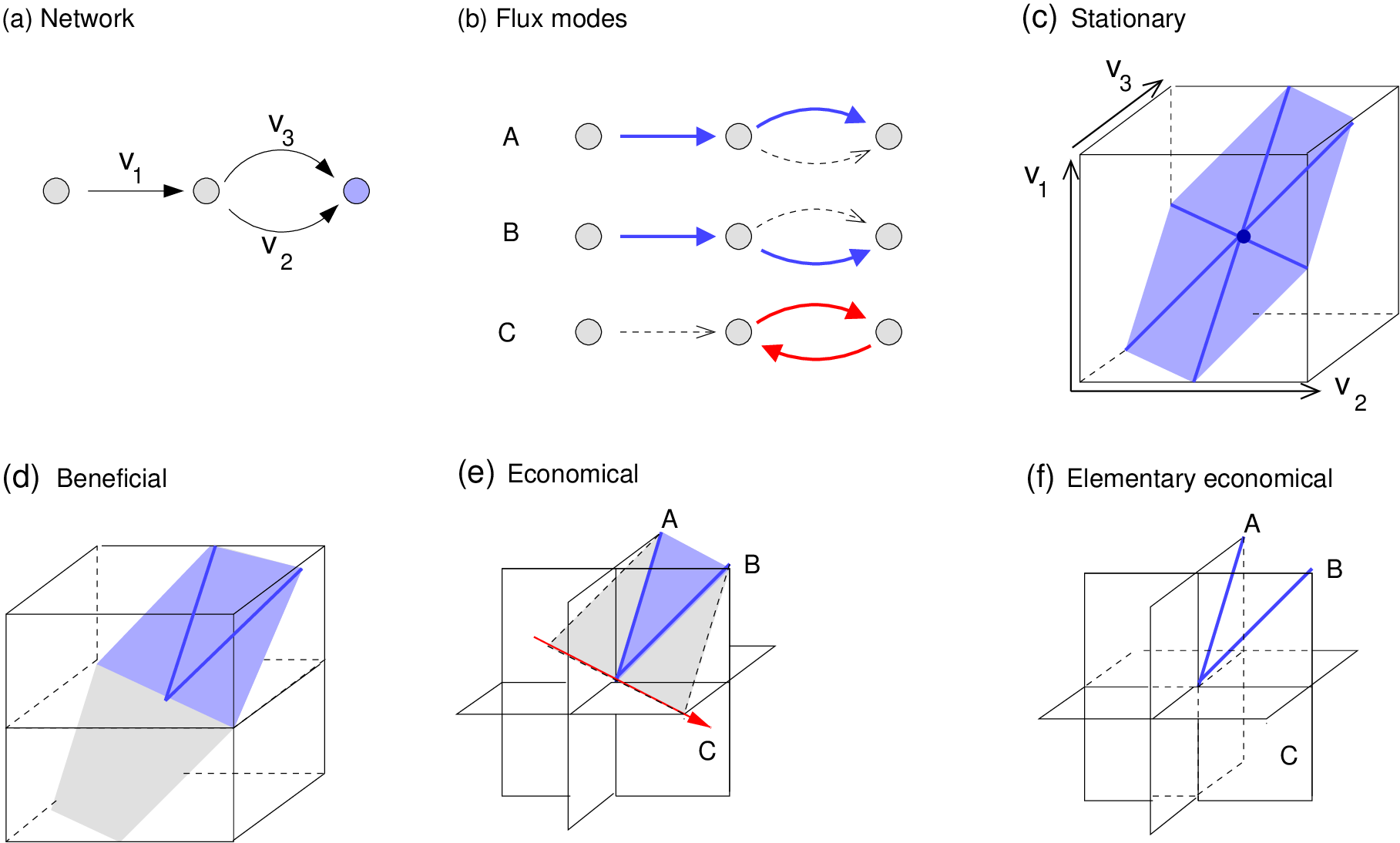}
 \caption{Metabolic {\flow}s.  (a) Example pathway with a production
   objective (production of blue compound) and flux bounds
   $-1<v_{l}<1$ (arbitrary units). (b) Elementary beneficial (blue)
   and {\nonbeneficial} {\flow}s (red).  (c) Flux profiles as points
   in flux space. By definition, flux profiles are stationary and
   therefore lie in a plane. Each reaction can have a positive,
   negative, or zero flux.  Out of the $3^{3}=27$ possible
   {\fluxpattern}s (i.e.~segments of flux space), only 13 can be
   realised by stationary fluxes (six patterns correspond to
   triangles, six to lines, and the central dot represents
   $\vv=[0,\,0,\,0]\trans$). The other flux space segments are not
   intersected by the plane of stationary {\flow}s. (d) A linear FBA
   objective (here, the rate of reaction 1) defines a flux {{\gain}}
   vector $\bvdir = [1,\, 0,\, 0]\trans$ and thereby a set of
   beneficial flux profiles.  Five of the {\fluxpattern}s (with
   $v_{1}>0$) are beneficial (a necessary condition for being
   economical), while all others can already be excluded. (e) Only
   three beneficial patterns (blue triangle and edges A and B) are
   sign-orthogonal on the futile cycle (red arrow) and thus economical
   (for an explanation of sign orthogonality see SI section
   \ref{sec:metabolicmodels}).  (f) Elementary economical flux
   modes. In FCM, optimal {\flow}s are usually corners of the flux
   polytope, indicating that alternative pathways should be used
   separately (polytope corners) and not as linear combinations
   (internal polytope points). For details see SI section
   \ref{sec:SIcombinationOrAlternative}.}
 \label{fig:fbacomparison2}
 \end{center}
\end{figure*}

\myparagraph{Economical fluxes in flux space} If flux profiles are
depicted as points in flux space, the stationary {\flow}s (nullvectors
of the stoichiometric matrix $\Nint$) form a linear subspace.  By
applying upper and lower flux bounds, we obtain the FBA flux polytope
(see Figure \ref{fig:fbacomparison2}). Thermodynamic constraints can
be used to restrict the solutions: by excluding {\flow}s containing
thermodynamic cycles, we obtain a collection of convex polytopes, each
representing a thermodynamically feasible flux pattern.  Economic
constraints have similar effects: {\nonbeneficial} flux patterns are
excluded and {\flow}s are restricted to feasible, i.e.~economical
segments of flux space (see Figure \ref{fig:fbacomparison2}).

\myparagraph{The interpretation of economic variable and value
  production balance depends on the choice of an underlying optimality
  problem} To understand the meaning of economic values and value
production, let us step back and consider flux profiles that are
solutions of an optimality problem, for example, a minimisation of
flux or enzyme costs. The optimality conditions, as mathematical
equations, contain fitness derivative and auxiliary variables called
shadow values that arise from constraints \cite{lieb:18lagrange}. For
example, mass balance constraints in a model lead to optimality
conditions that resemble our valure production equation while the
shadow values in these equations play the role of economic potentials.
Importantly, this holds for an entire class of (constraint-based or
kinetics-based) problems: we always obtain the same balance equations!
The economic potentials may vary, depending on kinetic constants,
enzyme cost weights, or the flux cost function assumed, but we know
that economic potentials exist and must be consistent with the
fluxes. Conversely, if no compatible economic potentials exist for a
given flux mode, this flux mode is uneconomical and cannot be the
solution of an underlying optimality problem of this sort. Since the
condition holds generally, we can use it in flux analysis, even
without specifying the underlying problem in detail -- we just need to
assume such a problem exists!  This also means: if we postulate a
value balance, economic potentials or enzyme investments may mean
different things depending on the underlying optimality problem
assumed\footnote{A similar logic exists in thermodynamics: if we apply
  a sign constraint $\Deltar \mu_{l}\,v_{l}<0$ for fluxes, this
  constraint stems from an underlying variational problem for Gibbs
  free energy dissipation (details see below). If these flux
  constraints are fulfilled, we know that there must be a
  (thermodynamically consistent) kinetic model that realises our flux
  distribution (even if we don't know this model in
  detail). Restrictions on (or estimates of) chemical potentials can
  make these unknown models more realistic.}. Assuming that such an
underlying optimality problem exists, we can search for economic
potentials that correspond to plausible models (that is, models with
plausible values of the kinetic constants and enzyme costs in line
with protein data).

\myparagraph{Interpretation of economic potentials and enzyme
  investment by flux cost minimisation or kinetic models} For the
underlying optimality problems, we consider two possibilities (for
other possibilities, see \cite{lieb:18lagrange}). One
possible interpretation is based on Flux Cost Minimisation (FCM) with
a flux cost function $\fluxcost(\vv)$. For plausibility reasons, flux
costs should increase with the absolute flux, so
$\zcostgeneric = \apointcostvs = \frac{\partial \fluxcost}{\partial
  v}\,v>0$. The shadow values in such models have been studied
\cite{rems:13}, compared to chemical potentials in thermodynamics
\cite{wajo:07}, and interpreted as economic potentials
\cite{lieb:18lagrange}.  The optimality condition of our FCM problem
yields a value production balance equation of the form (\ref{eq:reactionbalanceeqFlux}),
with a flux burden $a_{\rm v}=\frac{\partial \fluxcost}{\partial v}$ and
thus an investment
$\zcostgeneric = \frac{\partial \fluxcost}{\partial v}\,v$ (which, as
we assumed, will be positive). Any flux profile that satisfies this
balance equation will also be the solution of some FCM problem. In our
second interpretation, we consider a different underlying optimality
problem: a kinetic model with optimal enzyme levels $\esymbol$ and an
increasing cost function $\hminus(\esymbolv)$.  In this
interpretation, the  investment $\zcostgeneric$ is actually and enzyme investment, defined as
the {\enzymecost}
$\husdot = \frac{\partial \hminus}{\partial \esymbol}\,\esymbol$,
which is positive (or zero for inactive reactions), see appendix
\ref{sec:justification}.  For example, with  the size-weighted protein
concentration $\hminus(\esymbolv)= \sum m_{l} \,\esymbol_{l}$ (with
protein sizes $m_{l}$) as our enzyme cost, a reaction has an
investment $\zcostgenerics_{l}= m_{l} \,\esymbol_{l}$.  Below,
$\zcostgeneric$ will usually be called ``{\enzymeinvestment}''.

\myparagraph{Flux burden, flux value and economic potential
  difference} Which flux profiles allow cells to grow fast depends
largely on the flux burdens, that is, the enzyme cost per reaction
flux in every reaction. If Eqs (\ref{eq:reactionbalanceeq}) and
(\ref{eq:reactionbalanceeqFlux}) are based on an underlying FCM
problem, the flux burdens are defined by the derivatives
$\frac{\partial \fluxcost}{\partial v}$. If the flux cost function
$\fluxcost(\vv)$ represents enzymatic costs, \co{ref FCM} the
burdens describe the enzyme cost per flux. In kinetic models, we
define them differently: we consider enzymatic rate laws
$v=\esymbol\,\ratelaw(\cv)$ with enzyme efficiencies
$\ratelaw=v/\esymbol$ (called catalytic rates or apparent $\kcat$
values) which depend on metabolite concentrations. Now {\fluxburden}s
can be defined as
$\hvs = \frac{\zcostgenerics}{v} = \frac{\hus\,\esymbol}{v}$, i.e.~the
enzyme investment $\zcostgeneric$ per flux, or the enzyme {\price}
$\hus=\partial \hminus/\partial \esymbol$ divided by the catalytic
rate $v/\esymbol$.  In both modelling frameworks, economic potentials
are the shadow values related to mass-balance constraints
\cite{lieb:18lagrange}. The term in brackets in
Eq.~(\ref{eq:reactionbalanceeq}), called \emph{flux {{\myvalue}}}
$\gvtots$, is the sum of a flux {\gain} $\bvdirs$ and an indirect
flux {\myvalue} $\Deltar \wtots$. This second term represents benefits
anywhere in the network to which our reaction contributes indirectly
by supporting the steady state. Importantly, this term can be written
as a difference of product and substrate values\footnote{The symbol $\Deltar x_l$ \coout{doch
    einfaches delta, und strich nur wenn flux gain dabei ist?}
  denotes the difference of a metabolite-specific variable $x_i$ along
  reaction $l$.}  $\Deltar \wtotl = \sum_{i} n_{il}\, \wtoti$ in the
reaction. But how can we find the economic potentials?  Flux
{{\gain}}s $\bvdirs$ and external economic potentials $\wexts$ follow
from our benefit function, but the internal potentials $\wints$ and
{\enzymeinvestment}s $\zcostgenerics$ remain to be found. In any case,
the value production balance provides an important condition: since
{\enzymeinvestment}s $\zcostgenerics$ must be positive, fluxes $v$ and
flux {{\myvalue}}s $\Deltar \wtots + \bvdirs$ must have the same  signs:
therefore, known flux directions  constrain the economic
potentials $\wtots$.

\myparagraph{Flux \myvalue\ and catalytic rate} The economic
potentials are not constant but vary with the metabolic
state. Changing enzyme efficiencies will lead to changes in economic
potentials: to see this, let us think about reaction kinetics and
thermodynamics.  First, according to
Eq.~(\ref{eq:reactionbalanceeqFlux}), the economic potentials are
closely related to the flux burdens $\hvl$.  In kinetic models, with
{\enzymeinvestment}s given by $\zcostgenerics=\hus\,\us$ (with enzyme
{\price} $\hus$), a reaction {\fluxburden} reads
\begin{eqnarray}
 \label{eq:fluxBrudenEfficiency}
\hvs = \frac{\hus\,\esymbol}{v} =  \frac{\hus}{\ratelaw} =\hus\,\tau
\end{eqnarray}
with the catalytic rate $\ratelaw=v/\us$ and enzyme slowness \co{auch
  in tabelle in CBA kin!}  $\tau=1/\ratelaw = e/v$. At fixed enzyme
{\price}s $\hus$, this means: as an enzyme becomes less efficient
(e.g.~at lower substrate levels or close to chemical equilibrium),
more enzyme is needed to sustain the flux, and the flux {\burden}
$\hvs$ increases.  The higher investments (per flux) become embodied
in downstream metabolites, increasing their economic potentials.
How does thermodynamics come into play?
The metabolite concentrations determine the chemical potentials
$\mu_{i}=\mu_{i}^{\circ}+RT\, \ln c_{i}$ (assuming activity
coefficients of 1) and thermodynamic forces
$\theta = -\Deltar \mu/RT$.  The force in a reaction determines the
ratio of microscopic one-way fluxes \cite{beqi:07} and affects the net
catalytic rate. Reversible rate laws can be written
as\footnote{Formula (\ref{eq:ReversibleRateLaw}) holds for positive
  fluxes. The general formula reads
  $v=e\, \kcat^{\rm
    flux}\,\sign(\theta)\,(1-\exp^{-|\theta|})\,\eta^{\rm kin}(\cv)$
  where $\kcat^{\rm flux}$ is the $\kcat$ value in flux direction.}
\begin{eqnarray}
  \label{eq:ReversibleRateLaw}
  v = \esymbol\,\kcat\,(1-\exp^{-\theta})\,\eta^{\rm kin}(\cv)
\end{eqnarray}
with a reversibility factor
$\eta^{\rm rev}(\theta)=(1-\exp^{-\theta})$ and a kinetic factor
$\eta^{\rm kin}(\cv)$ that depends on the rate law \cite{nflb:13}. \co{JA! bei factorised rate law sagen, dass
  flussrichtung bekannt sein muss! (vorteil von modular rate laws: sie
  muss es nicht sein)}
According to Eq.~(\ref{eq:ReversibleRateLaw}), the flux {\burden}
depends on reaction kinetics and thermodynamics:
\begin{eqnarray}
\label{eq:fluxBurdenEfficiencyDetailed}
  \hvs =\frac{\hus}{\kcat\,(1-\exp^{-\theta})\,\eta^{\rm
      kin}(\cv)}.
\end{eqnarray}
By inserting this formula into Eq.~(\ref{eq:reactionbalanceeqFlux}),
we can see how flux {\myvalue}s and economic potentials $\wtots$  depend
on enzyme price $\hus$, turnover rate $\kcat$, driving force $\theta$,
and kinetic efficiency factor $\eta^{\rm kin}$. For a related
question, the opportunity cost of one-way backward fluxes, see SI
\ref{sec:thermoCostTerms}.

\co{JA! AUCH CBA LABOUR! Appendix: tabelle mit allen
  groessen, symbolen und kurzer erklaerung; ref wo? v, wr, wv, delta
  wv, z = hudot, av ..} 

\begin{figure*}[t!]
\begin{center}
\parbox{6.6cm}{\includegraphics[width=6.2cm]{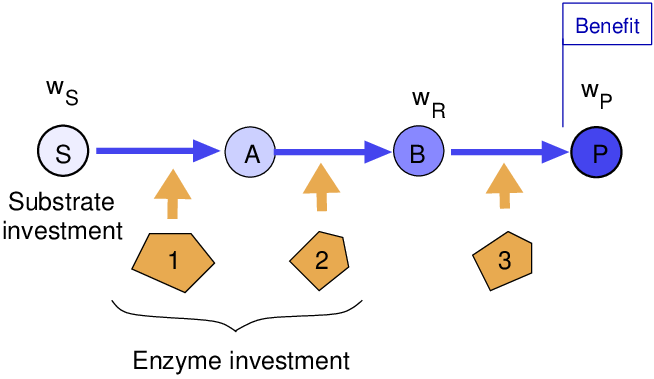}}
\parbox{9.4cm}{
  \caption{Economic potentials and enzyme investments.  A metabolite's
    economic potential reflects its use value, i.e.~its effect on the
    metabolic objective. In optimal states (defined by some underlying
    optimality problem), this use value must be equal to the embodied
    investments, divided by the production rate. \co{``gift of
      nature''} \co{say: the embodied value of the external metabolite
      S is the (predefined) potential itself. then for A, add the
      enzyme investment per flux WA = Ws + z/v. and so on ..} \co{Mit
      wert $> 0$ anfangen und das erklaeren:
      $wr = ws + 1/v \sum_{l} z_{l}$} In a linear pathway with a
    ``free'' pathway substrate, the value inflow
    $v_{\rm prod_{i}}\,\wtoti$ into a metabolite (i.e.~its total
    production rate, multiplied by the economic potential), is given
    by the substrate and {\enzymeinvestment}s upstream.  \co{JA nice!
      two types of logic; local enzyme + local substrate's value; or
      global logic: all previous enzymes, and initial substrate
      value!} \co{JA! evtl zeigen (zweites bild??): ineffizienter
      transporter (zb kleine substratkonzentration): hoehere
      potentiale!  (auch wieder value flow zeigen)} \co{JA!  noch eine
      reaktion mit kofaktoren zeigen?} \co{die abbildung is nicht
      wirklich hilfrecih .. mehr bilder, um verschiedene kinetische
      effekte (zB av) zu zeigen?  auch gleich pathway cost usw?}  }
    \label{fig:productionConsumption}}
\end{center}
\end{figure*}

\myparagraph{Economic potentials, embodied investment, and embodied
  value}  \co{JA! ueberall wo ``embodied'' zum ersten
  mal auftaucht (auch andere artikel); die def von ``embodied value''
  (derivative) und ``embodied investment'' (point derivative)
  klarmachen; jeweilige definitionen!  welche rolle spielt der fluss?
  welche probleme gibt es bei flux modes die keine einfache linear
  kette bilden? // auch klarmachen, dass investments durchaus auch
  substrat-investments umfassen koennen! das die investments in der
  kette ``weitergereicht'' werden.}  A metabolite's economic potential
$\wtoti$ describes the metabolites' use value, i.e.~its indirect
effect on the metabolic objective. Together with the metabolite's
total consumption rate $\rate_{i}$, it determines a value outflow
$\rate_{i}\,\wtoti$. In optimal states, this outflow must be equal to
a value inflow, descibring the nutrient and {\enzymeinvestment}s
embodied in the metabolite. We can see this in Figure
\ref{fig:productionConsumption}.  \co{Erst gut die beiden begriffe
  erklaeren! Def: $\wtots^{O}=\wtots \, v_{\rm prod}$ ``embodied
  investment'' // $wtots$ ``embodied value'' (im bsp lin kette
  beziehung von fluss und prodrate erklaeren; in diesem beispiel sind
  sie gleich! (sonst summe ueber all produzierenden fluesse)))} In a
linear pathway without flux {\gain}s, the enzyme investments
accumulate along the flux, and the economic potentials increase by
$\Deltar \wtots=\zcostgeneric/v$ in each reaction. The initial
substrate (with economic potential $w_{s}$) provides a substrate
investment $v\,w_{s}$; each enzyme provides an enzyme investment
$\hus\,\esymbol = \hvs\, v$.  Investments accumulate along the
pathway, leading to an embodied investment
$[w_{s} + \sum_{l=1}^{j}\hvl]\,v$ in each metabolite $j$,
corresponding to an embodied value
$w_{s} + \sum_{l=1}^{j}\hvl =w_{r_{j}}$.  Therefore, the economic
potential of a metabolite X,
$w_{\rm X} = w_{\rm s} + \frac{z_{s \rightarrow x}}{v}$, is given by
the economic potential of the pathway substrate S, plus all enzyme
investments between S and X, divided by the flux.

The ``steps'' in potential depend on the enzyme {\price}s (e.g.~their
molecular mass) and inversely on enzyme efficiencies. \co{auf
  abschnitt davor zurueckkommen .. embodied value depends on kinetics
  + thermodynamics along the chain, minimum von hu, kcat, usw
  .. vergroessert durch therm und st efficiencies .. BSP import of
  metabolite LEADS TO higher embodied value!} \co{FN: zur abschaetzung
  der embodied values die formeln von oben; am einfachsten mit
  (minimal) enzyme price und kcat; dann (wenn bekannt) korrekturen mit
  driving force oder concentrations.}  Inefficient enzymes
(e.g.~enzymes with low $\kcat$ values, substrate concentrations, or
driving forces) are costly, and varying costs (per flux) entail
varying embodied values in downstream metabolites!  For instance, as
the extracellular glucose level decreases, glucose transporters
becomes less efficient and more transporter molecules are required to
obtain a desired flux: this increases the embodied values of
intracellular metabolites.

\co{SORT AND EDIT the paragraph} \co{alles ausfuehrlicher: vor allem
  substrate investments and cofactor investments; and say clearly that
  each (positive) flux gain REDUCES the increase in economic potential
  (the embodied values in following metabolites)} Aside from
transporter and enzyme costs, embodied values may also reflect other
terms in the balance equations: pathway substrates may add to embodied
values downstream, while cofactors usage may add or absorb value
within the pathway. \co{FN briefly mention role of cofactors (also
  describe by flux gains!)  wasser weglassen und nur in FN sagen, dass
  man es mitbetrachten koennte? dass sein potential ungefaehr null
  gesetzt werden kann, weil die konzentration im vergleich zu anderen
  stoffen so hoch ist und biochem reakt die konznetration kaum
  aendern?} Along the pathway, the embodied value increases,
reflecting enzyme investments. If a reaction provides a direct
benefit, this flux {\gain} decreases the downstream embodied values,
and if cofactor pairs are involved in reactions, their economic
potentials are taken into account; due to such terms, economic
potentials may sometimes decrease along the flux\footnote{If
  potentials decrease along the flux, this may be due to non-enzymatic
  reactions or non-optimal enzyme levels.}.

\section{Thermodynamic and economic constraints}

\myparagraph{\ \\Thermodynamically feasible flux directions}
\co{thermodyn cycle: lieber ``loop''?}  Thermodynamic laws link the
metabolic flux directions to chemical potentials via the thermodynamic
driving forces (negative Gibbs free energies of reaction, in units of
RT) \cite{belq:02,bbcq:04,faqb:05}. These laws can be used to
constrain fluxes, to exclude infeasible flux cycles
\cite{sclp:11,nolm:12}, and to constrain metabolite concentrations
\cite{hjbh:06,kuph:06,hebh:07,hohh:07,fmsy:11}. They also play a
central role in the economy of the cell.  \co{SORT; definitionen
  klarer aufbauen} All metabolic reactions must be exergonic,
i.e.~dissipate Gibbs free energy.  \co{JA! in CBA opt: strong (linear)
  effect on flux for theta zwischen 0 und 1, little effect for theta
  bigger than 1} A reaction's energy dissipation rate (in units of
$RT$) is given by\footnote{If Gibbs free energy is dissipated in the
  form of heat and if other forms of energy can be neglected, $\sigma$
  is just the entropy (in units of $R$) production per volume and
  time.}  $\sigma_{l} = \theta_{l}\,v_{l}$ where
$\theta_{l}=-\Deltar \mu_{l}/RT$ is the thermodynamic driving force.
In a physically feasible model, every active reaction must dissipate
Gibbs free energy: the condition $\sigma_{l}>0$ implies that
$\mbox{sign}(v_{l}) = \mbox{sign}(\theta_{l})$ (whenever
$v_{l} \ne 0$), or in short $\vv \sqsubseteq \thetav$ (``the {\flow}
$\vv$ is conformal to the force vector $\thetav$'')
\cite{belq:02,bbcq:04,faqb:05,hjbh:06,hohh:07,fmsy:11}.  for our
models, we obtain a sign relation between chemical potentials and
fluxes\footnote{The inequality (\ref{eq:thermocondition}) follows 
  from thermodynamic laws.  In a well-mixed chemical solution at given
  pressure and temperature, each metabolite species (with mole number
  $n_{i}$) contributes an amount $G_{i}$ to the system's total Gibbs free energy $G$,
  and the contribution is given by $G_{i}=n_{i}\,\mu_{i}$, where
  $\mu_{i} = \partial G/\partial n_{i}$ (in kJ/mol) is called chemical
  potential.  Reaction fluxes must dissipate Gibbs free energy.
  For a positive dissipation rate $v_l\,\Deltar \mu_l$, a flux $v_{l}$
  must have the same sign as the thermodynamic force
  $\theta_{l} = -\Deltar \mu_{l}/RT$, that is, it must lead from higher
  to lower chemical potentials. Without the condition $v_{l}\ne 0$ in
  Eq.~\ref{eq:thermocondition}, we obtain the strong sign condition,
  requiring that non-zero  forces evoke a flux in
  the same direction (see SI \ref{sec:economicandthermodynamic}).
  \co{This implies that  reactions cannot be fully suppressed kinetically (e.g.~by
    enzyme inhibition), but can vanish only in chemical
    equilibrium. This means that Eq.~(\ref{eq:thermocondition}) must
    hold in all reactions (active and inactive ones).} \co{strong sign
    cond kommt auch nochmal unten nach kirchhoff's laws}} \co{hier
  schon direct term?}
\begin{eqnarray}
\label{eq:thermocondition}
-\Deltar \mu_{l} \, v_{l}\, >0,
\end{eqnarray}
which must hold for all active reactions.  \co{(``weak sign
  constraint'')} The chemical potentials are further related to
metabolite concentrations: assuming constant activity coefficients,
they are given by $\mu_{i} = \mu^{\circ}_{i} + R T \ln c_{i}$ and
therefore linear in $\ln \cv$.  Thus, thermodynamically feasible flux
mode $\vv$ requires a metabolite profile $\ln \cv$ for which
$-\Deltar \mu_{l}$ and $v_{l}$ have equal signs (in all active
reactions). \co{WO? say weak sign constraint. any non-zero flux
  requires a driving force in the same direction}

\myparagraph{Thermodynamically feasible flux profiles and flux cycles}
In kinetic models with consistent reversible rate
laws\footnote{``Consistent'' means that the kinetic constants must
  satisfy Haldane relationships \co{REF} and the equilibrium constants
  satisfy Wegscheider conditions.\co{REF}}, energy dissipation and
condition (\ref{eq:thermocondition}) are automatically satisfied. This
holds for all rate laws of the form Eq.~(\ref{eq:ReversibleRateLaw}).
In flux analysis, in contrast, the thermodynamic laws must be
guaranteed by imposing condition (\ref{eq:thermocondition}) an extra
constraint, which then allows us to detect and exclude infeasible flux
modes.  Given a {\flow} $\vv$, how can we see whether there exist
consistent chemical potentials, satisfying condition
(\ref{eq:thermocondition})?  At least it is easy to exclude this: in a
closed loop of unimolecular reactions (without any other metabolites
or cofactors being produced or consumed), a cycle flux cannot lead
from higher to lower potentials everywhere.  We can generalise this
logic to more general types of cycle fluxes.  A flux distribution
dissipates energy at a total rate
$r_{Q}=RT \,\sigma = -\Deltar \muv\cdot \vv = -\muv\trans\,\Ntot\,\vv =
-\muv \cdot \prodratev$, with the vector $\muv$ of chemical
potentials, the vector $\Deltar \muv = {\Ntot}\trans \,\muv$ of chemical
potential differences, and the vector $\prodratev = \Ntot\,\vv$ of
metabolite net rates (note that $\Ntot$ contains both internal and
external metabolites).  Cyclic flux profiles, without any net
conversion of metabolites ($\prodratev = \Ntot\,\vv=0$), do not
dissipate any Gibbs free energy and are thermodynamically infeasible
(unless there is some extra ``hidden'' energy dissipation,
e.g.~through compounds ignored in the model).  In general, this holds
for any ``flux cycles'', defined as nullvectors of the stoichiometric
matrix $\Ntot$ (including internal and external metabolites). If a
{\flow} $\vv$ contains a cyclic submode $\modevectorcyc$ (i.e.~if
$\vv$ and $\modevectorcyc$ share all flux directions in their active
reactions), the {\flow} is infeasible. Hence, to obtain a feasible
flux mode, such cycles must be excluded
\cite{qibe:05,sclp:11,pjmd:17}.  \coout{letzte referenz: sabines
  artikel (thermodyn feasible EFMs); konsequenzen erwaehnen?}

\co{zu speziell?  move this to SI and mention here in one sentence
  only?}  The formula $\theta_{l}=-\Deltar \mu_{l}/RT$ assumes that
our model actually describes all reactants. Sometimes, models are
simplified by omitting some cofactors, protons, or water, and to keep
the model correct, we use a formula
$\thetav = -\Deltar \muv/RT + \thetav^{\rm dir}$ with an extra force
term $\thetav^{\rm dir}$ that represents contributions of the
neglected compounds\footnote{Note that this extra term can enable
  cycle fluxes that would otherwise be impossible. This makes sense:
  for example, if the extra term stands for an omitted cofactor pair,
  in reality these cofactors would be able to drive the
  cycle.}. Similarly, a term $\theta^{\rm dir}$ in a model can always
be replaced by the chemical potential of a virtual reaction
product. This means that we can get rid of such terms, at least
formally, by adding extra products to our reaction sum formulae.

\co{WO?}  When looking for possible cell states, we need to
distinguish between what is thermodynamically possible \emph{in
  principle} (allowing for arbitrary metabolite concentrations) and
what is thermophysiologically possible (with metabolite concentrations
within physiological bounds). For example, in theory any thermodynamic
reaction can be reversed (by keeping the substrate concentration at
zero or making the product concentration large enough). However, in
real cells some reactions can never be reversed because the required
concentration ratios will never exist in a living cell (for example,
because of thermodynamic requirements in other reactions). To define
thermophysiologically feasible driving forces and fluxes , we need to
consider physiological concentration ranges (as, for example, in the
MDF method \cite{nbfr:14}.

\co{SORT the following three paragraphs and the material below them}

\myparagraph{Reasons for the similarity between economic and
  thermodynamic constraints} 
Thermodynamic and economic constraints have very different
justifications. While thermodynamic laws come from physics and must hold  in
any state of the cell, economic laws are based on the  extra assumption of an
optimal usage of enzyme. 
Given that thermodynamic and economic
constraints are different in nature, why do they look so similar?
Formally, both constraints can be derived from one
variational principle, the principle of Flux Cost Minimisation (FCM)
\cite{lieb:18fcm}.  In FCM, a {\flow} must be stationary and must
minimise a cost function at a predefined benefit.  Typically, flux
cost functions represent enzyme {{\investment}}s or substrate
consumption, while  flux benefits represent biomass production
or other production objectives.  But we may  use the same
formalism to describe thermodynamics: treating  negative energy dissipation
as a flux cost, \co{at a fixed external energy dissipation ?}, \co{AND what is the benefit?}
\co{check: woher kommt das minuszeichen?  negative, because
  dissipation is maximised?}  FCM yields the thermodynamic flux
condition Eq.~(\ref{eq:thermocondition}) as an optimality condition
(see SI \ref{sec:economicandthermodynamic}), with the negative
chemical potentials as ``economic potentials''.

\myparagraph{Economic and thermodynamic constraints are similar} The
economic and thermodynamic constraints on fluxes, Eqs
(\ref{eq:reactionbalanceeq2}) and (\ref{eq:thermocondition}), in the
  version with direct (economic or thermodynamic) force terms\footnote{A direct force term in thermodynamics resembles
  -- at least formally -- the flux gain in the reaction balance
  equation. On the contrary, in the economic balance equations, we can
  always replace all flux gain terms by economic potentials of
  hypothetical metabolites, introduced just for this purpose: then we
  obtain the general simple relationship $\gvtotl = \Deltar \wintl$.},
are mathematically similar. While thermodynamics requires a positive energy
dissipation $r_{\rm Q}=[-\Deltar \mu_{l}+\theta^{\rm dir}]\,v_{l}$,
value production Eq.~(\ref{eq:reactionbalanceeq2}) requires a positive
flux benefit $[\Deltar \wintl + \bvtotl] \,v_{l}$.  \co{The main
  difference, the plus sign with $\Deltar \wtot$) is just a matter of
  convention: it stems from the definition of economic values and
  makes them increase along fluxes (whereas chemical potentials
  decrease along fluxes).}  \co{clear distinction between cycle fluxes
  and {\nonbeneficial} modes!  (T-CYCLES and E-Cycles)} 

The similarity between these conditions leads to various other
similarities between economic and thermodynamic constraints. In both
cases, the flux directions depend on (chemical or economic) potential
differences (plus, possibly, extra direct terms). Endergonic submodes
in thermodynamics (which ``absorb'' Gibbs free energy) and {\wasteful}
submodes in enzyme economy (which ``absorb'' benefit) play a similar
role. Also, in both cases, a condition for feasible flux profiles is
that there exists a consistent choice of (chemical or economic)
potentials.  Likewise, uneconomical submodes can be found by
considering {\futile} flux variations: if a {\flow} contains a futile
submode it is uneconomical. \co{math. zusammenhaenge besser
  erklaeren:} And geometrically, when {\flow}s are seen as points in
flux space, economic constraints, just like thermodynamic constraints,
exclude some segments in the solution space (see Figure
\ref{fig:fbacomparison2}).

\co{AB HIER alles analog zu thermo beschreiben; dann unten erst wieder beide direkt vergleichen}

\co{role of flux gain (analog zu thetaDIR)}

\myparagraph{Economic and thermodynamic cycles} Du eto their similar
structure, formulae from thermodynamics can be transferred to
metabolic economics.  This concerns, for example, the role of cycles.
Both types of conditions exclude flux profiles that would require
potentials to increase in a circle \cite{pfbp:02,nolm:12}, which can
be tested by searching for cyclic (in thermodynamics) or {\futile}
flux motifs (in metabolic economics). All flux profiles with such
motifs can be excluded. \co{below, note that cycles are defined with
  respect to a matrix, and say what these matrices are} \todo{To
  formally define these cycles, we first define infeasible submodes:
  in thermodynamics, infeasible submodes are elementary cyclic flux
  modes which represent thermodynamically infeasible flux variations.
  \co{rest des abschnitts einfacher!} If a {\flow} $\vv$ contains such
  a submode, the {\flow} is thermodynamically infeasible. \co{also def
    here: flux motif} \co{These submode criteria directly reflect the
    weak thermodynamic and economic sign constraints, stating that
    positive driving forces or {\fluxvalue}s are needed for positive
    fluxes, but that fluxes can always vanish.}  More details on
  thermodynamic and economic constraints can be found in SI
  \ref{sec:economicandthermodynamic}.

\co{hier auch sagen: cyclic and futile modes are nullvectors of
  matrices, Ntot and Nbene; which leads to various other analogies.}

\co{However, in the definition of ``cycles'', direct forces or direct
  flux gains need to be taken into account.}  \co{Just like in
  thermodynamics, cycles in metabolic economics can become feasible if
  there are additional ``forces'', e.g.~by an extra gain due to the
  conversion of cofactors (see SI
  Fig.~\ref{fig:analogiesFBAEBAVBA1}).}

\co{JA - SCHREIBEN! analogous definitions! different stoich. matrices!
  \co{Just like null vectors of $\Nint$ are stationary, null vectors
    of $\Ntot$ are thermodynamically infeasible (loopful) and null
    vectors of $Nbene$ are futile. The latter two can be used as
    ``test'' modes for infeasible flux distributions. If such a test
    mode, as a variation, is conformal with $\vv$, then $\vv$ must be
    endergonic (in the thermodynamic case) or futile (in the economic
    case). \co{auch Multi-objective modes $\kv$ (which must satisfy
      ${\Amat \choose \Nint}\,\kv = 0$ with a given matrix $\Amat$ of
      flux gain vectors)} \co{Discuss bezug zu potentialen!}}
  \co{hier abschnitt ueber loopless / thermo feasibility /
    physio-thermo feasibility; dann entsprechend das gleich fuer
    oekonomie: hergeleitet von Nben!  Bezug zu erlementarmoden (von
    Nint), Ntot)} dann: Elementary!}
}

\co{JA!! XXX discuss, (analog zu thermo: thermofeasible vs thermo-phys
  feasible) distinguish futile (in general), given ext ec pot (or some
  known ec pot); and cyclic-futile (containing cycles that would be
  infeasible for ANY external ec pot.); def loopless; ``economically
  feasible'' vs ``eco-physiologically feasibl'', with some given
  (bounded) ec pots}
\co{in analogie zum vorschreiben von cext und bounds on cint (bzw
  entwprechend fuer mu): vorschreiben von w, bounds on w internal?:
  eco-physiologically feasible?}

\co{KURZ ZUSAMMENFASSEN! NUR der eigentlich beweis zaehlt! ELEMENTARY CYCLES:
  potentials and cycles: arne sagt: beard 2004 (with matroids) is
  still flawed. there is also something by mavrovouniotis
  (``thermodyanmic bottleneck''?); final proof (unflawed, and allowing
  for bounds, and therefore bottlenecks in addition to cycles, by
  elad. In seiner masterarbeit: insbesondere Section 2.2 und dort
  Prop. 8 und Thm. 3. Excluding elementary \nonbeneficial cycles is
  enough because: (i) flux distribution v is unfeasible if + only if
  it contains any unfeasible cycle; (ii) if it contains an unfeasible
  cycle, it also contains the elementary unfeasible cycles included in (and conformal with)
  this cycle.  (iii) so if a flux distribution contains no unfeasible
  elementary cycle, it contains no unfeasible cycle, and is therefore
  feasible. BEWEIS, dass ELEMENTARY futile cycles fuer beweise
  genuegen?  ok .. to be proven: if a flux distribution contains a
  futile cycle, it also contains a elementary \nonbeneficial
  cycle. \co{JA! FN: for proof, use the fact that $wv \trans$ in the
  gain/stoichiometric matrix can be seen as the row for a virtual
  metabolite, which must be balanced. that is, everything that can be
  proven for stationary flux distributions can be transferred to
  {\futile} flux distributions -- that is, proofs for elementary modes
  translate into proofs for elementary \nonbeneficial flux
  distributions.} (BTW mention this somewhere CLEARLY). So we only need
  a proof that, if a flux distribution is stationary, then there is
  some (non-zero) elementary flux distribution with the same signs at
  the overlap. STEFAN FRAGEN!  BZW WIE GEHT DER BEWEIS IN THERMO
  FluxAn?}

\myparagraph{Chemical and economic potentials can be  coupled
  through flux directions} Chemical and economic potentials describe
different aspects  of metabolism, but they are linked by the fluxes  in optimal states. Fluxes
must not only run from higher to lower chemical potentials, but also  from lower to higher
economic potentials. This   effectively couples the two  types of potentials.
The thermodynamic condition $-\frac{\Deltar \mu}{RT}\,v=\sigma>0$ and
economic conditions $(\Deltar w +\bvdirs)\cdot v = \hus\,u>0$ imply
that flux {\myvalue} $\gvtots= \Deltar \wtots + \bvdirs$ and
thermodynamic force $\theta = \frac{-1}{RT}\,\Deltar \mu$ must have
opposite signs in all active reactions. By combining the two
equations, we obtain a relation  between flux {\myvalue}s and thermodynamic
forces: 
\begin{eqnarray}
\label{eq:potentialProduct0}
v = \frac{\gvtots}{\hus\,\esymbol} = \frac{\theta}{\sigma},
\qquad \mbox{and thus} \qquad \frac{\gvtots}{\theta} =
\frac{\hus\,\esymbol}{\sigma}.
\end{eqnarray}
 Flux {\myvalue} and thermodynamic force must show the same ratio as {\enzymeinvestment} and energy dissipation. So for example, if a
reaction with predefined  flux comes close to  chemical equilibrium
($\theta = \frac{-\Deltar \mu}{RT} \approx 0$), the flux {\burden}
becomes infinite and (in an optimal state) its flux {\myvalue}
$\gvtots$ becomes infinite too.

\co{LOGIC (in graphics?)
zero external production
  -> no energy dissipation
    -> perpetuum mobile 2nd kind
  -> no flux benefit
    -> futile (in model with production objective)
}

Finally, there are models in which thermodynamic and economic
constraints coincide: in models with a production objective (and
without concentration bounds\co{wirklich? nicht flux bounds?}), cyclic flux modes are not only
thermodynamically infeasible, but also  futile. In this case, \co{REF to
  paper with proof; ask elad?}  economic and thermodynamic
constraints on fluxes are fully redundant \co{but only in one reaction: some
  futile modes may still be thermo-feasible!}  (see SI
\ref{sec:proofEquiEnergEcon}).  This explains why FCM, originally
designed to avoid futile and wasteful fluxes, can repress
thermodynamic loops.  \co{Can we say that (for production objectives)
  that $\{FCM(w) | w \in R^n_+\}$ solutions = VBA solutions $\subset$
  loopless FBA solutions $\subset$ FBA solutions and also TMFA
  solutions $\subset$ FBA solutions? Here by VBA I mean without the
  thermodynamic constraints. JA!}  \coout{Correct me if I'm wrong, but
  adding thermodynamic constraints without concentration bounds (AKA
  loopless-cobra constraints) will be redundant if you have the
  economical constraints as well. I think it would be nice to explain
  this relationship}

\myparagraph{Kinetics and thermodynamics put lower bounds on
  {\enzymeinvestment}s and flux values} An important constraint on the 
economic potentials follows from molecule properties of the
enzymes.  In each reaction, the {\fluxburden} must be larger than
$\hvsmin = \frac{\husmin}{k^{\rm cat}}$ where $\kcat$ is the maximal catalytic rate of the
enzyme\footnote{To derive the lower bound, we  assume (without loss of generality) positive
  fluxes. \co{mention ``invariant'' definition with respect to flux
    directions.} In reactions with negative fluxes, $\hvsmin$ would be
  negative and the inequality must be reversed. This also concerns all
  following inequalities.  However, if we model a single metabolic
  state, this complication can be avoided by reorienting all reactions
  to have positive fluxes.} and  $\husmin$ is the minimal
{\price}.  \co{kam schon vorher:} By considering
reaction thermodynamics (as in
Eq.~(\ref{eq:fluxBurdenEfficiencyDetailed})), we obtain a 
stricter essential value
$\hvsmin = \frac{1}{(1-\e^{-\theta})} \frac{\husmin}{k^{\rm cat}}$ that
depends on the driving force $\theta_{l}$.  In enzyme-optimal
states, {\fluxvalue} $\gvtots = \Deltar \wtots + \bvdirs$ (value
production per flux) and {\fluxburden} $\hvs$ (enzyme investment per
flux) must be equal in each active reaction. So for a positive
  flux, due to enzyme kinetics the flux {\myvalue}
  $\gvtots =\Deltar \wtots+\bvdirs$ must exceed the essential flux
  value, which is  given by
\begin{eqnarray}
\label{eq:potentialProduct}
  \gvtots > \gvtots^{\rm min} = \frac{\hus}{k^{\rm cat}}\,\frac{1}{\theta}
\end{eqnarray}
and therefore inversely proportional\footnote{The proof of
  Eq.~(\ref{eq:potentialProduct}) is simple: due to
  Eq.~(\ref{eq:fluxBurdenEfficiencyDetailed}), with
  $\eta^{\rm kin}\le 1$ and $\theta \le 1-\e^{-\theta}$, the
  {\fluxburden} has a lower bound
  $\hvs \ge \frac{\hus}{\kcat\,\theta}$. In cases where
  $\eta^{\rm kin} \approx 1$ (full substrate saturation, no
  inhibition) and $\theta \gg 1$ (large force), the flux burden hits
  this bound.  Otherwise, if enzymes are inhibited or if substrate
  level is low, we obtain $\hvs \gg \frac{\hus}{\kcat\,\theta}$. Since
  flux {\burden} $\hvs$ and flux value
  $\gvtots = \Deltar \wtots+\bvdirs$ must be balanced
  (Eq.~(\ref{eq:reactionbalanceeqFlux})), we obtain the same bound for
  the flux value.}  to the force (see SI
\ref{sec:SIpotentialProductProof}).  The resulting limits on
{\fluxvalue}s\footnote{Note that inequality
  (\ref{eq:potentialProduct}) is an approximation of this bound.}
($\gvtots \ge \hvsmin$) and economic potentials
($\Deltar \wtots > \hvsmin- \bvdirs$) (see SI
\ref{sec:ProofMinimalFluxDemand}) must hold for all active enzymatic
reaction: if a {\fluxvalue} is below the essential value, the reaction
must be shut off. \co{FN: note that ``the flux value must be higher
  than the minimal flux price'' is always meant as a necessary, not as
  a sufficient condition!}

\begin{figure*}[t!]
\begin{center}
  \parbox{8cm}{(a) Mass balance
    $\frac{\md c_{i}}{\md t} = \sum_{l} n_{il}\,v_{l} = 0$ \\[2mm]
    \includegraphics[width=7.2cm]{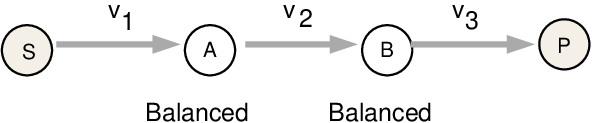}\\[8mm]
    (b) Thermodynamics
    $ \sigma_{l} = - \frac{\Deltar \mu_{l}}{RT} \,v_{l} > 0$ \\[2mm]
    \includegraphics[width=7.2cm]{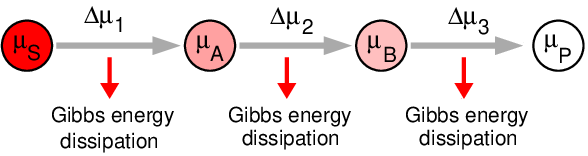}\\[8mm]
    (c) Enzyme economy
    $ \zcostgenericl = [\Deltar \wtotl + \bvdirl]\,v_{l} >0$ \\[2mm]
    \includegraphics[width=7.2cm]{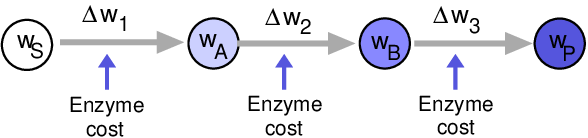}
  }\hspace{5mm} \parbox{7.5cm}{\caption{In Value Balance Analysis, metabolic
    fluxes are shaped by three basic principles: stationary fluxes, dissipation of Gibbs free energy in each reaction,
    and production of positive value in each reaction to balance the
    enzyme investment. \coout{weniger woerter in bild} (a) In a steady
    state, production and consumption of internal metabolites must be
    balanced. In the example, all fluxes $v_{l}$ must be equal. (b) To
    produce entropy (entropy production density $\sigma_{l}>0$), each
    reaction must dissipate Gibbs free energy, so fluxes must follow
    the thermodynamic forces $\theta_{l} = -\Deltar \mu_{l}/RT$ from
    higher to lower chemical potentials $\mu_{i}$.  (c) In
    enzyme-optimal states, flux point benefits must be balanced with
    the {\enzymeinvestment}s $\zcostgenericl = \hul\,\ul$. Since the
    enzyme investments are positive, fluxes and flux {{\myvalue}}s
    $\gvtotl = \Deltar \wtotl + \bvdirl$ must have equal signs.  In
    reactions without flux {\gain}s $\bvdirl$, fluxes must lead
    towards higher economic
    potentials $\wtoti$.}    \label{fig:analogiesFBAEBAVBA}}
\end{center}
\end{figure*}

\myparagraph{Analogiy to electric circuits: Kirchhoff's rules and sign
  constraints} The laws for metabolic fluxes and reaction
thermodynamics resemble Kirchoff's rules for electric circuits
\cite{kirc:45,edgi:07}. Metabolic fluxes correspond to electric
currents, and thermodynamic driving forces correspond to
voltages. Kichhoff's node rule (for currents) states that charge is
conserved, so incoming and outgoing (steady) currents must cancel out
for any region of an electric circuit (and in particular, for each
node). The loop rule (for voltages) states that voltages, given by
potential differences, must sum to zero over any closed loop.  \co{FN:
  may sound obvious; but unlike, for example, heat exchange, summed
  over a thermodynamic cycle process! also unlike things in some
  drawings by escher!} Similar rules exist in reaction thermodynamics:
steady metabolic fluxes must satisfy mass conservation (like in the
``node rule'' for charge conservation), and thermodynamic driving
forces, as potential differences, must sum to zero over a cycle (where
``cycles'' are defined more generally, as null space vectors of the
total stoichiometric matrix $\Ntot$). Finally, like their electric counterparts, the
thermodynamic forces determine the flux directions (sign constraint).

If we ask about the precise relation between forces and fluxes, or
voltages and currents, we find some remarkable differences.  In
electric circuits, Ohm's law states that a current is proportional to
the voltage, with a constant current/voltage ratio (conductivity) and
voltage/current ratio (resistance). Ohm's law implies that nonzero
voltages always cause currents (entailing a ``strong sign
constraint'').  While Ohm's law holds well for many conductors, it is
just an empirical law and does not hold generally. In diodes, for
example, the relationship is nonlinear and a threshold voltage must be
exceeded to obtain a current.  \co{(not compatible with a ``strong
  sign constraint'', but still compatible with a ``weak sign
  constraint'')} In reaction thermodynamics, the situation is even
more complicated. The metabolic flux does not depend on the force
directly, but on substrate and product concentration (via kinetic
laws).  Nevertheless, rate laws of the form $v(\theta)$ exist as
approximations.  Near chemical equilibrium, a linear force-flux
relationship is sometimes assumed, \co{in a Taylor expansion around
  the equilibrium state)} like in Ohm's law. More generally, even
without an exact force-flux formula, we may expect that ``everything
else being equal''\co{difficult to say what this means ,assuming
  kinetic rate laws}, a higher force causes a higher flux. An example
is provided by the factorised rate laws, where
$v \sim (1-\exp{-\theta})$ if the kinetic efficiency term is
approximated to be constant. How are these force-flux relationships
related to sign constraints?  The weak sign constraint always holds
due to basic thermodynamics. In addition, in a linear flux-force
relationship, even the smallest force will lead to a flux (in line
with the assumption of a strong sign constraint). But more generally,
if  enzymes can be completely inhibited allosterically,
even a positive force will not always lead to  a flux, so the strong sign
constraint does not hold. 

  \myparagraph{Economic force-flux relationships?}  \co{JA! also
    called ``economic driving force'' UEA for flux value? what word
    then for potential difference?}  Now let's compare this to value
  balance analysis. In fact, things are quite similar to
  thermodynamics: all three laws -- node rule, loop rule, and rule for
  flux directions -- hold, \emph{mutatis mutandis}, for metabolic
  fluxes and  ``forces'', i.e.~the flux values resulting from flux
  gains and economic potentials.  \co{WO? The analogies to Kirchhoff's rules
  are further discussed in SI \ref{sec:economicandthermodynamic}.}
  What about sign constraints and ``force-flux relationships''?  While
  a positive flux value (or ``economic force'') is necessary for a
  positive flux (hence the weak sign constraint always holds), the
  flux is not a function of the flux value.  There is no Ohm's law,
  and not even a nonlinear law like for diodes. However, like in
  thermodynamics, we may consider such dependencies metaphorically or
  as approximations, assuming that higher economic forces tend to go
  with higher fluxes. The ``economic force'' is typically given by an
  economic potential difference, but there may also be flux gains or
  values of cofactor pair that serve as an extra ``voltage source''
  for the flux.  An ``economic flux-force relationship'' is not based
  on a strict argument, but assuming a gradual dependence: starting
  from the facts that ``without force, there is no flux'' and ``at
  some positive force, there will be a flux'', we claim that ``the
  higher the force, the higher the flux''. In contrast to reaction
  thermodynamics, this relationship must for sure be nonlinear.  We
  already learned that for a reaction to be active, the flux value
  must exceed some positive ``essential'' value given by enzyme
  molecule properties. Therefore, a force-flux relationship can only
  be a nonlinear one with a ``threshold force'', like in electric
  diodes. This means that an economic ``conductivity'' or
  ``resistance'' cannot be defined, unless one defines it as the local
  derivative of the assumed force-flux relationship.  It also means
  that the strong sign condition cannot hold.

  \co{diskutieren: analogie zu strom; zv wirkt wie batterie; muesste
    unten in ``choices bwteeen metabolic strategies'' kommen)}
  \co{jetzt noch alles, was aus dem ``ohmschen gesetz'' folgt:
    festklemmen von spannung // value production (fluss mal tension)
    waere dann so etwas wie leistung im elektrischen stromkreis}
  \co{``economic TKM'' vorschlagen?}

  \co{In summary, the electric analogy fails in two main points:
    first, economic sign constraints hold only for active enzymatic
    reactions; second, to justify a non-zero flux the {\fluxvalue}
    must not only by non-zero, but exceed some positive value, which
    depends on enzyme properties (I come back to this question
    below).We saw that a forward flux requires a flux value above a
    positive threshold value, given by the essential flux
    {\burden}. This threshold value has no analogue in
    thermodynamics. According to our flux sign constraint
    (\ref{eq:thermocondition}), even a small driving force may already
    cause a flux.  If the strong sign condition is used, any positive
    force will cause a flux.  With economic forces this is different:
    like the weak flux sign condition, the economic constraint
    (\ref{eq:potentialProduct0}) assumes that even a non-zero force
    does not necessarily lead to a flux.  }

\section{Value balance analysis}

\myparagraph{\ \\Value Balance Analysis} In this section, we shall use
economic and thermodynamic \co{energetic UEA?  einmal am anfang
  erklären?}  constraints together in a flux modelling framework
called Value Balance Analysis (VBA). Similar to Energy Balance
Analysis (EBA), VBA is not an optimality problem itself, but an
algebraic framework to describe flux distributions that are
stationary, thermodynamically feasible, and economical at the same
time. We know that such flux distributions are exactly the potential
solutions of some (unknown, underlying) metabolic optimality
problems. Thus, our main aim is not to find a single solution, but
rather sets of plausible solutions, which may stem from plausible
underlying models (which, themselves, may be FBA-like, kinrtic, or
part of hypothetical whole-cell models). This leaves space for
sampling or for finding specific solutions by applying extra knowledge
(in the form of constraints or heuristic selection criteria). In all
these cases, VBA incorporates the idea of resource allocation by
requiring that each feasible state (in VBA) must be an optimal state
according to some underyling resource allocation problem!  \co{Explain
  this well, and say why we go in this direction (instead of
  introducing yet another optimality problem!)- that this is supposed
  to be an abstraction that brings generality and clarity.}

\myparagraph{Value Balance Analysis is based on three types of
  constraints} Above we learned that flux directions are governed by
thermodynamic and economic constraints.  For flux analysis, it makes
sense to use these constraints simultaneously.  Consider a metabolic
network with flux {{\gain}} vector $\bvtot$.  To be economical, a
{\flow} must satisfy a value production balance
(\ref{eq:reactionbalanceeq}) with suitable economic potentials
$\wtoti$, and since {\enzymeinvestment}s $\zcostgenericl$ are
positive, the fluxes $v_l$ and flux {{\myvalue}}s
$\hvl = \Deltar \wtotl + \bvdirl$ must have equal signs.  If all
economic potentials $\wtoti$ were known, they would therefore
determine the flux directions, and FBA could be used to find
economical flux modes with these directions. However, a metabolic
objective determines only the \emph{external} economic potentials,
while the internal potentials $\wintm$ remain to be found.  To do
this, we consider a variant of FBA called Value Balance Analysis (VBA)
in which fluxes, chemical potentials, and economic potentials appear
as model variables. A VBA problem is defined by a metabolic network, a
linear flux objective $\fluxbene(\vv) = \bvtot\cdot \vv$, and perhaps
other data such as external potentials. For example, with biomass
production as a single objective ($\fluxbene=v_{\rm BM}$), we obtain
an economic potential $w_{\rm BM}=1$ for biomass and zero potentials
of all other external compounds, and zero flux gains for all
reactions. By splitting the effective flux gain $\bvtot$ into
$\bvtot = {\Next}\trans \wext + \bvdir$ (see section
\ref{sec:EcPotEcFlux}), we obtain the external economic potentials
$\wextj$ and the flux {{\gain}}s $\bvdirs$ as model parameters.  The
value production balance (\ref{eq:reactionbalanceeq}) provides a
constraint besides the usual stationarity and thermodynamic
constraints. Unlike the chemical potentials, which depend on
metabolite concentrations\footnote{We use the common approximation
  $\mu = \mu^{(0)} + RT \,\ln c$, assuming constant activity
  coefficients.}, the economic potentials can be treated as separate
variables (see SI \ref{sec:economicandthermodynamic}). Altogether, VBA
describes fluxes and chemical and economic potentials (in vectors
$\vv, \muv^{\rm int}$, and $\wint$) by the following three constraints
(see Figure \ref{fig:analogiesFBAEBAVBA}):
\begin{eqnarray}
\label{eFBAconstraints}
\Nint\,\vv&=&0 \qquad  \mbox{Mass balance and stationary fluxes} \nonumber \\
\vv &\sqsubseteq& \thetav \quad\;\,\,\, \mbox{Energy dissipation in all active reactions} \nonumber \\
\vv &\stackrel{\rm enz}{\sqsubseteq} & \gvtot \quad \mbox{Value production in all active enzymatic reactions},
\end{eqnarray}
with thermodynamic driving forces $\theta = -\Deltar \mu$ and flux
{{\myvalue}}s $\gvtots = \Deltar \wtots + \bvdirs$. The
symbol\footnote{Generally, $\xv \sqsubseteq \yv$ (``$\xv$ is conformal
  to $\yv$'') states that all non-zero components in $\xv$ have the
  same signs as the corresponding components in $\yv$.}  $\sqsubseteq$
denotes ``equal signs in all active reactions'', and
$\stackrel{\rm enz}{\sqsubseteq}$ denotes ``equal signs in all active
enzymatic reactions''.  \co{satisfiability problem! can be for
  sampling, choosing, fitting rather than optimisation; (potential)
  optimality is guaranteed by the economic constraints! in these satisfiability problems, ..}  External
chemical and economic potentials (and maybe uptake and excretion
rates) may be predefined, while all other variables need to be found.

\myparagraph{Computing metabolic fluxes, economic potentials, and
  {\enzymeinvestment}s} The VBA conditions Eq.~(\ref{eFBAconstraints})
state that our fluxes (at the chemical and economic potentials chosen)
must be stationary, exergonic, and must produce positive value. All
triples $(\vv, \muv,\wtot)$ that satisfy these constraints are
allowed. To further restrict the solutions, we may predefine some
variables or constrain them to known or assumed physiological
ranges. Together with a linear objective, this would yield
mixed-integer linear programming (MILP) problems \cite{hafb:96}. In
contrast to the linear optimisation problems in classical FBA, such
problems are typically non-convex and much harder to solve (see Figure
\ref{fig:fbacomparison2}).  To keep the calculations simple, we do not
use a MILP solver but choose our variables step by step. We first
choose a pattern of feasible flux directions. To be (thermodynamically
and economically) feasible, a {\fluxpattern} must allow for a
stationary {\flow} and for a choice of economic and chemical
potentials (that it must be free of cyclic or {\nonbeneficial}
submodes).  Once we have such a flux pattern, we can easily determine
all other model variables (stationary fluxes, chemical potentials, and
economic potentials) by Linear Programming, separately for each type
of variable. Thus, the only difficult step  is to choose
feasible flux signs: we may do this by choosing an (economically and
thermodynamically) feasible flux mode, either by employing FCM or
choosing a non-feasible {\flow} and by eliminating all cyclic \co{besseres wort! production neutral?} \co{what
  about endergonic?} or {\nonbeneficial} motifs (see SI
\ref{sec:removefutilecycles}).  Given our flux pattern, feasible
fluxes, chemical potentials, and economic potentials can be chosen
independently. To choose a specific solution, we may use sampling or
optimisation (with extra assumptions and usage of data).

\myparagraph{Choosing plausible economic potentials} How can we
determine economic potentials in practice?  In a state with known flux
directions, Eq.~(\ref{eq:reactionbalanceeq2}) puts constraints on the
potentials (see Eq.~(\ref{eq:reactionbalanceeq2a})), just like
thermodynamics laws put constraints on the chemical potentials (and
thus on log-metabolite concentrations) \cite{hohh:07}.  If this is the
only information we have, the economic potentials may be sampled,
optimised, fitted to data, or chosen by heuristic rules within these
constraints.  The resulting {\enzymeinvestment}s can be computed from
the value production balance.  However, to realise these economic
potentials by kinetic models \cite{lieb:14a}, sampling our economic
potentials at random may not be very wise: the resulting kinetic
constants may be unrealistic.  Anticipating this, we should choose
realistic economic potentials from the start, for example by fitting
them to measured enzyme investments.  In fact, each set of economic
potentials corresponds to an underlying kinetic models with specific
kinetics and enzyme cost functions.  If we are happy with arbitrary
VBA solutions, we can freely choose the economic potentials within the
given constraints. But if our aim is to construct kinetic models, all
relevant data and constraints must be considered already when choosing
the potentials. In particular, if we desire a \emph{realistic} kinetic
model, we need to use, at least, some good rules of thumb for
integrating avilable data and constraints.
  
\myparagraph{Plausible economic potentials inferred from flux and
  proteome data} If the fluxes, metabolic objective, and
{\enzymeinvestment}s in a network are known, it is easy to compute the
economic potentials.  In a linear pathway in an optimal state, they
can be found like this: for each metabolite, we sum all upstream
enzyme (and possibly substrate) investments; this yields the
``embodied investment''.  Then, by dividing by the metabolite's
production rate (in this case, the pathway flux), we obtain the
metabolite's embodied value. In an optimal state, this embodied value
is equal to the use value (or ``economic potential'').  In networks
with branches and cycles, this simple calculation method does not work
because we cannot just sum the investments along a one path.  Instead,
we rely on the value production balance
$(\Deltar \wtots+\bvdirs)\,\,v = \hus\,\esymbol$ and the value balance
$\Deltar \wtots+\bvdirs = \frac{\hus \, \esymbol}{v} =
\frac{\hus}{\ratelaw} = \hvs$, which define relationships between
economic potentials $\wtots$, flux {\gain}s $\gvtots$, fluxes $v$,
enzyme {\price}s $\hus$, enzyme levels $\esymbol$, enzyme efficiencies
$\ratelaw$, and flux {\burden}s $\hvs$.  Here is one possible
calculation method. We use molecular weights as proxies for enzyme
prices\footnote{Of course, also other measures for prices can be used,
  such as total synthesis rate, or ATP investment.}  $\hus$ (with a
scaling factor to be determined later) and estimate the
{\enzymeinvestment}s $\husdot = \hus\,\esymbol$ from proteomics data
and enzyme prices.  If fluxes or catalytic rates are known, we can
estimate the {\fluxburden}s $\hvl$ either by $\hvl = \hus\,\esymbol/v$
from enzyme prices, proteomics data, and fluxes, or directly by
$\hvl = \hus/\ratelaw$ from and enzyme prices and catalytic rates.
These estimates can further be corrected\footnote{The
  {\summationcondition} demands that the vector $\hv-\bvdir$ lies in
  the image space of ${\Nint}\trans$, that is, in the nullspace of
  $\Kmat\trans$. Hence, to adjust a given vector $\hv$, we can project
  $\hv-\bvdir$ onto this space (proof in \cite{lieb:14a}).}  to comply
with the {\summationcondition} (see SI
\ref{sec:potentaislfromenzymepar}), and be used to obtain the economic
potential differences $\Deltar \wtots$. From these potential
differences, together with the known economic potentials of external
metabolites and conserved moieties, we can compute the individual
internal economic potentials.\co{FN: explain how, with
  $\Deltar \wint = {\Nint}\trans\,\wint$}
    
\myparagraph{Guessing economic potentials from data and simplifying
  assumptions} If data are missing or uncertain, a ``blind'' choice of
$\wint$ is not very practical, and we should use some extra
information to obtain plausible values: for example, we consider how
economic potentials are constrained (e.g.~by flux directions) and then
choose them to match measured enzyme investments (that is, enzyme
investments reflecting proteomics data and assumed enzyme prices) as
closely as possible.  In practice, we can combine different
measurement quantities, including flux and protein data, enzyme
kinetic constants, and molecule sizes (as proxies for enzyme
prices). If enough data are given and optimal states are assumed, the
potential differences can be computed from our equations, and if data
are missing, we replace them by plausible general assumptions.  The
aim is to choose economic variables that would also appear in
realistic kinetic models. Some methods for computing economic
potentials, relying on different data or plausible
assumptions\footnote{Reasonable bounds on conversion values \co{UEA,
    auch CBA lag, CBA kin usw} $\Deltar \wint$ can be obtained from
  experimental data \cite{horh:11} (see SI
  \ref{sec:potentaislfromenzymepar}).}, are described in SI
\ref{sec:algorithmsflux}. We may assume, for example, that all
reactions have equal {\enzymeinvestment}s (and therefore an equal
value production
\co{https://en.wikipedia.org/wiki/Principle\_of\_indifference}
\co{erklaert?}
$\gvtotl\partialder= (\Deltar \wtotl + \bvdirl)\,v_{l}$) or that they
have equal {\fluxburden}s (and therefore equal flux {{\myvalue}}s
$\gvtotl=\Deltar \wtotl + \bvdirl$), or that (uncertain) measurement
data for some of these values are given.  In all of these cases, we
obtain simple ``educated guess'' formulae for the economic potentials
and other economic variables (SI \ref{sec:princdistribinvestment}).

\co{also propose an ``MDF like'' method for computing the ec pot (or
  rather a flux-weighted (or even flux/kcat-weighted) MDF method,
  assuming that very small enzyme investments are unrealistic and
  should be avoided in models)} \co{FN, WO?  In FBA we may first
  determine mu; then, given estimated $\kcat$ values, determine the w
  under constraints!}

\co{WO? FN: if the elasticities and gc values were also known, the
  economic potentials could be directly computed (formula from CBA
  kin), but this is usually not the case.}
  
\co{noetig? besser anfangen: also was wuerden wir zB konkret machen,
  um die ATP-ADP differenz zu bestimmen?}

\co{IN ABSCHNITTE DAVOR EINBAUEN? Even if we cannot determine the
  potentials directly, we can establish constraints (e.g.~by finding
  the $\Deltar \wtoti$) and use extra information to determine
  plausible potentials within these constraints, for example by
  fitting them, to proteomics or enzyme kinetics data. From the
  reaction balance equation, we obtain the potential difference
  $\Deltar \wtotl = \hul/v_{l}-\bvdirl$. If the external potentials are
  known, we obtain $\Deltar \wint = \hu/\vv - \bvtot$. If this does not
  uniquely determine the vector $\wint$, we can find a unique solution
  by setting the economic potentials of conserved
  moieties. \co{erklaeren!}  This logic also applies to entire
  metabolic networks (consisting of pathways, connected by central
  precursors and cofactors) and even to cell models comprising
  metabolism, enzyme production, and ribosome production.  Aside from
  the reaction balance equation, we can also use the variation
  conditions, to obtain potential differences along entire pathways:
  the variation condition for fluxes \cite{lieb:14a} relates the
  {\enzymeinvestment}s along any flux profile to the value production,
  and thus to the $\Deltar w$ along this profile\footnote{Here are some
    details. The reaction balance (\ref{eq:reactionbalanceeq}) relates
    the enzyme investments to value production, i.e.~to the economic
    potential difference, multiplied by the flux.  For a pathway
    within a larger network this means: the potentials on the pathway
    boundary are linked to the total {\enzymeinvestment} along the
    pathway. To see this, we consider a flux profile
    $\modevector^{\rm sub}$ only within the pathway (and zero
    outside).  The enzyme cost for catalysing this flux profile must
    be balanced by the flux benefit from substance conversion
    (i.e.~between ``input'' and ``output'' compounds on the pathway
    boundary). For example, consider a local flux profile with a few
    inflowing substrates $S_{1}, .., S_{n}$ with zero economic
    potentials and a single outflowing product $P$. In this case, the
    value production (the economic potential of $P$, times its
    production rate) is equal to the sum of {\enzymeinvestment}s. This
    can be generalised: if a pathway produces several products, it is
    the sum of production benefits that counts. If the substrates have
    non-zero potentials (e.g.~for a subsystem within a larger
    metabolic network), they appear in the balance, and the value
    production and the corresponding {\enzymeinvestment} will be
    smaller. Similarly, external cofactor pairs will change in the
    balance. Similar to the cost/benefit balance, we may also consider
    the {\myvalue}/{\price} balance: we just divide benefit and cost
    by a reference flux (e.g.~the production rate of $P$). \co{Dann
      ergibt sich der wert (potential) von P als summe aller
      flusslasten, jeweils geteilt durch den produktionsfluss von P,
      und evtl die summe aller eingehenden oek potentiale, geteilt
      durch den jeweiligen yield.}  Using this, we can consider any
    subregion of the network, choose a flux mode within this region,
    compute the {\enzymeinvestment}s along the flux mode, and equate
    it to the economic potential difference at the region's boundaries
    (the difference between metabolites produced and consumed by this
    mode). By applying this logic systematically, e.g.~to elementary
    flux modes, economic potential differences between distant
    metabolites can be inferred. \co{REF to example in CBA kin - oder
      das beispiel hierher?}}  }

\begin{figure*}[t!]
  \begin{center}
    \begin{tabular}{lll}
      (a) Chemical potentials & (b) Economic potentials
   &  (c) Enzyme {\benefitshade}s \\[2mm]
   \includegraphics[width=5.5cm]{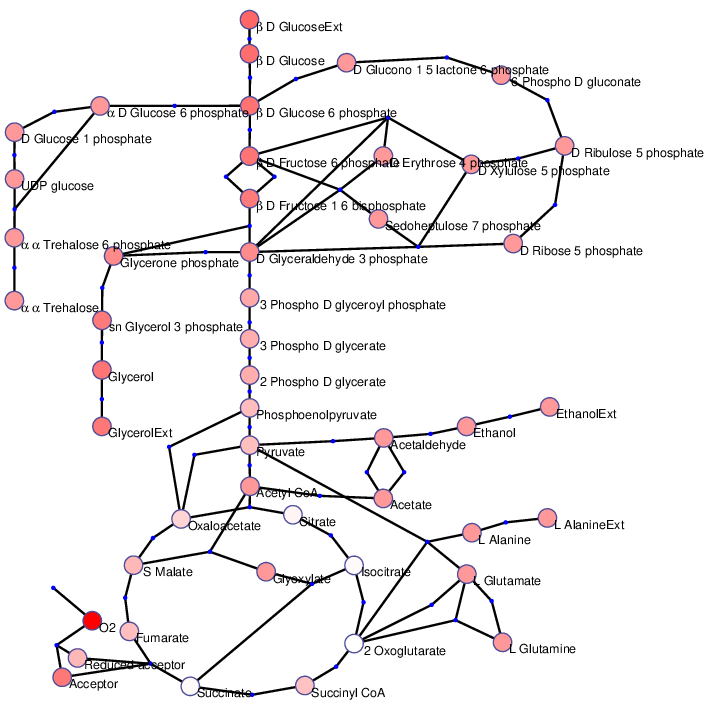}&
   \includegraphics[width=5.5cm]{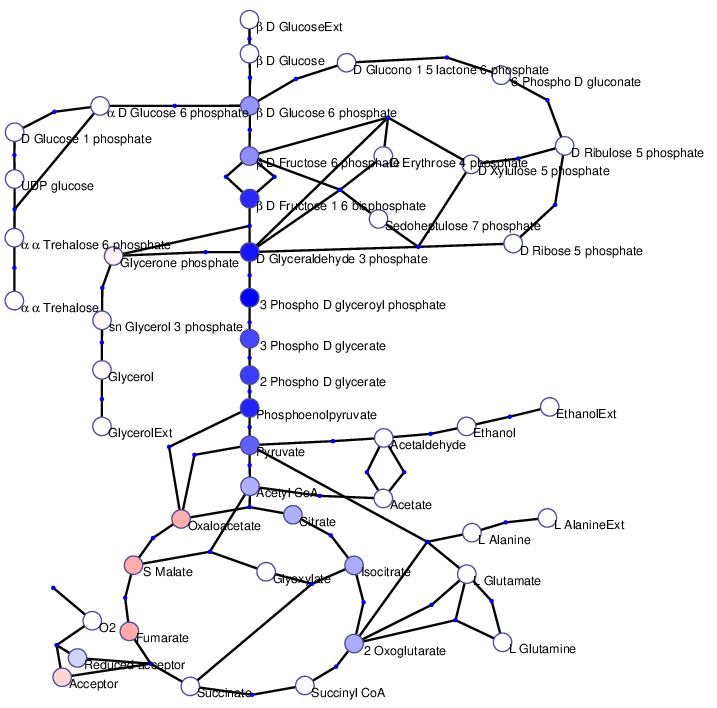}&
   \includegraphics[width=5.5cm]{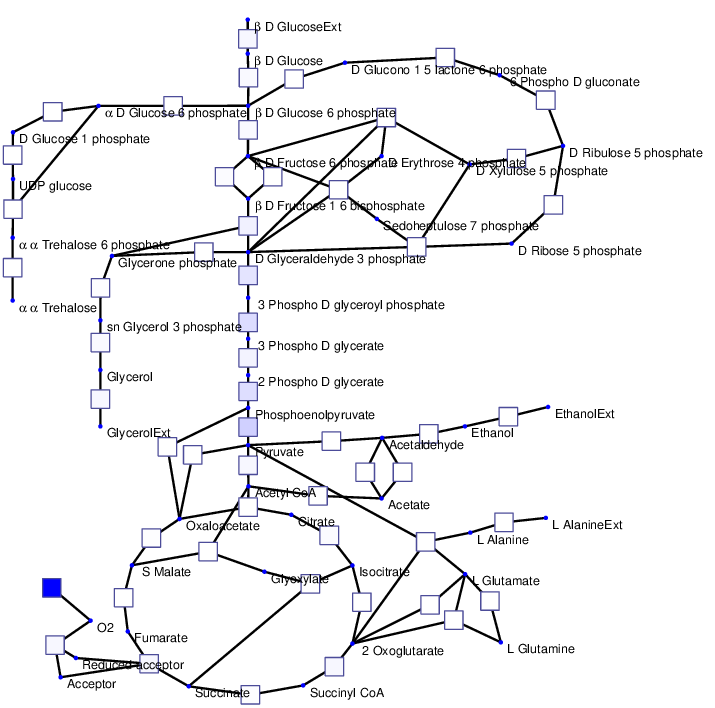}
    \end{tabular}
  \end{center}
  \caption{\co{JA! doch fluesse im ersten bild zeigen und dan
      2x2-anordnung?} \co{JA! doch bilder mit kofaktoren in SI
      zeigen} \co{woher kommen die Negativen werte?}  \co{COLORBAR;
      pfeile groesser} VBA for central metabolism of yeast
    (\emph{S.~cerevisiae}).  (a) Network model of glycolysis, TCA
    cycle, and pentose phosphate pathway, with ATP production as the
    metabolic objective. Cofactors are not
    displayed. (a) Flux profile.  Thermodynamically and economically
    feasible fluxes (arrows) \co{show them!} were obtained by using
    Flux Cost Minimisation (FCM).  \co{use MDF method; say this here
      // with heuristics/proteomics data: refer to proteomaps (to be
      added to abb 1?). what data set?}  Chemical potentials (white:
    low, red: high) were chosen in agreement with the flux directions
    (for details, see text). \co{IN TEXT! how? use par bal instead?}
    (b) Economic potentials (blue: high.  white: low.  pink: negative)
    were inferred from  {\enzymeinvestment}s estimated from
    proteomics data \cite{nkcn:12}.  (c) Enzyme investments $\zcostgenericl$, which match the 
    {{\enzymebenefit}}s $(\bvdirs + \Deltar \wtots)\, v$). The highest
    \enzymeinvestment (corresponding to the highest benefit) occurs in
    oxidative phosphorylation (bottom left) because of its high enzyme
    {\price}.  \co{stimmt das?}  \coout{show GFE production and
      {\enzymeinvestment} as bar diagrams?}  \coout{WEG?  (for a
      detailed picture, see SI Figure
      \ref{fig:yeastpotentialssplit}).} }
    \label{fig:yeastpotentials}
\end{figure*}

\myparagraph{Constructing kinetic models in enzyme-beneficial states}
\co{mention ``systematic layered modelling, ref CBA opt} Given a
metabolic model with known kinetics and a metabolic objective, we can
determine an optimal state, which yields economical fluxes and
economic potentials.  Can we turn this around? That is, can any
possible value structure -- a set of compatible economical fluxes and
economic potentials -- be realised by some kinetic model in an
enzyme-optimal state? We know \co{from where?} that the answer is yes:
such models can be constructed systematically. Figure
\ref{fig:yeastpotentials} shows an example, a model of yeast central
metabolism with ATP production as the benefit function. To reconstruct
a kinetic model, we first determine a VBA-feasible (thermodynamically
and economically feasible) {\flow} (arrows) \co{show them!}. We can do
this by using linear FCM.  \co{how were fluxes defined precisely?}
Then, chemical and economical potentials are chosen by heuristic
assumptions \cite{lieb:14a}.  \co{klarmachen, welche infos in
  rekonstruierte potentiale geflossen sind und wie (ergibt das
  gesamtbild sinn?). dazu abschnitt in SI?}  Data about enzyme levels,
protein sizes, and catalytic constants can be used to estimate
realistic economic potentials. \co{ref CBA kin?}  Next, we search for
kinetic models that realise our fluxes and economic potentials. There
are many such models, and to construct one of them, we need to
determine consistent economic loads and reaction elasticities
satisfying the {\loadbalance}.  From the elasticities, we can then
reconstruct the kinetic constants \cite{lieb:14a}.  This procedure
leads to kinetically and economically plausible models that can be
used for studying optimal metabolic behaviour, e.g.~optimal enzyme
adaptation to static or periodic external perturbations \cite{lksh:04,
  lieb:14c}.

\myparagraph{Multiple objectives} Metabolic strategies arise from
trade-offs between opposing objectives, e.g.~achieving a high ATP or
biomass production, at low metabolic fluxes
\cite{szzh:12}. Mathematically, trade-offs can be described by
multi-objective problems: in a model with several flux objectives
$\fluxbene^{(n)}$ (e.g.~production of different target compounds),
each objective defines a pattern of flux {\gain}s $\bvtotn$, and each
metabolite carries different economic potentials $\wtoti^{(n)}$ for
the different objectives (or briefly a vectorial economic
potential). By comparing the different potentials across the network,
we can learn if the underlying objectives are compatible and where
they clash (i.e.~e.~where there require different flux
directions). How can trade-offs between the objectives be described?
\coout{pareto: equivalenz zeigt dass beide paretoansaetze aequiv
  sind?; CBA II: doppelte interpretation: ueber lagrange und direkt
  ueber met econ, DORT: fuer fluess; HIER: fuer KIN refer to shoval et
  al.~ logik: pareto for fluxes and ME are different; but related
  (through equivalence ...). likewise, ...}  First, the individual sub
objectives \co{auch cba opt: ``subobjectives'' verwenden statt
  individual objectives?} can be combined into a single objective
function, for instance by taking a linear combination\footnote{Note
  that this also works for nonlinear combinations of objectives: in
  this case, the economic variables will be linear combinations with
  state dependent prefactors.}
$\fluxbene=\sum \alpha_{n} \fluxbene^{(n)}$: we obtain a single flux
{\gain} $\bvtot=\sum_{n} \alpha_{n}\,\bvtotn$, and each metabolite has
a single potential $\wtoti=\sum \alpha_{n} \wtoti^{(n)}$. Second, if
subobjectives are uncertain or change rapidly, cells may adapt their
enzyme levels to the average or expected objective (where
subobjectives are weighted with probabilities or relative durations).
\co{WEG? Again, the multiple objectives, flux gains, and economic
  potentials are replaced by linear combinations.}  Third, we may keep
the subobjectives separate and search for \co{explain that convex
  combination is possible and leads to the same convex combination of
  potentials!}  Pareto-optimal states, i.e.~states in which no
subobjective can be improved without compromising the others
\cite{szzh:12}.  \co{Details in SI, satz hier kurz!}  Importantly, any
Pareto-optimal state is also an extremal point\co{FN: not necessarily
  an optimum; ref to CBA opt} of a single-objective problem, when the
single objective is a convex mixture of the subobjectives (where the
linear combination of objectives depends on the point on the Pareto
front)\co{REFERENCE?}.  This also means: any state on the Pareto front
can be treated by VBA, assuming a combined single objective. Thus, in
all three cases -- combinations of objectives, uncertain objectives,
and Pareto-optimal states -- VBA can be applied, and the economic
variables are given by linear combinations of the economic variables
for the individual objectives.

\begin{figure*}[t!]
  \begin{center}
 \includegraphics[width=10.5cm]{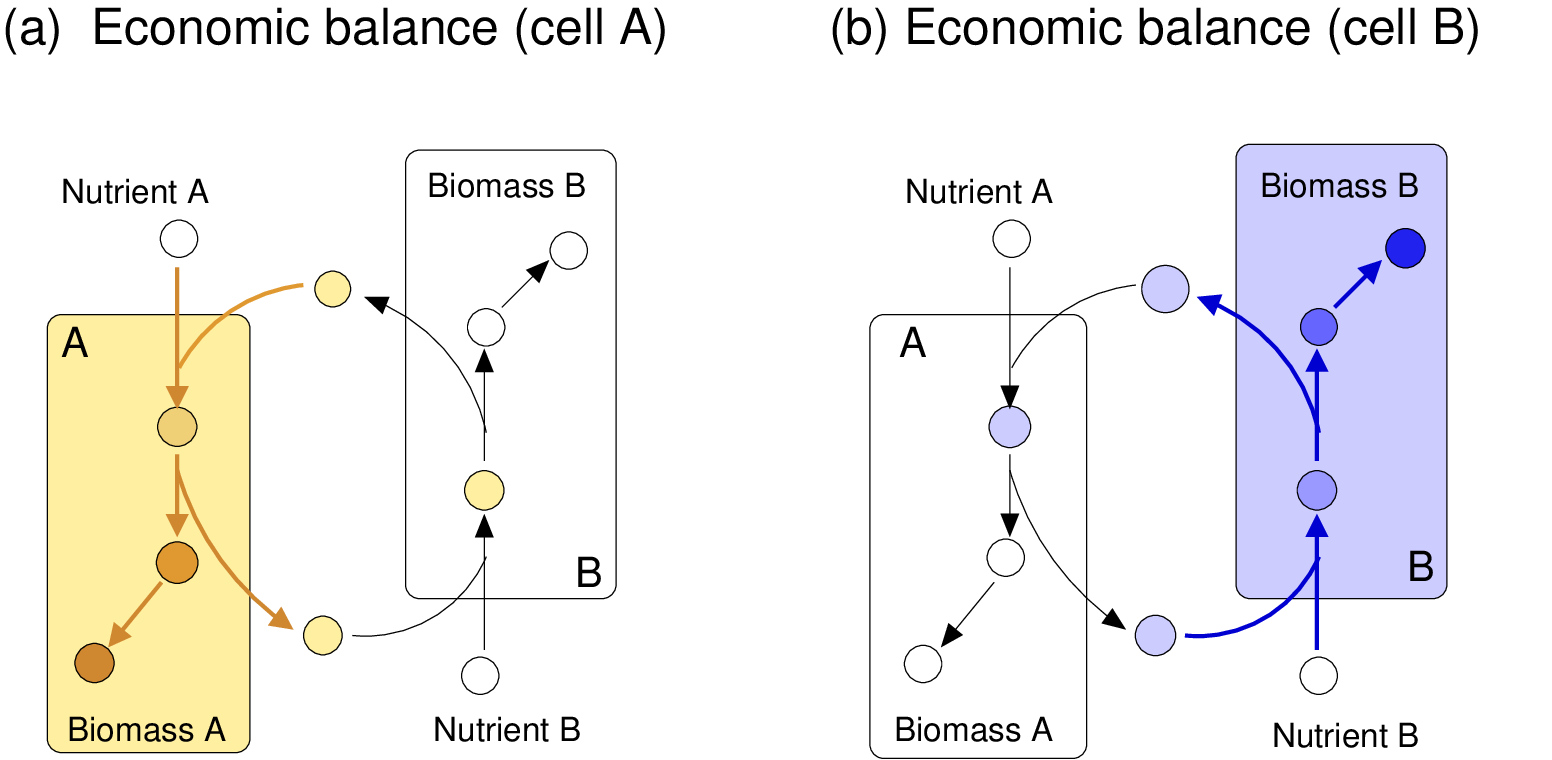}
 \caption{\co{erklaeren: ohne B weaere der fluss in A nicht moeglich!}
   \co{for precise quantification (usefulnes of B for A) see CBA
     communities} \co{In FN; what about example producer / cleaner? should also
     be possible! that's maybe the more relevant bcs well-known  example? nee, komplizierter! if actetate accumulates, (assuming production + slow dilution) thermodynamics makes enzymes in cell A inefficient + therefore very expensive! that is, no solution is possible, because total investment is larger than total value production (ref to FCM changing fluxes)}
   Cross-feeding cells share a {\flow} that is economical for each of
   the cells. In the schematic, cells A and B provide each other with
   essential compounds.  Left: Economic potentials derived from
   biomass production in cell A, as a metabolic objective (shades of
   brown).  Reactions in cell A (brown arrows) must satisfy a value
   production balance for these potentials.  Right: Biomass production
   in cell B leads to another set of economic potentials (shades of
   blue), defining value production balances to be satisfied within
   cell B. A symbiosis requires a Nash equilibrium, in which neither
   of the cells would benefit from changing its enzyme profile. As a
   necessary condition, each cell must satisfy the value production
   balance for its own economic potentials, and within ins own
   enzymatic reactions. }
 \label{fig:symbiosis}
 \end{center}
\end{figure*}

\myparagraph{Cell community models} If several objectives are given,
it is unlikely that all of them can be optimised by a single metabolic
state (i.e.~by the same fluxes, metabolite concentrations and enzyme
concentrations).  However, if we do not require optimal flux profiles,
but just enzyme-beneficial ones, \co{why would this be interesting?} a
solution for several objectives may well exist. The situation
resembles the case of a flux profile that simultaneously satisfies
energetic and economic constraints (derived from ``two objectives'',
represented by chemical and economic potentials, and with right-hand
sides of the balance equations representing energy dissipation and
{\enzymeinvestment}). \co{WD?:} In a multi-objective optimisation,
each objective defines a set of economic potentials (a vector of
economic potentials, as it were). A ``lucky'' {\flow} that satisfies a
value production balance for each type of potentials is extremely
unlikely, but a variant of this can be used to model cell communities.
Figure \ref{fig:symbiosis} shows an example, two cross-feeding cells
which exchange compounds through a common {\flow}.  Each cell has its
own objective (maximising its own biomass rate), defining two sets of
economic potentials in the network containing both cells.  To reach a
maximal fitness, each cell must satisfy the value production
balancesin its own enzymes, and with its own economic potentials: so
as usually we obtain exactly one balance equation for each reaction.

\myparagraph{Nash equilibrium states} \co{FN: An ESS is a special case
  of a Nash equilibrium.  instead, it's a single strategy that
  dominates all possible ``mutant'' stategies, i.e.~if all cells use
  this strategy, there is no beneficial niche for a small number of
  cells with another strategy.} In a cell community model, a Nash
equilibrium is a state in which none of the cells would benefit from
changing their fluxes and enzyme levels, given the behaviour of all
other cells. All balance equations must be satisfied in cell A for the
economic potentials of A, and in cell B for the economic potentials of
B. To find such states, we may first run a multi-species VBA (in which
each cell needs to satisfy its own balance equations, and with respect
to its own benefit function; this is a necessary condition for a Nash
equilibrium!  The approach can be extended to cases with more than two
species.  Multi-species FBA models already exist, but these models
typically employ a single ``community objective'', that is, they
presuppose symbiosis instead of explaining how symbiosis arises from
cells pursuing their own benefit: there must be a Nash equilibrium in
which each cell maximises its species-specific objective, given the
other cell's behaviour. VBA takes this into account.  This approach --
assuming ``selfish'' cell with different objectives, but sharing a
common flux distribution, and searching for VBA solutions -- describes
not only cross-feeding, but also other Nash equilibrium states: for
example, between a parasite and a host or between cells that are
trapped in a ``tragedy of the commons'' dilemma, in which they consume
nutrients fast, and inefficiently, just to gain a speed advantage over
the other cells.

\co{\textbf{Communities and Pareto optimality} say that this was
  pareto's original usage of the concept! GIVE DEF! stefan
  schuster/markus koebis fragen wegen "partial pareto optimality"
  (meine eigene idee: für jede variable gibt es eine der
  zielfunktionen, sodass die variable bezueglich dieser zielfunktion,
  und gegeben alle anderen variablen, optimal ist?)}  \co{nee, hier
  eigentlich: es ist ja vorgegeben, welche variable welche
  zielfunktion optimieren muss! } \co{uebergang - relation to general
  pareto problems unklar; allg beziehung pareto / multi-species
  unklar} \co{say how this may relate to a Pareto front treatment of
  ecosystems} \co{i.e.~to find a linear combination of the objectives
  such that the economic potentials (or: for a linear combination of
  the objectives) the balance equations are satisfied.}  \co{FN was
  ist mit variante von pareto, wo jede einzelne variable optimal ist
  bezueglich eines (der mehreren) kriterien). // ausfuehren in FN!
  generalise from multi-species scenarios to completely other cases?}
\coout{o syntrophy growth Stolyar et al 2007 MSB // Mee et al, PNAS
  2014 "syntrophic exchange ..."  synthetic ecosystem discovery: A and
  B cannot grow, but A+B can} \co{REF zu cba bact comm! ziel:
  symbiose, die effizient ist (und sich gegen ineffinizente
  verteidigen kann) // In chemostat (bei gleichem Wachstum): //
  kooperative waere moeglichst geringe importrate (d.h. viele zellen
  koennen bestehen); yield! // egoistisch waere: bestehen bei
  moeglichst geringer auesserer konzentration: (rate?)  (max catalytic
  rate? (which might entail max yield in some cases))}

\section{The principle of minimal fluxes}

\myparagraph{\ \\Flux cost minimisation} \co{SORT parag! logik: since
  solving Eq (11) by milp solvers may be difficult, it is good to have
  a practical method that generates solutions. in fact, ... FCM and
  mcFBA and such methods do this! .. dann methoden erklaeren} A
feasible metabolic state, according to VBA, consists of an
economically and thermodynamically feasible {\fluxpattern} and of
fluxes $\vv$, chemical potentials $\muv$, and economic potentials
$\wint$ compatible with this pattern. In fact, finding the flux
pattern is the most difficult step: all other steps consist of linear
problems (which can be satisfiability or optimisation problems, or
variability analyis, depending on the puptpose of modelling)
require. A feasile flux pattern can be obtained from a given feasible
flux profile.  In models with a production objective, \co{we know that
  economical fluxes are thermo-feasible!} we can compute such flux
profiles by FBA with flux cost minimisation or molecular crowding
\co{letzteres bewiesen? in CBA lag zeigen!}.  The principle of minimal
fluxes \cite{holz:04}, a heuristic rule for flux prediction,
postulates that cells realise a given flux benefit at a minimal sum of
(possibly weighted) absolute fluxes. \co{JA! FN: originally, apply
  after flux benefit maximisation to avoid loop fluxes and highly
  underdetermined solutions; but also possible as the first step (with
  an invented benefit)! also ref to dill paper ..!}  Flux Cost
Minimisation \cite{lieb:18fcm} generalises this principle to other
flux cost functions, which can be linear (e.g.~a weighted sum of
fluxes in FBA) or nonlinear.  Realistic flux cost functions
(e.g.~representing the cost of catalysing enzymes) follow from kinetic
models \cite{nfbd:16,wnfb:18}: given a {\flow} $\vv$, we search for
the enzyme and metabolite profiles that realise these fluxes at a
minimal cost. This minimal cost, as a function of $\vv$, yields an
effective flux cost function\co{JA! REF Wortel, FCM}.  \co{The
  derivatives of this enzymatic flux cost function (``flux burdens'')
  are given by avl = hu/kapp = hu/k(copt))} \co{the same procedure,
  based on enzyme plus metabolite cost, yields the ``kinetic flux cost
  function'' REF FCM}

\begin{figure*}[t!]
  \begin{center} 
    \includegraphics[width=14.5cm]{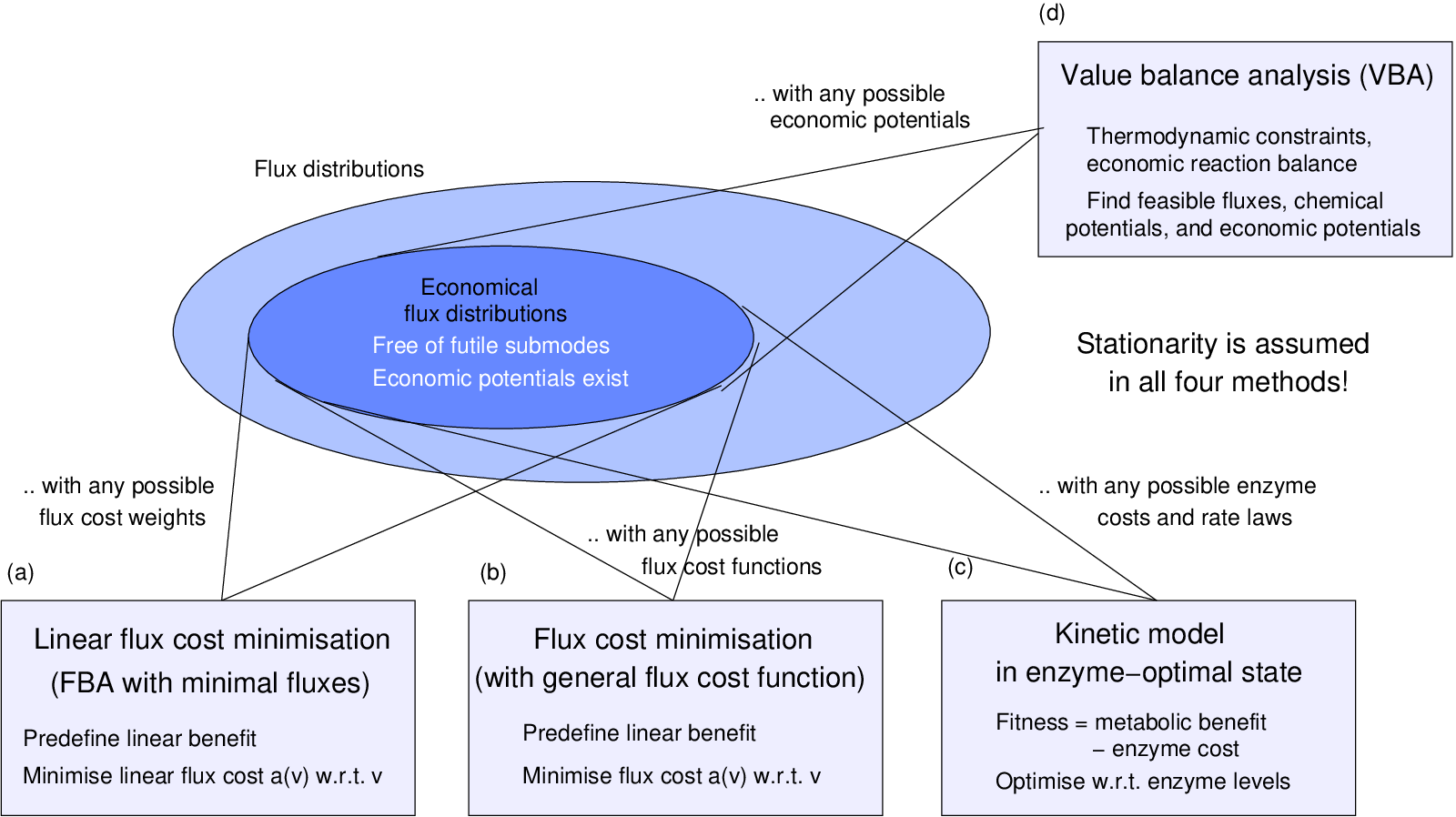}
    \caption{Different optimality principles yield the same set of
        flux solutions.  Value Balance Analysis (VBA) is a variant of FBA
      that restricts  flux profiles to potential solutions of
      underlying resource allocation problems (including  linear FCM, FCM,
      and kinetic models).   (a) Linear FCM minimises a flux cost
      $\fluxcost(\vv)$ at a given flux benefit $\fluxbene(\vv)$, with
      linear cost and benefit functions
      $\fluxcost(\vv) = \av_{\rm v} \cdot \vv$ and
      $\fluxbene(\vv) = \bv_{\rm v} \cdot \vv$ and reaction
      orientations such that $\vv \ge 0$.  The optimality condition
      reads $(\Deltar \wtot + \bvtot) = \frac{1}{\xi}\, \av_{\rm v}$,
      with Lagrange multipliers $\wtot$ and a positive scaling factor
      $\xi$) \cite{lieb:18lagrange}.  (b) Nonlinear FCM works
      similarly, but with nonlinear flux cost functions. (c) In enzyme
      optimisation, the aim is to  maximise a fitness
      $\ffit(\vv,\cv,\esymbolv)= \bbenefit(\vv)-
      \metcost(\cv)-\hminus(\esymbolv)$ while requiring $\Nint\,\vv=0$
      (stationarity) and $\vv=\ratev(\cv,\esymbolv)$ (rate
      laws). \co{With a flux gain vector \co{btotalv?}
        $\bvdir = \partial \bbenefit/\partial \vv$, enzyme price
        vector $\huv = \partial \hminus/\partial \esymbolv$, and
        enzyme elasticity matrix $\Eunu$, the optimality condition
        reads $(\Deltar \wtot + \bvdir) =\Eunu \inv\, \huv$.} (d) All four modelling frameworks \co{framework
        UEA} predict economical flux profiles, and each of them can
      predict any economical flux profile depending on model
      parameters.  While  the economic reaction balance is an optimality condition all
      of the previous methods, in
      VBA it is employed as a constraint.}
    \label{fig:CBA_FCM_venn_diagram}
  \end{center}
\end{figure*}

\myparagraph{VBA, flux cost minimisation, and enzyme optimisation have
  the same flux solutions} Flux cost minimisation is closely related
to VBA. With the value production balance as an optimality condition,
it is no surprise that FCM predicts economical {\flow}s\footnote{In
  FCM without flux bounds (e.g.~upper bounds to restrict fluxes
  or lower bounds to enforce them), the logical connection  to VBA is easy to
  see. Flux constraints in an FCM problem lead to extra shadow
  values that act as effective flux gains and appear in the flux
  benefit function in VBA and in the economic reaction
  balance. \co{nebenbei: in FCM with such constraints, the solutions
    need not be EFMs, but may be combinations of these efms and
    extreme rays are identical on signed flux polyt. in contrast, flux
    min solutions may still be convex combs of efms} SIEHE ENDE SECTION 5 (``principle of min fluxes'')}! In fact,
different optimality problems -- Flux Cost Minimisation (including FBA
with weighted flux minimisation) and enzyme optimisation in kinetic
models -- yield the same sets of solutions, which are also the
solutions of VBA (see Figure \ref{fig:CBA_FCM_venn_diagram}).  In each
method, the solutions depend on model parameters (e.g.~flux cost
weights in linear FCM or enzyme cost functions in kinetic models), but
the \emph{range of possible solutions} (obtained by screening all
model parameters) is always the same.  An economical flux mode $\vv$
(with some flux benefit function $\fluxbene(\vv)$) is the solution of
a linear FCM (with the right flux cost weights), of a nonlinear FCM
(with the right flux cost function), and of an enzyme optimisation
(with the right enzymatic rate laws and enzyme cost function): all
these methods predict econonomical {\flow}s, and their solutions also
can also be described by VBA with the right potentials and
{\enzymeinvestment}s.

\myparagraph{Why different modelling approaches predict economical
  flux modes} Why do all these optimality problems predict economical
flux modes? The reason is that all of them share an optimality
condition of the same form, a value production balance. In VBA, this
value production balance is imposed as a dogma. In all cases,  value production
balance arises from mass-balance constraints. In optimality
problems, these constraints  give rise to shadow values. \co{FN: To define shadow values, we
imagine that a constraint mass balance were relaxed (e.g.~if a
mass-balance constraint could be violated by virtual in- or outfluxes)
and ask by how much the metabolic objective would increase.  In other
words, shadow values describe an incentive to break these
constraints\footnote{In our models, constraint-violating variations
  serve as a mathematical tool for formulating optimality
  conditions. However, variations may also be real: if a pathway is
  coupled to other parts of the cell, changes outside the pathway may
  lead to extra in- or outfluxes, which can be described by variations
  and are associated with costs and benefits outside the
  pathway.}. Any optimality problem that contains mass balance
constraints (e.g.~in enzyme optimisation, flux cost minimisation,
resource balance analysis, or other optimality problems) leads to
shadow values (our ``economic potentials''), satisfying the same
balance equations and defining economical {\flow}s as the possible set
of solutions.} \co{JA! ERKLAEREN! mention ``production economics'' and
  ``concentration economics''; evtl schon in Intro? sagen, dass VBA
  genau den ``production economics''-teil von metabolic value theory
  betrifft (waehrend der zusaetzliche ``concentration economics'' teil
  in kinetischen modellen hier offengelassen wird, und daher viele
  kinetische modelle im einklang mit VBA moeglich sind).}  There is another
explanation for the strange agreement between modelling frameworks.
Any enzyme-optimal kinetic model can be converted into a linear
FCM problem with the same solution: given an enzyme-optimal state, we
constrain our model to the optimal metabolite concentrations; fluxes
and enzyme levels are now proportional, and enzyme cost can be
minimised by solving a linear FCM problem! On the contrary, given the
solution of a linear FCM problem (with flux prices $\hvl$ and solution
$\vv$) and a kinetic model that can realise the same fluxes (with a
enzyme profile $\esymbolv$), we can adjust the enzyme prices such that
$\hul = \frac{v}{\esymbol}\, \hvl$, thus putting the kinetic model
into an enzyme-balanced state. On a more theoretical level,
a kinetic model (with an optimal state  $(\vv,\cv,\esymbolv)$)
can always be replaced by an FCM problem (with an optimal flux profile $\vv$),
where the  given benefit function $b(\vv)$ stems directly from the
kinetic model and the flux cost function $\acost(\vv)$ is the
enzymatic flux cost function for the kinetic model, $\min_{\cv,\esymbolv} \hminus(\esymbolv)$
subject to $\ratev(\cv,\esymbolv)=\vv$.  The FCM problem recovers the
kinetic problem, but restricted to a submanifold in state space (which
contains the optimal point). This shows that FCM \co{das in extrsatz sagen:(and even linear FCM)}
is a good starting point for a systematic kinetic model construction
as described above (also called ``layered modelling''). \co{REF CBA
  local\cite{lieb:cbalocal} / CBA field\cite{lieb:cbafield}}

\myparagraph{Linear FCM as a method for finding enzyme-optimal states}
We already saw that economical, thermodynamically feasible {\flow}s
can be easily found by linear FCM. By varying the flux cost weights,
any economical flux profile can be obtained, and by sampling cost
weights at random, we can construct an ensemble of economical flux
profiles, with given flux directions.  \co{FN Note that the linear FCM
  problem must not contain flux-enforcing constraints.  Otherwise, in
  VBA we need to consider flux constraints on the same reactions,
  which leads to extra shadow values in the flux gain vector!}
\co{uebergang!} Each of these profiles yields a feasible flux pattern,
\co{JA! FN: Even if in FCM the flux DIRECTIONS were already
  predefined, this would not completely determine the flux pattern
  (because in the optimal solution fluxes can still be shut down).}
and for each of these patterns, other economical flux modes can be
found by classical FBA or by flux sampling. As a side results, the FBA
solutions yield shadow prices for mass-balance constraints, which can
be seen as economic potentials. All the resulting solutions satisfy
the constraints of VBA and can be realised by kinetic models in
enzyme-economical states.

\co{wie solllte man nun die FCM-parameter waehlen?}  \co{explain this
  further up bei interpretation; paragraph: VBA and elementary flux
  modes} \co{Briefly ref to Steuer etc results, say that EFM result
  would mean: solution only on polytope vertices! (move some text from
  further down (``Pure or combined..'') to here?)}

\coout{In comparison to FBA, FCM reduces the solution space (possibly
  to a single {\flow}) and suppresses fluxes that do not contribute to
  the metabolic benefit.  Unlike FBA, it can predict fluxes of
  suboptimal yield \cite{bvem:07}. A simple flux minimisation, with
  costs given by the sum of absolute fluxes, favours short pathways,
  for instance the direct route from X to B in Figure
  \ref{fig:fbacomparison} (d). Weighting the fluxes differently, for
  example, by assigning higher weights to lumped reactions, makes the
  cost function more realistic. If flux costs are meant to be proxies
  for {\enzymeinvestment}s, the flux weights may be chosen to be
  proportional to enzyme chain length $L_{l}$ and effective rate
  constant $\enzdegrate_{l}$ for enzyme degradation
  \cite{bnlm:10,horh:11}, and inversely proportional to the catalytic
  constant $k^{\rm cat}_{l}$. \coout{and a simple formula would be
    $\acostvl = L_{l}\,\enzdegrate_{l}/k^{\rm cat}_{l}$.} In reactions
  close to chemical equilibrium, the net flux is considerably reduced
  by backward fluxes \cite{fnbl:13,nflb:13}; this burden can also be
  taken into account in the flux cost (see Eq.~7 in \cite{holz:04}).

  \coout{If we compare different models (no matter which!)  with
    linear flux objectives that have their optimum at the same flux
    benefit: for all of them, the {{\enzymebenefit}} $\sum_{l}\hul\,\ul$
    must be equal! If they have their optima at different flux
    benefits, then $\sum_l \bvtotl\,v^{(\alpha)}_{l} = \sum
    ´{\hul}^{(\alpha)} \,\ul^{(\alpha)} $}}

\coout{zur bedeutung der flusskosten und oek pot, in FCM.  Search for
  cases where both models describe the same system, so that
  $\vv'=\vv$, $\hat \bvtot= \bvtot$. In order for $\wint'=\wint$ to
  hold, it has to hold that $\frac{1}{\xi}\,q_l=y_l$. In order for
  $\frac{1}{\xi}\,q_l=y_l$ to hold, $\wint'=\wint$ has to hold except
  for gauging. Given an ME optimum, we can construct an FCM model
  which will have the same optimum. Given an FCM optimum, we can
  construct an ME model with the same state as an extremum.}

\coout{Cite Steuer + Teusink papers} \coout{explain again (or
  elsewhere) how  regulation by effector molecules and post-translational
  modifications are treated.}  

\coout{Weitere beziehungen (kurz erwaehnen): Def $\loadint =
  {\Eunint}\trans\,\gvtot$. Es gilt $\gc= \loadint - \hc$ where $\hc = -
  \partial \bbenefit/\partial \cv$ and where $\gc$ must have the form $\gc =
  \Gmat \trans \, \loadint_{\rm cm}$ or 0 (if there is no moiety
  conservation).}

\section{Futile cycles} 

\co{WO? kleine illustrationen zu futile and wasteful

  \includegraphics[width=5.5cm]{/home/wolfram/projekte/cba/zeichnungen/A22.jpg}\\
unmöglich! Futile

\includegraphics[width=5.5cm]{/home/wolfram/projekte/cba/zeichnungen/A23.jpg}\\
unmöglich! wasteful}

\begin{figure*}[t!]
\centerline{\includegraphics[width=16.5cm]{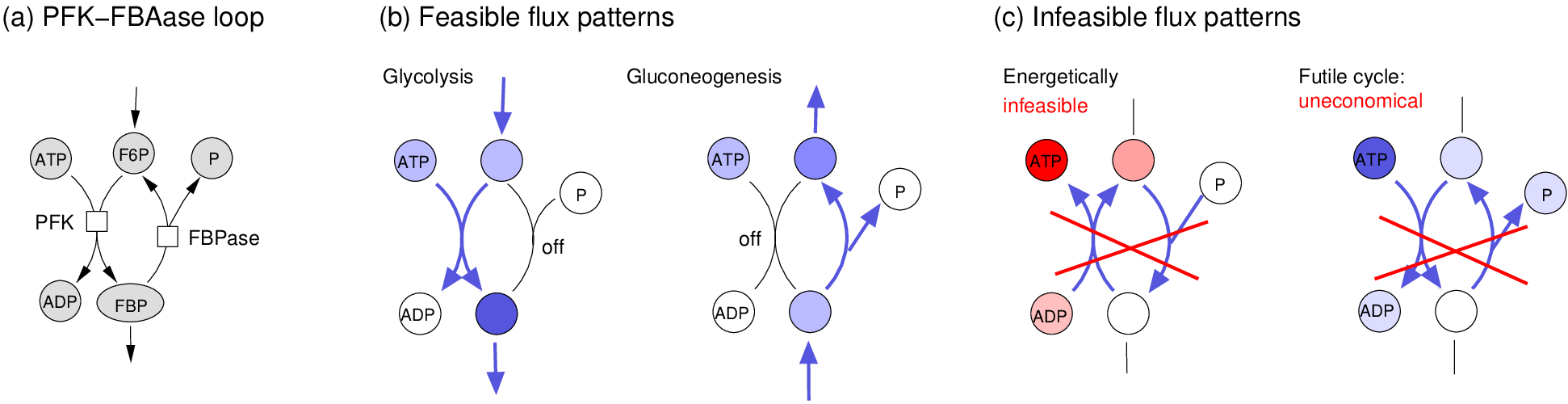}}
\caption{Feasible fluxes in upper glycolysis.  obtained by
  excluding infeasible cycles. (a) Phosphofructokinase (PFK) and
  fructose bisphosphatase (FBPase), two enzymes in upper glycolysis,
  form a cycle that can burn ATP.  In cells this cycle is often
  suppressed. (b) Feasible {\fluxpattern}s.  PFK is used in glycolysis
  (left), and FBPase is used in gluconeogenesis (right). The economic
   condition for a forward PFK flux reads
  $w_{\rm FBP} + w_{\rm ADP} > w_{\rm F6P} + w_{\rm ATP}$, and for
  FBPase it reads $w_{\rm F6P} + w_{\rm P} > w_{\rm FBP}$. In each
  single reaction, there can be consistent economic potentials (where
  $w_{\rm ATP} > w_{\rm ADP} + w_{\rm P}$) (shades of blue) and
  chemical potentials (not shown) compatible with the flux
  direction. (c) Infeasible flux cycle.  Left: in the ATP-producing
  (counterclockwise) direction, the cycle is thermodynamically
  infeasible at physiological metabolite concentrations (chemical
  potentials shown in red).  Right: in clockwise direction, the cycle
  consumes ATP and is thermodynamically feasible.  However, the cycle
  is \wasteful\ if the economic potentials decrease from ATP to ADP
  and phosphate, and if there is no flux {{\gain}} (e.g.~no advantage
  from heat production). To exclude cycles in either direction, one
  the enzyme must be repressed. Abbreviations: F6P:
  fructose 6-phosphate; FBP: fructose 1,6-bisphosphate; P: inorganic
  phosphate.}
    \label{fig:pfk_example}
\end{figure*}

\myparagraph{\ \\ Futile cycles} We can  now address some questions
raised in the introduction. Are there {\flow}s that entail a waste of
enzyme in \emph{any} kinetic model? And can we recognise them by
typical ``futile patterns''?  A good example is futile cycles that
degrade valuable compounds without an obvious benefit.  Figure
\ref{fig:pfk_example} shows an example, a cycle formed by two enzymes
in upper glycolysis. Phosphofructokinase (PFK) transfers a phosphate
group from ATP to fructose 6-phosphate (F6P), converting it into
fructose 1,6-bisphosphate (FBP).  Fructose bisphosphatase (FBPase)
catalyses the backward reaction, but instead of generating ATP it
releases phosphate.  Together, the two enzymes can drive a cycle that
splits ATP into ADP and phosphate and releases heat.  To save valuable
ATP, cells can interrupt the cycle by inhibiting or repressing at
least one of the enzymes. In FBA, this cycle is often excluded
manually by applying a constraint.  But can we justify this more
generally, based on a principle of optimal enzyme usage?

\co{Aims below: 1. nutzen fuers verstaendnis von wegen,
  netzwerkreonstrukition; regulation. 2. klarere definition von
  ``nutzlos'', abgrenzung von and kombination mit thermodynamischen
  zyklen.}

\myparagraph{Futile or wasteful flux motifs indicate a waste of enzyme
  or substrate} In VBA, \todo{instead of spotting futile cycles
  intuitively,} the notion of ``futile cycles'' is given a precise
meaning.  Futile or wasteful flux motifs are hallmarks of uneconomical
enzyme usage.  As we learned above, economical flux distributions must
be free of futile or wasteful motifs!  \co{Non-beneficial motifs show
  that a flux distribution is uneconomical, and we can recognise
  uneconomical flux modes by finding elementary non-beneficial
  submodes.} In a \flow\ $\vv$, a {\futile} motif is a set of active
reactions that can support, by itself, a {\futile} flux profile with
the same flux directions as in $\vv$.  Similarly, a wasteful motif is
a reaction subset that can support, by itself, a wasteful flux profile
(with the same flux directions as in $\vv$). Futile submodes may look
like cycles, but may also arise in linear pathways without a valuable
product.  \co{which of the following results hold only for production
  objectives?  sicherheitshalber am anfang: production objective
  annehmen? which of the conclusions hold for futile, resp wasteful
  cycles?}  Unlike existing verbal or topological definitions of
futile cycles \cite{kdus:10}, \co{JA! FN: sagen: hier etwas anderer
  sprachgebrauch, mit distinction between futile und wasteful} this
algebraic definition allows for clear statements about optimal
metabolic states: futile or wasteful motifs make a flux mode
uneconomical, i.e.~incompatible with a feasible choice of economic
potentials or optimal enzyme profiles in kinetic models.  Moreover, a
futile motif tells us in which reactions resources are wasted. \co{A
  futile motif shows that enzyme is wasted (i.e.~used without
  generating a sufficient benefit). A wasteful motif shows that the
  current enzyme usage even decreases the benefit, so enzyme is not
  only wasted, but it contributes to decreasing metabolic performance
  (where performance refers to flux benefit minus metabolite cost).}
\co{FN: can the two types of cycles be associated with rate and yield?
  are they related to rate-yield trade-offs? In a model with
  production objective, a flux profile with a wasteful submode cannot
  have maximal product yield (and thus benefit); the yield can be
  imporved by subtracting the submode. Analog fuer futile mode +
  enzyme cost? ICH denke ja .. decreasing th value yield vs decreasing
  the rate ..}

\co{nochmal kriterium mit ec pot wiederholen and ganz
  klarmachen (einschleisslich bv)\co{FN? note that flux bounds can
    give rise to shadow flux gains that appear  in the
    definition of futile and wasteful motifs, and can be represented in the form of
    virtual extra metabolites}// Note that wasteful modes are just
  beneficial modes in reverse! so avoiding wastful modes means: given
  the (changing) boundary ec potential, either make sure htat the flux
  goes in the right direction, or interrupt the mode by regulation!
  For (structurally) futile modes, it just means: interrupt the mode!
  (allg distinguish between structurally and momentarily futile
  modes?)}

\co{
\co{\textbf{Futile and wasteful cycles}} \co{(FN wasteful modes are
  basically beneficial modes in reverse direction)} \co{JA!  NOCHMAL
  kurz beide kriterien sagen!  distinguish between futile and wasteful
  when explaining futile cycles: futile: direct sign of non-economical
  state (even if no potentials are known); wasteful auch, bei
  production objective?  nachdenken!// what about {{\wasteful}}
  cycles?  can they be found?  they are the ones that are mostly
  considered when it comes to ``futile cycles'' // check: can we
  exclude wasteful cycles, but using a similar argument as for futile
  cycles?}  While futile submodes waste enzyme only, {\wasteful}
submodes also waste metabolite value (e.g.~negative value production
by consumption of ATP without other benefits).  \co{WEG?  Cells may
  avoid a waste of metabolites but still waste enzyme, e.g.~by
  expressing and inhibiting FBPase simultaneously to facilitate flux
  reversals.  The expressed subnetwork would allow for a {\wasteful}
  submode, \co{REF oliver+me} but this submode does not appear in the
  kinetically active subnetwork and in the actual {\flow}.}
}

\co{\textbf{Futile motifs and thermodynamics} JA! SCHREIBEN! How are
  futile cycles related to reaction thermodynamics? Note that futile
  motifs need not be thermodynamically infeasible (see Fig \todo{3}
  for an example). In models with a production objective, all
  thermodynamic cycles are also futile. However, since
  thermodynamically infeasible modes cannot be active, cells do not
  have to actively suppress the by regulation. Hence, if enzyme
  economics is assumed: in a model (assuming optimal states), cycles
  that are nonbeneficial or thermodynamically infeasible must be
  excluded; in a real cell, all nonbeneficial BUT THERMODYNAMICALLY
  FEASIBLE cycles must be avoided (eg by regulation) to avoid
  non-optimal states.}

\myparagraph{Futile cycle in upper glycolysis} Let us come back to the
PFK-FBPase cycle. To see why cycle fluxes should be repressed, we
analyse all feasible and all infeasible submodes, considering
thermodynamic and economic constraints \todo{(Figure 10)}.  \todo{The
  choice between glycolysis and gluconeogenesis (without cycle flux)
  depends on the economic potentials. Generally, a high economic
  potential of FBP favours glycolysis, while a low economic potential
  favors gluconeogenesis: however, there can be economic potentials
  for which both flux are uneconomical: if both
  $w_{FBP}+w_{ADP}<w_{F6P}+w_{ATP}$ and $w_{F6P}+w_{P}<w_{FBP}$ (which
  implies that $w_{ADP}+w_{P}<w_{ATP}$ (note that water has been
  omitted for simplicity.),} we find that the cycle flux is
economically infeasible in forward direction and thermodynamically
infeasible in reverse.  If ATP (plus water) has a higher value than
ADP (plus phosphate) and if there is not direct flux beneficial
(e.g.~heat production being beneficial), the PFK-FBPase cycle in
forward direction is not beneficial, and {\flow}s with this cycle are
uneconomical. We can tell this just from the economic potentials,
which represent fitness demands in the entire cell: given these
potentials, it does not matter whether ATP, ADP, and phosphate are
modelled as external or internal metabolites, and not even whether the
rest of the network is known.  Running the cycle in reverse
\todo{(Figure 10c, left)} would produce ATP, but it would require a
drop in chemical potential between FBP and F6P, implying
unphysiological concentrations.  Since both cycle cirections are
infeasible, to avoid an uneconomical flux the cell needs to interrupt
the cycle by repressing an enzyme. \co{FN inhibiting; avoid waste of
  metabolite value, but not waste of enzyme; repressing: save waste of
  both metabolites and enzymes} This is no surprise, but now we have
found a way to formally derive this very generally from metabolic
models. Of course, this does not exclude futile cycles in
reality. Here we just say that these cycles would contradict certain
simple optimality principles, and how this can be checked.
  
\begin{figure*}[t!]
  \begin{center}
    \begin{tabular}{lll}
      (a) Futile PFK/FBPase cycle & (b) Number of {\futile} submodes per reaction\\[2mm]
      \includegraphics[width=7.5cm]{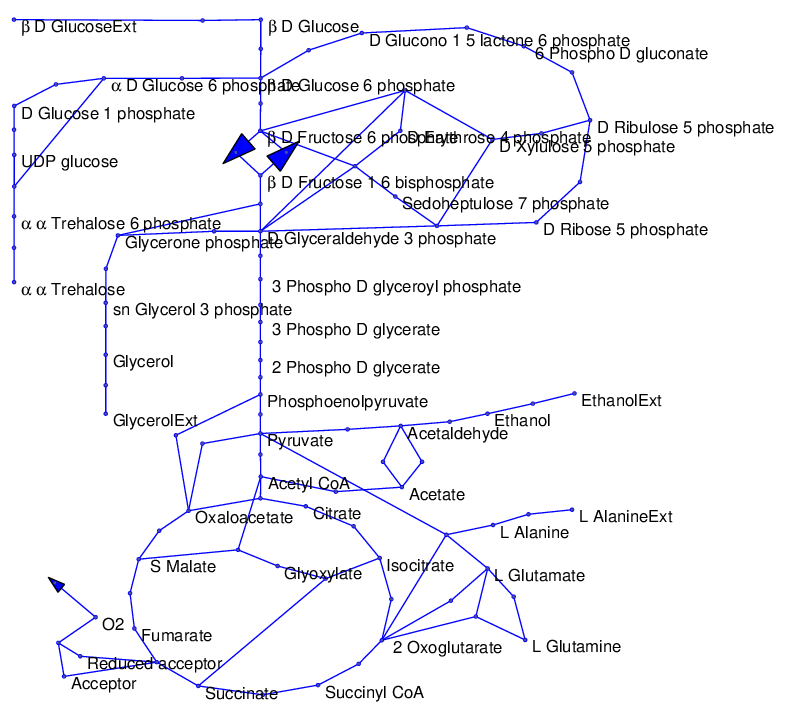} & 
      \includegraphics[width=7.5cm]{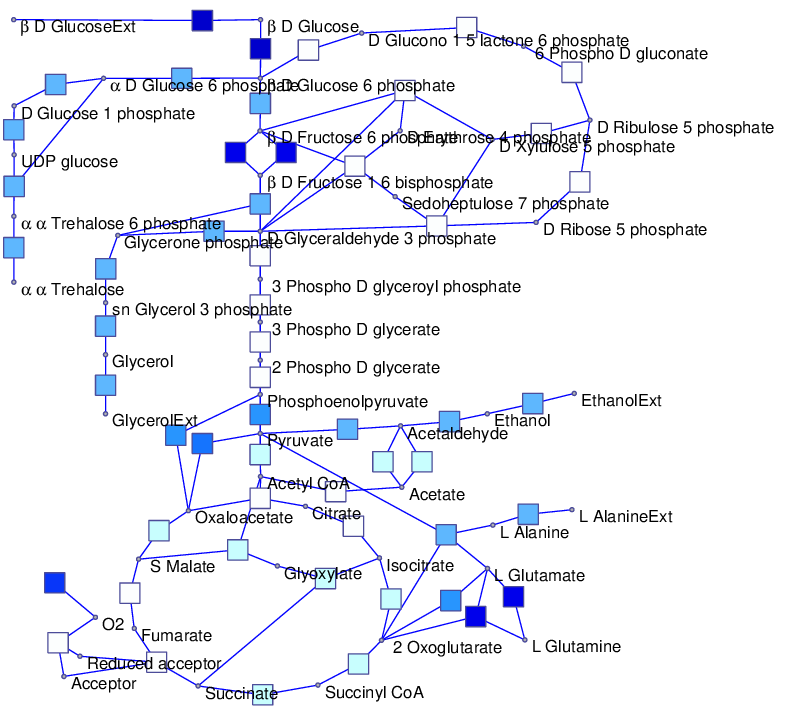}
    \end{tabular}
    \caption{\co{alles etwas gerader zeichnen (auch in CBA kin)}
      \co{in a kreise etwas groesser (wie in fig 6); in b show
        reaction names instead of met names?}\co{Pfeile sch\"oner
        (dicker, spitzen vor); show and discuss larger futile cycles
        (less obvious ones); try to show something surprising with
        futile cycles} Futile cycles in central metabolism of yeast
      (model from Figure \ref{fig:yeastpotentials}).  The network
      structure contains two elementary cyclic modes (which are
      thermodynamically infeasible) and 303894 elementary
      {\nonbeneficial} \co{stimmt das? oder futile?}  submodes (all of
      which are thermodynamically feasible).  (a) Elementary futile
      flux mode (shown by arrows), comprising the PFK/FBPase cycle
      (see Figure \ref{fig:pfk_example}). The mode produces ATP
      through oxidative phosphorylation (bottom left) and consumes it
      again in the PFK/FBPase cycle.  (b) Involvement of reactions in
      {\nonbeneficial} \co{futile??}  modes.  For each reaction, the
      count number of such modes is shown (white: 0; dark blue: 10);
      to focus on smaller cycles, large flux modes (comprising more
      than 10 reactions) were not considered. \co{wieviele sind das?}
      \co{give justification! and produce graphics where the are not
        neglected!} Reactions with large count numbers (dark blue)
      would be plausible targets for regulation.}
    \label{fig:yeastfluxanalysis}
  \end{center}
\end{figure*}

\myparagraph{How to detect futile cycles} If all enzymes in a cell
were expressed simultaneously, this could easily lead to futile or
{\wasteful} cycles.  To disrupt futile cycles, cells need to express
enzymes selectively, i.e.~repress some enzymes. How can we describe
this in models?  To disrupt futile cycles, we first need to find all
such cycles in a given flux distribution.  To see if a given flux
distribution is economical, we may search for compatible economic
potentials (by solving a linear programming problem): if no solution
exists, the flux distribution is uneconomical. However, this does not
tell us where the problem is localised, i.e.~what reactions we need to
suppress to make the {\flow} economical. \co{sagen: hier focus on
  FUTILE modes} In theory, we can detect and remove futile cycles by
enumerating and subtracting all {\nonbeneficial} submodes, but in
practice this would be impossible: any given pair of flux
distributions can be linearly combined to yield a {\futile} {\flow}!
Luckily, we only need to consider \emph{elementary} {\futile} test
modes\footnote{A {\futile} submode is elementary if it
  does not contain any smaller {\futile} submodes
  \cite{scdf:99,faqb:05}.  Elementary {{\wasteful}} modes are defined
  similarly.}, \co{proof???} which can still be enumerated in medium-sized networks.
The yeast metabolic model in Figure \ref{fig:yeastfluxanalysis}
contains 303894 elementary {\futile} modes (calculated by efmtool
\cite{efmtool, test:08}).  \co{evtl wichtig: direkte berechnung
  thermodynamisch korrekter efms:
  http://www.ncbi.nlm.nih.gov/pmc/articles/PMC4354105/} Many of these
modes are large (the average size is 32 active reactions).  \co{WO?
  ja, ist interessant, dass sie so lang sind!  man wuerde sie alleine
  nicht sehen} By comparing a {\flow} to each of these modes, we can
find its futile motifs, and thus the reactions that make it
uneconomical.

\co{Check
  examples of predicted regulation in yeast more thoroughly.  ELAD:
  maybe think of other less trivial examples, e.g. the combination of
  PEP carboxylase and the malic enzyme (although I think that
  {\nonbeneficial} mode actually is happening), or combinations of
  amino acid synthesis and catabolism.}

\myparagraph{Eliminating the futile cycles in a given {\flow}} \co{REF
  to nidan + liang 2007 (they probably have a problem though)} If a
flux profile contains known futile motifs, we can use them to correct
the fluxes.  By using a flux motif as a flux variation $\modevector$,
we can substract it from the {\flow} and interrupt the cycle.  In the
corrected {\flow} $\vv - \alpha\, \modevector$, the prefactor $\alpha$
is chosen to cancel one reaction flux in the submode, but without
reverting the flux directions, which leaves it thermodynamically
feasible.  By repeating this procedure, any infeasibile submodes can
be removed (see SI \ref{sec:removefutilecycles}). An example is shown
in Figure \ref{fig:examplefutilecycle}. To remove the cycle from flux
mode (c), we subtract from it the elementary cycle from Figure (b) and
obtain the {\flow} shown in (a).  Thermodynamically infeasible modes
can be removed similarly and along with the futile modes
\cite{sclp:11}.  \co{(aehnlich price (2006))}

\myparagraph{Futile cycles can be suppressed by enzyme regulation} To
ensure economical {\flow}s, cells need to suppress {\nonbeneficial},
thermodynamically feasible submodes. They can do this by repressing
one enzyme in each of these submodes\footnote{If enzymes are repressed
  transcriptionally, protein production costs are saved.
  Posttranscriptional inhibition does not save enzyme costs, but can
  prevent a waste of metabolites. Therefore, {\futile} submodes should
  be disrupted transcriptionally, while {{\wasteful}} submodes may be
  disrupted by posttranscriptional inhibition.}.  If an enzyme appears
in multiple {\nonbeneficial} submodes, this makes it a plausible
target for regulation. By this criterion, PFK and FBPase rank among
the top regulation targets in the yeast model in Figure
\ref{fig:yeastfluxanalysis}. But how can the switch between glycolysis
and gluconeogenesis be realised biochemically? If both PFK and FBPase
expression were dependent on a common regulation parameter, there
might be a parameter range in which both enzymes are partially active,
causing a futile cycle. To prevent this, the two enzymes should rather
be controlled by a bistable switch (implemented, for instance, by a
positive feedback in transcriptional or post-transcriptional
regulation).  \coout{mention strongly repressed PFK /FBPase; as
  preemptive expression. CBA yields only the first, simple
  expectation} \coout{vor regulation?}

\myparagraph{Suppression of futile cycles by network structure}
\co{NOTE: 1. beneficial, futile, wasteful. 2. wasteful modes are
  beneficial modes in opposite direction. If we want to understand how
  avoidance of non-beneficial modes shapes networks, we need to see
  how a given network structure translates into possible flux
  distributions. 1. in reality (with ``typical'' physiological
  concentrations), some flux directions are given; take this as a
  given fact. 2. any existing reaction may be suppressed, so
  non-beneficial fluxes can always be avoided. HOWEVER, a reaction
  that contributes ONLY to non-beneficial fluxes would not be useful,
  and would be selected against. In other words, a reaction should
  only exist if it is part of at least one beneficial flux mode.}
\co{MOVE TO SI! genauer schreiben. futile and wastful
  unterscheiden. say that there are varying (external) objectives,
  leading to varying internal econ pot (ande therefore varying
  ``internal pathway objectives''), and that being futile or wasteful
  depends on these objectives - need for flexible regulation! same
  logic comes again later, with usage of pathways!} \co{mention that
  all predictions also rely on steady state assumption! (dont forget
  that some enzymes could be especially used in non-steady states
  .. wrong predictions!)}  \co{explain more clearly: reactions and
  regulation should be such that each flux mode is either useful,
  thermodynamically infeasible, or controllable by regulation
  (e.g.~enzyme repression). Structures that are ALWAYS useless should
  not exist to begin with! (e.g.~a flux that leads, for thermodynamic
  reasons, always into a ``wasteful'' direction.)} We saw that cells can block 
{\nonbeneficial} submodes  by enzyme repression.
But what can we learn from VBA about network structures in general? Obviously,
network structures that enforce non-beneficial submodes
should not even exist: metabolic networks should be such that
(thermodynamically feasible, but) non-beneficial submodes should
either be avoided (by suitable network structures) or be subject to
regulation. \co{ (or in which certain reactions can be active only
  in non-beneficial submodes)} \co{was heisst das konkret? zumal jede
  wastful mode das gegenteil einer beneficial mode ist! konkreter
  schreiben, was passiert mit futile bzw wasteful; those that are
  (likely) thermo-infeasible are not a problem!  was heisst das
  konkret zb fuer membrane leakage, import/export of metabolites etc?
  usage of cofactor-inefficient pathways?}  According to VBA,
{\nonbeneficial} flux modes -- such as potential futile cycles or
unnecessary biosynthesis pathways -- will waste economic value. Unless
they are thermodynamically infeasible (and therefore, be excluded by
physics), they should be selected against\footnote{For the sake of the
  argument, we consider a non-beneficial flux mode that actually
  provides no benefit at all, under any reasonable constraints or side
  objectives.}.  What will the resulting networks look like?  First,
each reaction or pathway should contribute to at least one beneficial
flux mode. \co{stimmt nicht:} Second, {\nonbeneficial} submodes should
contain two repressible enzymes\footnote{In a {\nonbeneficial} submode
  with only one repressible enzyme, this enzyme would always be
  repressed and would not be conserved in evolution.}: the PFK-FBPase
system is an example. \co{discuss a linear export pathway with 1
  repressible enzyme bcs of varying external potentials!}  These rules
can also be used as sanity checks in automatic network construction.
\co{and as a criterion for the selection of good pathways in metabolic
  engineering?}

\co{FN: The exclusion of futile cycles (and network structures that
  favour such cycles) may also explain some patterns in phylogenetic
  profiles. Phylogenetic profiles describe the co-occurrence of
  enzymes across organisms. \co{REF} \co{Mathematically, futile
    submodes should be ``sign-orthogonal'' on phylogenetic profiles
    ..}, but this works only if the flux directions (arising from
  phylogenetic profiles) are known ..}  \co{concrete predictions about
  network structures?}

\myparagraph{Why do cells employ futile cycles?}  It seems like futile
(or wasteful) cycles are relatively common in cells.  During
autophagy, for example, proteins are degraded and produced at the same
time.  Kinase-phosphatase cycles in signaling systems constantly burn
ATP. How can such cycles be reconciled with our theory? It may be that
some cell processer are simply not optimal (in fact, there is no
reason to believe that cells work precisely optimally in reality). But
maybe some of this depends on what objectives we attribute to a
cell. If a cycle looks futile, it may still have beneficial side
effects or may be enforced by constraints. For example, flux cycles
may enable cells to rapidly change their fluxes or to escape from
unfavourable metabolic states \cite{vwbh:14}. A ``futile'' consumption
of ATP may improve information processing (in signaling systems),
increase precision in DNA replication, or speed up responses in gene
expression \cite{alon:06}. Trehalose cycling, a ``futile'' cycle in
yeast, can prevent a breakdown of metabolic fluxes caused by the
``turbo design'' of glycolysis \cite{vwbh:14}.  Other cycle fluxes may
be beneficial due to the ensuing heat production: \co{FN: if gibbs
  free energy is dissipated (proportional to flux and driving force)
  this occurs typically in the form of heat. Hence, most reactions
  produce heat which increases temperature and can further speed up
  reaction, i.e.~make enzymes more efficient, but can also be harmful
  or may even serve as a weapon.} in a compost pile, for example,
higher temperatures increase the enzymatic rates and therefore the
growth of bacteria.  \co{JA!  FN: heat production for body temperature
  (neanderthals); CITE heat production in bees /defense against
  hornets REF!  } If we include them in a model as side benefits, this
leads to extra terms in the economical balance equation, and a
``\nonbeneficial'' cycle may become beneficial.  \co{FN: For example,
  in a model with biomass production as the objective, cyclic flux
  variations $\modevectorcyc$ (satisfying
  ${\bpsi}\trans \ \Next\,\modevectorcyc=0$) are {\nonbeneficial}.
  But if we add to our objective a benefit term for heat production,
  each reaction acquires a flux {{\gain}} $-\gamma\,\Deltar \mu\,v$
  (with $\gamma$ being the benefit weight for heat production), and
  ATP-consuming, heat-producing flux cycles may become beneficial
  because.}  \coout{Ron: heat production in bees /defense against
  hornets. but heat can also perturb enzymes} A particular reason for
apparently futile cycles are non-enzymatic reactions: if a degradation
reaction cannot be suppressed, cells need to reproduce the degraded
metabolites to keep them at a desired concentration.  In this case,
the economic potentials will show an unusual pattern with a drop in
the degradation reaction. \co{see Figure in CBA I appendix} Normally,
such a drop would indicate an uneconomical flux mode, but if a
non-enzymatic reaction (or in a dilution flux) is involved, a certain
value loss cannot be avoided, not even in optimal states.  In
metabolic value theory, non-enzymatic reactions can be taken into
account and such {\flow}s are correctly classified as economical
\cite{lieb:14a}.

\section{The choice between metabolic strategies}

\myparagraph{\ \\The choice between pathway fluxes} \co{Show examples
  with realistic numbers} Once we we can which flux profiles are
economical, we may ask about choices\footnote{\co{also mention this
    elsewhere / in other articles?}  ``Choosing'' is of course used
  metaphorically, similar to ``machine learning'', ``automatic
  reasoning'', or ``evolutionary games'' in other contexts. In all
  these cases, we hypothetically replace a physical system (a cell, an
  organism, or a computer) by a conscious being and ask how this being
  should behave, if it were in the place of our system and had to
  perform the same task.}  between metabolic pathways and how these
choices depend on enzyme efficiencies (and thus, on metabolite
concentrations).  Cells may choose between metabolic strategies
(e.g.~between different carbon sources or between different excretion
products) by expressing different sets of enzymes. Which choices are
best, and how does this on details of enzyme kinetics?  If an enzyme
is inhibited or knocked down, should other enzymes in the pathway be
upregulated (to compensate the decrease in activity) or should the
entire pathway be switched off (and the flux be rerouted to other
pathways)?  \co{evtl die ganze high-yield/low-yield-sache nach sec
  8?}Finally, should cells produce ATP by high-yield or low-yield
pathways (see Figure \ref{fig:rateversusyield})?  \co{JA! Different
  criteria for the choice of pathways have been proposed, including
  pathways length, protein mass, or thermodynamic forces. REF Arren,
  MDF usw} \co{Also in evolution or metabolic engineering. mention
  criteria: size, kcat, thermo usw; \co{WO? cite \cite{horh:11}
    A. Hoppe and C. Richter and H.-G. Holzh\"utter, Enzyme maintenance
    effort as criterion for the characterization of alternative
    pathways and length distribution of isofunctional enzymes}
  \co{also cite arren, carbon fixation pathways} implicitly most of
  them lead to enzyme cost per flux! ref ECM} Eventually, all these
criteria are related to the pathway's effective catalytic rate,
i.e.~the enzyme amount per pathway flux. \co{or more generally, its
  benefit/cost ratio, i.e. the flux benefit per enzyme cost. FN:
  \co{wo? auch in CBA kin + CBA labour! sagen dass hel (pragmatisch
    bestimmt werden kann als (i) proportional zu protein sizes and
    (ii) scaled such that the total enzyme investment (sum hel el) is
    equal to the total benefit (eg wbm * biomass prod); dh hel = prot
    size(l) * wbm * biomass prod / sum prot size(l) * el, and assuming
    that total protein mass (sum prot size(l) * el) is known for the
    cell.}  erstmal was zu kapp sagen, wie es sich aus kcat und
  efficiency terms bestimmt (s.o.): dann} More generally, evolution
may favour pathways with high enzyme productivity (e.g.~the target
flux per enzyme investment) or with high substrate productivity
(target flux per substrate consumption). In some cases, the two
criteria coincide: a low yield (substrate productivity) may imply a
low enzyme productivity (e.g.~due to the larger import fluxes, and
therefore higher enzyme investments).  In other cases, a low yield may
cause a higher enzyme productivity (because more energy is dissipated,
allowing enzymes to work more efficiently). \co{Rate-yield tradeoffs:
  the two arguments and why neither of them holds generally.}
\co{explizit unterscheiden zwischen near-equilibrium and
  far-from-equilibrium situations! explain what are the issues; alles
  danach sortieren? (auch in FCM usw)} \co{Hier diese ganzen logiken;
  dann unten wie VBA das bescrhreibt. auch: near equilibrium vs far
  from equilibrium} \co{this (and the equivalence of the two!)  is
  automatically captured here by embodied values, which may either
  stem from the original substrate or from the investment in its
  import)} VBA does not only consider absolute catalytic rates but
provides a new perspective based on values (i.e.~marginal cost and
benefit), relates economic potentials to {\enzymeinvestment}s and
fluxes, and explains the choice of fluxes depends on prior protein
investments ``embodied'' in the metabolites.

\co{Distinguish two questions: 1. Under what conditions can a given
  {\flow} be realised? 2. Given certain conditions, which of the
  possible {\flow}s will be realised?// Assume that the cell can run
  the profiles separately (e.g.~in different compartments) and can
  reallocate protein between them.// For given external conditions:
  always choose the one with the lower {\enzymeinvestment} (at a given
  point benefit) -- because for simultaneous use, the ratio point
  benefit / {\enzymeinvestment} (and thus the return on investment)
  would have to be the same.// For varying external glucose: in each
  {\flow}, rearrangement of the metabolite and enzyme concentrations,
  such that cost is minimised.// Then: scaling of absolute flux
  changes the $\sum\, bv\ v / sum\, hu\, u$ ratio!}

\co{auch: how econ potentials determine the overall fluxes (switching pathways on or off!)}

\co{in the following: general principles: (i) cost/benefit balance for
  pathways or any part of the network); (ii) lower bounds on value
  differences; (iii) importance of some key economic potentials; (iv)
  exclusive use  of isoenzymes; (v) high-yield and low-yield
  strategies; (vi) usage of pure vs combined metabolic strategies;}

\myparagraph{Economic potential differences, \co{wort?tension? dann fuer mismatch lieber ``stress''?} fluxes, and {\enzymeinvestment}}
In VBA,  flux profiles must be compatible with patterns
of economic potentials: if the economic potentials and {\fluxgain}s
are known, they  directly determine the flux directions.  In turn,
known flux directions put constraints on the economic
potentials. But all this is also related  to enzyme kinetics.  Known enzyme
{\price}s $\hus$, enzyme levels $\us$, and fluxes $v$ determine the
potential differences $\Deltar \wtots = \frac{\hus}{v} \us$: in
reactions without direct {\fluxgain}s, the potential  differences are
proportional to the enzyme levels (with $\frac{\hus}{v}$ as a
prefactor) and to the inverse enzyme efficiencies (or ``enzyme
slowness'') \co{uea!} $\tau = 1/\ratelaw = \us/v$ (with the enzyme
price $\hus$ as a prefactor). \co{FN: If  enzyme prices $\hul$
  (e.g. from protein sizes),  active reactions  and flux directions were
  known, jetzt daraus schluesse ziehen, die oekon potentiale mit
  effizienzen verbinden). zB using $k\,\Deltar \wtots = \hus$ and
  estimated $k^{\rm guess}$ and $\hus^{\rm guess}$, guess $\wtoti$
  such that $k^{\rm guess}\,\Deltar \wtots \approx \hus^{\rm guess}$}
Furthermore, since enzyme efficiencies $k=v/e$ depend on metabolite
concentrations, varying concentrations can make pathways either economical or
uneconomical.  To see how all this shapes metabolic states, we need
only two equations: the value production balance \co{REF} (for
reactions and pathways\co{see below}) and the relation \co{REF}
between flux {\burden}, kinetics, and enzyme {\price} (which defines a
minimum flux {\myvalue}).  Using the equations, we now ask whether a
given pathway should be used at all, which out of several alternative
pathways should be used, and whether fluxes through alternative
pathways should be combined.

\myparagraph{Metabolic pathways and economic potentials on pathway
  boundaries} If we split a metabolic network into
pathways\footnote{For the present purpose, the term ``metabolic
  pathway'' is used in a general, rather formal way: it may refer to
  any set of reactions, a single reaction, or the entire network.},
the cofactors and precursors on their boundaries serve as
``interfaces'' between the pathways \cite{spmk:02}. In metabolic
economics, these boundary (or ``connecting'') metabolites play an
important role.  First, their economic potentials and production rates
(by a given pathway) determine the total {\enzymeinvestment} inside
the pathway\footnote{The {\enzymeinvestment} in a pathway is equal to
  the value production at the pathway boundaries, which depends on
  boundary potentials and fluxes. This equality resembles Gauss'
  theorem in vector analysis, which equates the flow across a closed
  surface to the sum of sources of this flow in the enclosed region
  \cite{lieb:14a,lieb:cbafield}. In electrostatics, it describes the
  relation between charges and fields.  Instead of treating enzyme
  investments as ``sources'' of value, we may count enzyme investments
  as a ``value consumption'' or value flowing in from the rest of the
  cell, and matching metabolic value production. Described in this
  way, the overall value flow across the boundary of a pathway must
  vanish, and we obtain a conservation relation for value (in optimal
  states) without sources and sinks, but with enzyme investments
  counted as inflowing value!  The analogy suggests that metabolic
  values can be described by a conserved flow \cite{lieb:20a}. \co{in
    CBA field: die gesetze dort so nennen; enzyminvestment is wie
    ladung, metabolische fluss wie elektrischer fluss. dh .. ist
    ladung so etwas wie ein einstrom von elektrischen feld aus einer
    anderen dimension??}}.  Second, under some heuristic assumptions,
the economic potentials and rates on the pathway boundary determine
all economic potentials and {\enzymeinvestment}s inside the
pathway. Third, when modelling a single pathway, all relevant
information about the outside system is encoded in the economic
potentials on the pathway boundary. This facilitates modular
modelling: if a cell state changes, and if we know the changes of
variables in the connecting metabolites, this sufficies to understand
the economic adaptations inside each pathway, independently of the
rest of the network.

\myparagraph{Value production balance for pathways}
\label{sec:balancefor pathways} To quantify enzyme investments in
pathways, we first note that a pathway, just like a single reaction,
satisfies a value production balance. For a pathway $L$ (or for that
matter, any set of enzymatic reactions), we can sum
Eq.~(\ref{eq:reactionbalanceeq}) over the reactions $l$ to obtain the
pathway balance
\begin{eqnarray}
 \label{eq:gedankenexpsum}
 \sum_{l \in L} \Deltar \wtotl\,v_{l} + \sum_{l \in L} \bvdirl\,v_{l}
 = \sum_{l \in L} \hul\,\ul.
\end{eqnarray}
 Internal metabolites must be  mass-balanced, so their value production
and consumption cancel out and  the
first term depends only on value production on the pathway
boundary\footnote{\co{appendix/SI?} Let us see this in detail.  Pathway metabolites
  that participate in reactions outside  the pathway are called boundary
  metabolites. They are produced or consumed by the pathway. Pathway
  intermediates, in contrast, are mass-balanced in steady state, so
  their value production and consumption cancels out. Therefore the
  first term in Eq.~(\ref{eq:gedankenexpsum}), describing value
  production, depends only on the economic potentials and production
  rates of boundary metabolites\co{JA!
    $\Deltar \wtotl\cdot\,v_{l}=\wtot\trans\, \Ntot\,\vv = \wtot^{\rm
      bnd}\cdot\prodratev^{\rm bnd}$}. \co{bnd ist verwirrend wegen
    boundary / bound! lieber hier ``ext'' und sagen, dass damit
    pathway boundary gemeint ist?}  Therefore, value production can be
  split into terms for internal metabolites, pathway substrates, and
  pathway products
\begin{eqnarray}
  \label{eq:PathwayBalanceEquation}
  \sum_{l} \Deltar \wtotl\,v_{l} &=&
  \sum_{l}  \wtoti\,n_{il}\,v_{l} =
        \underbrace{\sum_{l}  \wtoti^{\rm int}\,n^{\rm int}_{il}\,v_{l}}_{=0} + \sum_{l}  \wtoti^{\rm prod}\,n^{\rm prod}_{il}\,v_{l}
        +\sum_{l}  \wtoti^{\rm cons}\,n^{\rm cons}_{il}\,v_{l} \nonumber \\
        &=&
  \underbrace{\sum_{j} n_{\rm prod,j}\,w_{\rm prod,j}}_{{\mbox{Value production}}}
-  \underbrace{\sum_{j} n_{\rm cons,j}\,w_{\rm cons,j}}_{{\mbox{Value consumption}}},
\end{eqnarray}
where the first term vanishes due to mass balance: the production and
consumption of internal metabolites cancels out, and so does the
associated value production and value consumption.}. 
\co{If
  our pathway produces a single compound (such as biomass), does not involve cofactors,  and if  substrates
  are ``free gifts of nature'' (assuming $w_{\rm substrate}=0$), this formula 
  simplifies to $n_{\rm prod}\,w_{\rm prod}$, where
$n_{\rm prod}$ is the product yield (product production rate per pathway flux) and  $w_{\rm prod}$ is the
 economic potential of the product.}
The pathway's total enzyme investment, on the right of the pathway
value production balance Eq.~(\ref{eq:gedankenexpsum}), is given by
\begin{eqnarray}
  \label{eq:PathwayBalanceEquation2}
  \sum_{l} \hul\,\ul =
  \underbrace{\sum_{l} \frac{\hul/\kcatl}{(1-\e^{-\theta_{l}})\,\eta^{\rm kin}(\cv)}}_{{\mbox{Enzyme \investment}}},
\end{eqnarray}
Treating the pathway as a single effective reaction \todo{(with a rate
  $v_{L}$ describing the pathway flux)}, this investment can be
written as $h^{\rm e}_{L} \,\u_{\rm L}$, with the total enzyme
concentration $\u_{\rm L} = \sum_{l \in L} \u_l$ and the
(state-dependent) {\enzymeprice} \co{swap indices?:}
$h^{\rm e}_{L} = \sum_{l \in L} \hul \,\frac{\ul}{\u_{\rm L}}$.
\co{flux burden: ..; flux value: .. EQ XXX!}
Eqs~(\ref{eq:PathwayBalanceEquation}) and
(\ref{eq:PathwayBalanceEquation2}) apply not only to ``dedicated''
pathways, but to any part of a network.  Applied to the network as a
whole, they state that total {\enzymeinvestment} and total metabolic
benefit (e.g.~of biomass production) in the network must be
equal. \co{hier die normierung von enzyme prices in Footnote
  erklaeren? This can be used to define a unit for enzyme prices that
  matches the unit for flux benefits.  \co{(see proteomaps
    Fig.~\ref{fig:proteomaps} (b))}}

\myparagraph{Essential flux values} As mentioned before, each
enzymatic reaction (with a given direction), has an essential flux
burden \co{see new eq OO on page 17} defined by enzyme molecule
properties. Via the {\myvalue/\price} balance
(\ref{eq:reactionbalanceeqFlux}) $\gvtots = \hvs$, this burden defines
a lower bound $\gvtots \ge \hvsmin$ on the possible flux {\myvalue}s. Below this
value, the balance cannot be satisfied and the reaction must be
inactive.  Eq.~(\ref{eq:PathwayBalanceEquation}) implies a similar
condition for any flux mode in a network (and for any pathway defined
by a localised flux mode). For example, in a linear pathway a positive
flux requires that
$ w_{\rm product} - w_{\rm substrate} \ge \sum_{l} a^{\rm min}_{l}$:
the potential difference (left) must exceed the essential flux
{\burden} (right) of the pathway. \co{aber was gilt allgemein, zb fuer
  verzweigte pathways? assume scalable pathway flux
  $v_{\rm PW}\,\hat{\vv}$ (with unitless flux distribution $\hat{\vv}$
  and $\hat{v_{l}}>0$ by convention). then
  $r_{prod}\,w_{prod} - r_{sub}\,w_{sub} =\sum_{l} \hvl\,v_{l}$, so
  $\hat{r}_{prod}\,w_{prod} - \hat{r}_{sub}\,w_{sub}
  =\underbrace{\sum_{l} \hvl\,\hat{v_{l}}}_{a_{v_{PW}}} \ge
  \underbrace{\sum_{l} \hvl^{min}\,\hat{v_{l}}}_{a_{v_{PW}}^{min}}$}

\co{alles was wir oben zu ohmschem gesetz, dioden etc   gelernt haben, gilt auch fuer ganze pathways! }

\myparagraph{Isoenzymes and isopathways} How does the choice between
alternative pathways -- for example, importing a compound or producing
it in the cell -- depend on model parameters such as the composition
of the growth medium \cite{lieb:18fcm}? In VBA, which fluxes are
possible depends on the economic potentials of pathway substrates and
products: the pathway's flux burden $a_{v_{PW}}$ (which depends on the
catalytic rate)\footnote{\co{WO?} Metabolic fluxes
  and enzyme levels in cells are not simply proportional.  Instead,
  when enzyme levels are changing, this leads to a new steady
  state with new metabolite concentrations, enzyme efficiencies, and
  steady-state fluxes.  The relation between enzyme levels and
  steady-state fluxes is complicated and depends on many model details including 
  rate laws and small-molecule regulation of enzymes. The resulting
  changes in enzyme efficiencies  also cause changes in flux
  burdens.}  must match the external potential difference.  If a
pathway is active (i.e.~if it satisfies the balance condition), then
costlier alternative pathways (with a higher flux burden) cannot
satisfy the same condition: instead they are uneconomical and must be
suppressed. We can see this for a single reaction: isoenzymes, which
catalyse the same reaction, have the same flux {{\myvalue}} $\gvtot$.
This means: in an enzyme-optimal state, isoenzymes cannot be active
simultaneously unless their flux {\burden}s
$\hvl = \sum_{l} \hul \ul/v_{l}$ are exactly equal.  In a given model,
this will not happen by chance.  \co{except for toy models with
  unifrom parameter values or for models at rare ``bifurcation points''} Thus,
to minimise its {\enzymeinvestment}, a cell should use only the
cheapest isoenzyme, and in each moment there is probably only one.
FCM (with randomly chosen cost weights) also predicts this, but
without giving a general explanation. VBA provides a local criterion
for pathways that should be suppressed: a cost-efficient enzyme
``clamps'' the flux {{\myvalue}} by equating it to a low flux burden,
so costlier isoenzymes must be inactive.  When the enzyme is inhibited
or knocked out, the flux {{\myvalue}} can increase and another
isoenzyme may be used.  This logic holds also for isopathways,
i.e.~alternative pathways with the same substrate/product
stoichiometry (or with the same single production objective, e.t.~ATP
production).

\myparagraph{Resource allocation between pathways depends on the
  economic potentials of key metabolites.}  How can we understand
ecnomical fluxes, economic potentials, and enzyme investments in
larger networks? If we split a network into separate pathways, the
economic potentials of the connecting metabolites can tell us which
pathways are economical under which conditions.  In particular, they
tell us which pathway fluxes are beneficial (i.e.~able to balance a
positive flux burden) and whether the flux {\myvalue}s are above the
essential {\fluxburden}s (otherwise pathways must be inactive).
Second, given the fluxes, known boundary potentials determine the
total {\enzymeinvestment} in each pathway
(\ref{eq:gedankenexpsum}). Third, by employing simple heuristic
assumptions (e.g.~uniform {\enzymeinvestment}s, see SI
\ref{sec:princdistribinvestment}) we may estimate the economic
potentials inside the pathways from potentials on the pathway
boundaries.  If the boundary potentials are changing (e.g.~the
potentials of key precursors or cofactors), the internal economic
potentials in each pathway will change as well\footnote{Under simple
  heuristic assumptions (e.g.~assuming that all unknown enzyme prices
  should be identical), an increase of all economomic potentials on a
  pathway boundary increases the potentials inside the pathway (SI
  \ref{sec:SIscalingProperties}). In contrast, if the potential
  difference between pathway substrates and products increases, the
  extra potential difference will be distributed along the
  pathway. This leads to higher flux {\myvalue}s within the pathway,
  and in optimal states, higher flux burdens. If the pathway flux
  remains constant, this entails a higher {\enzymeinvestment}, and
  according to Eq.~(\ref{eq:fluxBurdenEfficiencyDetailed}), lower
  enzyme efficiencies.}, and it may be profitable to switch different
pathways on or off. \co{FN: REF zu elads aehnlichem paper ueber
  thermodynamik?}  \co{For example, the ATP-ADP potential difference
  appears in many reactions as an effective flux gain. Any change in
  this difference will affect many reactions; \co{in the balance (2)
    we may move this terms to the right. if we do that, the term
    effectively ``tunes'' the flux burdens (like a variable extra
    voltage source in many reactions, lso adjusting the essential flux
    burden). there is a related paragraph in the discussion!}
  \coout{das alles ausfuehrlicher bestrechen in modular models}}

\section{High-yield and low-yield pathways}

\co{in 5 wurde erwaehnt, dass ``reine'' strategien am
  wahrscheinlichsten sind. deshalb hier wahl nur zwischen reinen
  strategies (und kurz darauf zurueckkommen, dass gemischte stratgien
  auch moeglich sind)}

  \co{hier oder CBA labour? analyse von EMP-ED pathway, wie in
    flamholz et al SI, nur mit begriffen der MVT

    vergleiche ATP/glucose yield

    vergleiche ATP flux / enzyme cost

    aber: gesamt benefit haengt von gesamtwertdiferenz ab, nicht nur
    von ATP. wert von pyruvat? (niedriger oder hoeher als glucose?)
    ueberlegen!}
  
\co{
\todo{For simplicity, consider an entire pathway leading from an external substrate to some end product, and described by substrate uptake rate $v_{\rm S}$, product production rate $v_{\rm P}$, and total enzyme amount $e_{\rm tot}$. \co{FN: for generality: statt vp aich flux benefit, statt etot auch enzyme cost h(e)} We can characterise it by yield $\eta=\frac{v_{\rm P}}{v_{\rm S}}$ (which is a stoichiometric property of the flux distribution) and enzyme-specific production (or ``enzyme productivity'') $\vartheta$ \co{symbole uea?}. If we maximise $\vartheta$, we need to keep track of enzyme investments $e_{\rm tot}$: ... If we try to maximise yield, we need to keep track of substrate investments $v_{\rm S}$.. (in both cases, for a flux mode with given product rate $v_{\rm P}$)}
}

  \myparagraph{\ \\The choice between high-yield or low-yield
    pathways} \co{kommt auch in (CBA opt, CBA lagrange): bei
    rate-yield trade-off: o dill paper: \cite{madi:15} claim: "cells
    are optimised for enzyme efficiency of the fast-growing cells and
    not just growth rate or efficiency alone"; very simple "RBA
    version" of molenaar model} \co{say directly: substrate
    productivity (or yield): production per substrate; enzyme
    productivity (production per enzyme, or, at a fixed enzyme amount,
    production per cell, and therefore cell growth rate); dann: was
    die methoden machen. DAS so auch in FCM erklaeren!}  In an
  environment that favours fast growth, cells are expected to maximise
  enzyme productivity (biomass production per {\enzymeinvestment}), a
  proxy for growth rate \cite{sgmz:10}. In an environment with limited
  substrate, cells are expected to maximise cell yield (biomass
  production per uptake rate). \co{Yield, or substrate productivity,
    describes the biomass production per substrate investment
    (e.g.~amount of glucose).} \co{JA! (kam schon bisschen weiter
    oben) hier klarmachen: hier say: substrate efficiency vs catalytic
    rate; enzyme efficiency (bm production / total metab enzyme) and
    growth rate; rate vs yield usw} Depending on {\enzymeinvestment}s,
  kinetics, and external conditions, high growth rates may be achieved
  by either high-yield or low-yield flux modes \cite{wnfb:18}.  This concerns, for
  example, fermentation or respiration \co{REF?}  or variants of
  glycolysis with different ATP/substrate yield \cite{fnbl:13}. How
  can we predict such choices by models? Different flux prediction
  methods favour different objectives. \co{"rate" in FBA is often
    yield, multiplied with an uptake rate.  ref scpf:08 Is
    maximization of molar yield in metabolic networks favoured by
    evolution?}  Classical FBA, with bounds on individual fluxes,
  predicts substrate-efficient behaviour (high yield), which may not
  be enzyme-efficient.  In contrast, FBA with flux minimisation
  \cite{holz:04} or molecular crowding \cite{bvem:07} are made to
  predict enzyme-efficient flux modes, which may or may not be
  substrate-efficient.  \co{FN: Compromises between high yield and
    high rate have been studied in a multi-objective approach, applied
    to all possible elementary flux modes \cite{wnfb:18}.  While these
    methods can predict {\enzymeinvestment}s and fluxes, they do not
    generally explain how these choices depend on kinetic details.
    VBA, in contrast, provides explanations (see Figure
    \ref{fig:rateversusyield}).}  \co{note correspondence to satFBA
    logic!}

\begin{figure*}[t!]
\centerline{\includegraphics[width=10.5cm]{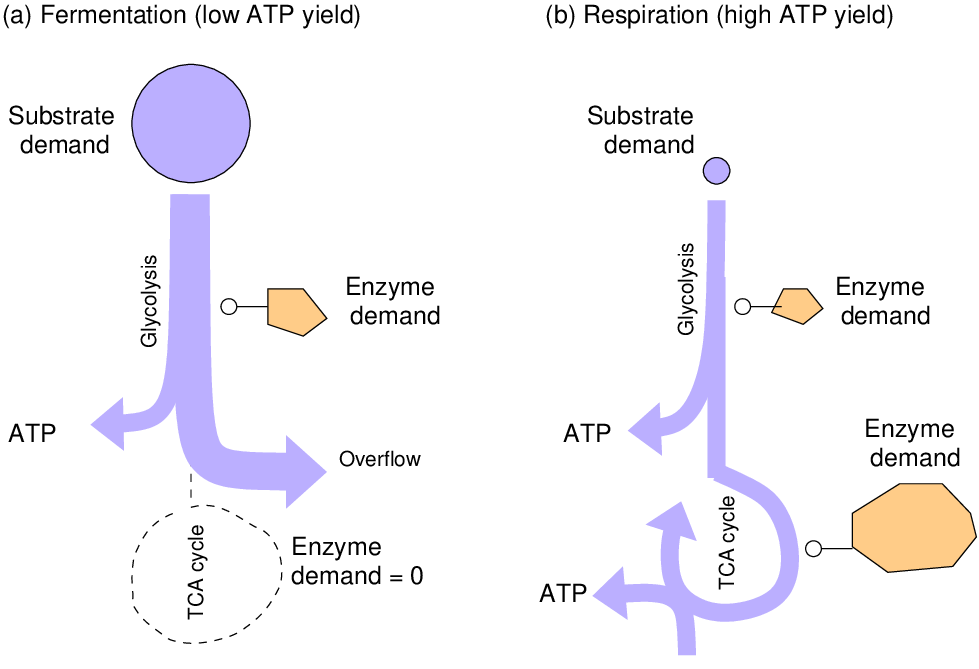}}
\caption{\co{SCHOENER! noch zwei bilder fuer low glucose? oder soll
    die abbildung lieber nach FCM paper? // pfeile schoener!}
  \co{should the figure show demands at equal ATP output? adjust
    arrows widths!}  Two flux modes in
  central metabolism (representing low yield and high yield, respectively).  Fermentation (Figure a) and respiration fluxes
  (Figure b) are compared at a given ATP production rate and scored
  enzyme demands (schematic drawing).  For each mode, we assume a
  configuration of metabolite and enzyme concentrations that realises
  the fluxes at a minimal enzyme cost).  \co{(for example, by FCM with
    enzymatic flux cost function)} (a) Fermentation. Due to its lower
  yield, the {\flow} consumes more substrate (glucose), but without
  investments in respiration, the overall enzyme demand can be low.
  (b) Respiration. Due to its higher ATP yield, substrate uptake (at a
  given ATP production rate) is smaller, so the enzyme demand in
  glycolysis tends to be lower. In the example, the overall enzyme
  demand is higher due to enzyme investments in respiration. In the
  comparison, fermentation will be preferred. However, this result
  depends on parameters: \co{3. bild zeigen!} at low glucose
  concentrations, glycolytic enzymes become inefficient, requiring
  higher investments in glycolysis; in the respiration strategy, which
  generally requries less glycolytic enzymes, this increase in
  investment is lower, and so at low glucose levels, respiration will
  be preferred.}
    \label{fig:cba_flux_respiration_fermentation}
\end{figure*}

\coout{FBA at a fixed uptake rate (and no other flux constraints),
  favours high-yield fluxes. Flux cost minimisation with uniform cost
  weights favours short pathways, possibly at the expense of a lower
  yield. A usage of individual cost weights can lead to other --
  high-yield or low-yield -- solutions, covering all {\flow}s that
  satisfy the value production balance.}

\coout{\textbf{Detailed explanation of Figure
    \ref{fig:rateversusyield}} The objective is to produce a the
  external compound C (respresenting ATP). After reaction 1, the
  intermediate X can either be used for production of C (reaction 2)
  or be converted to the sink metabolite D (e.g.~excreted from the
  cell, reaction 3). There are two possible strategies: using reaction
  2 provides the highest yield (i.e.~amount of C produced per amount
  of A consumed) and is therefore an efficient strategy. Nevertheless,
  using reaction 3, an inefficient strategy, may increase the
  production rate (i.e.~amount of C produced per time) or can come at
  a much lower {\enzymeinvestment} (if the enzyme for reaction 2 is
  expensive) and therefore be preferable. Each of the two solutions
  could be optimal for the cell,
  depending on {\enzymeinvestment}s and kinetic rate laws.\\
  The choice between these strategies is reflected by different flux
  predictions: if we maximise the production of C with a bound on
  reaction 1 and an irreversibility constraint on the reaction X
  $\rightarrow$ D, FBA will predict the efficient strategy. In
  contrast, if we use the same metabolic objective as CBA constraints,
  but without a linear optimisation, the resulting flux polytope will
  contain both the efficient and the inefficient flux mode, as well as
  linear combinations of them, so none of the potential kinetic
  solutions is excluded. Nevertheless, CBA does restrict the solution
  space: to ensure a positive metabolic benefit, it requires reaction
  1 to have a positive value in any case. Furthermore, since the flux
  B $\rightarrow$ X $\rightarrow$ D is a {{\wasteful}} mode, reaction
  2 can run in reverse direction.}

\myparagraph{The choice of metabolic strategies depends on enzyme cost
  at a given flux benefit} \co{gilt erstmal nur bei growth maximisation} \co{cite also stefan + kaleta} Let us
consider an example, the usage of overflow. In
aerobic conditions,  cells typically  respire, but they may further  increase their
glycolytic flux by adding  extre fermentation.  In experiments,
the choice between respiration and (extra) fermentation depends on the
extracellular glucose concentration, suggesting that glucose
concentration determines the relative advantage of the two
strategies. In FCM (including FBA with a minimal sum of fluxes, EFCM
\cite{lieb:18fcm} or satFBA \cite{murs:15}) different strategies are
scored by their enzyme costs at a given production
rate, e.g.~ATP production. Given an enzyme cost function, each flux
profile will have a (minimal possible) enzyme cost which depends on kinetic constants
and extracellular concentrations.  Such enzymatic flux costs can be defined for
single reactions, pathways, or entire flux distributions. Let us see
an example (Fig.~\ref{fig:rateversusyield}), where we compare
(low-yield) fermentation to (high-yield) respiration at a given ATP
production flux.  Fermentation needs a higher glucose influx and
therefore uses a higher amount of glucose transporters and glycolytic
enzymes (at a given glucose concentration).  Respiration, in contrast,
uses the costly TCA cycle and oxidative phosphorylation. At high
glucose levels, fermentation will be relatively cheap.  At low glucose
levels, the cost of transporters and glycolytic enzymes increases:
both strategies become costlier, but since fermentation relies on
these enzymes more strongly, at some point it will lose its advantage:
(high-yield) respiration becomes comparably cheaper and allows for
faster growth. This effect has been compared to Giffen behaviour in
economics \cite{yaha:2019}.

\co{JA! previous argument ``global'', from enzyme cost per flux for the
  entire flux mode. now a ``local argument, focusing on particular
  metabolite and comparing how they were made (resulting in embodied
  investment), and what is made from them (choice of usage strategy).}

\myparagraph{The choice of metabolic strategies depends on substrate
  and {\enzymeinvestment}s} \co{at low nutrient levels, transporter
  efficiency is low (slowness is high) and transporter demand (per
  flux!) is high. if additionally, the yield is low, then the
  transporter demand per production rate (of the end product, i.e. the
  metabolic benefit) is even higher.  the high transporter investments
  embodied in intermediate metabolites lead to a requirement for high
  yield!} If external metabolite consumption is ``costly'' (i.e.~if
uptake is directly penalised by the metabolic objective), metabolite
uptake should be kept low, and the metabolite should be used
efficiently, i.e.~with a high yield. The same holds for a metabolite
that is taken up via a costly transporter: also in this case, a
high-yield strategy will be preferred. This means: what exactly makes
the uptake costly (metabolite usage, investments in transporter, or
\todo{anything else}) does not matter for the choice of the downstream
metabolic strategy, it is only the cost that counts. In VBA, this
logic holds not only to extracellular substrates, but to any
metabolites: in a reaction A $\rightarrow$ B, if the concentration of
A is low, a large amount of enzyme is needed; this makes B effectively
costly, and in the rest of the system, B should be used economically!
Let us describe this in the language of VBA. The economic potentials
in optimal states arise from embodied investments, e.g.~usage of
extracellular substrates, transporters, and enzymes needed to produce
a metabolite.  Exporting a metabolite with a positive economic
potential (i.e.~converting it into a worthless extracellular
metabolite) would be uneconomical; instead, this metabolite should be
converted into a valuable product (even if this requires additional
{\enzymeinvestment}s).  This means: in enzyme-optimal states, high
investment along a pathway, leading to valuable intermediates, also
require a high yield\footnote{Metabolites that embody investments can
  be compared to backgammon stones which become more valuable the more
  they advance. The more points have been ``invested'' in a stone, the
  less a player would typically sacrifice this stone. \co{JA! discuss
    ``sunk cost'' fallacy: say that the theory describes optimal
    steady states, not what should be done AFTER a wrong investment;
    here there is no sunken cost fallacy because we describe an
    optimal state (where by definition this fallacy cannot obtain). In
    reality, there may be such fallacies (see nuclear power). we also: the more advanced
    stone is actuallyl more useful for winning the game, because it is
    closer to the end point; this means: it also has a higher
    theoretical use value, unlike in synthesis cost}}. This explains
why the choice of metabolic strategies depends on extracellular
concentrations.

\begin{figure*}[t!]
\centerline{\includegraphics[width=15.5cm]{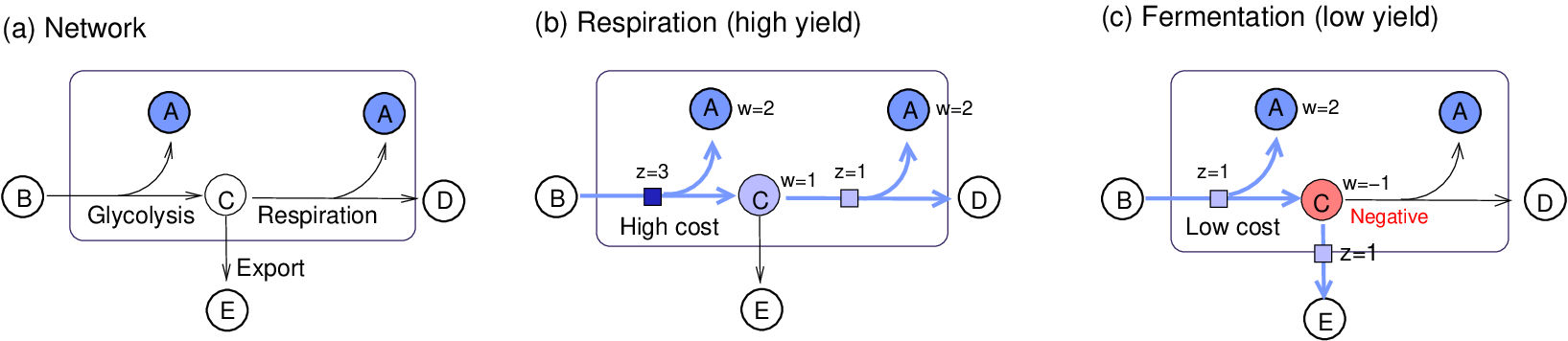}}
\caption{\co{dunkelblau etwas heller} The choice between a high-yield
  and a low-yield flux mode depends on upstream enzyme {\investment}s.
  (a) Schematic model of glycolysis, respiration, and overflow
  metabolism (export of incompletely oxidised compounds such as
  lactate or ethanol). The metabolic objective scores the production
  of A (representing ATP). We compare two possible flux modes, respiration (in (b)) and overflow metabolism (in (c)). Classical FBA would favour respiration because of its higher yield (i.e.~a higher
  production of A per uptake of B). In VBA, either strategy can be
  economical.  (b) Respiration flux. Enzyme investments $\zcostgenericl$ and
  economic potentials $\wtoti$ are shown by numbers and shades of
  blue.  An overflow flux  from C to E (zero potential) would not be
  economical. (c) Fermentation flux.  We assume a lower enzyme
  investment in glycolysis than in (b), caused by other kinetic
  constants or a higher glucose concentration. The low enzyme
  investment and the high value of ATP produced lead to a negative
  economic potential in C (in red), and C can be exported with a
  positive {{\investment}} in the export reaction. If the {\flow}s in
  (b) and (c) are scaled to equal ATP production rates, the glucose
  influx in (c) will be higher. \todo{However, if we choose the profile with the lower
  enzyme cost (at a given flux benefit), then  despite its lower yield,
  fermentation can be more cost-efficient than respiration.}  \co{FN
    changing the substrate value (e.g.~to model that cells need to
    save substrate for the future) will have the same effect as making
    the transporter more expensive!}  \coout{However, this does not
    mean that they are equally costly \emph{as such}; it only means
    that a further increase of the ATP production would lead to the
    same additional {\enzymeinvestment}s in both cases.}}
    \label{fig:rateversusyield}
\end{figure*}

\co{combined usage of certain pathways; arren: could different
  variants in central metabolism favor certain variants of other
  pathways for economic reasons?}
\co{say that substrate efficiency may coincide with enzyme
  efficiency if cost for substrate is mostly due to transporter (and
  first enzyme) investments - in fact, for the value balance this is
  the same!}

\myparagraph{The choice between fermentation and respiration}
Economical states may either show a balance between high value
production \co{flux times difference of pathway product and substrate
  values} and high enzyme investment (high-yield strategy), or between
low value production and low enzyme investment (low-yield strategy).
Which type of balance we find \co{depends on the enzyme investment per
  flux benefit \co{y/b} (or on the ``return on investment''
  $(b-y)/y$). \co{(or the overall enzyme cost productivity b/y.}} will
depend on external concentrations, kinetic constants, or enzyme
prices.  \co{erklaeren, dass high yield die kosten der ersten enzyme
  senkt und daher vorteilhaft ist, wenn die teuer waeren (zb bei
  kleiner aeusseren konzentration; dass man das hier aber als embodied
  value beschreiben kann!)} Figure \ref{fig:rateversusyield} shows an
example. In central metabolism, glucose is converted into an
intermediate C (for example pyruvate, lactate, or ethanol) which can
either be exported (overflow with a low ATP yield) or further oxidised
(by respiration, with a high ATP yield). Classical FBA favours
respiration because of its higher yield.  In VBA, either of the
strategies can be economical, depending on their ATP/enzyme
productivity. In kinetic models, the ATP production per enzyme
{\investment} is hard to determin because it depends on model
parameters. However, there is an interesting rule of thumb: how a
metabolite should be used depends on the enzyme or substrate values
embodied in the metabolite. We can see this in
Fig.~\ref{fig:cba_flux_respiration_fermentation}. If external glucose
levels are low, glucose transporters are inefficient, and many
transporter molecules are needed to achieve a desired influx.  This
transporter investment is embodied in downstream metabolites, leading
to high economic potentials, and the cell cannot waste these costly
metabolites in a low-yield strategy.  In contrast, high glucose levels
make the metabolite's embodied value decreases, and as its economic
potential becomes negative, a low-yield overflow strategy may become
economical. We can also see this more formally. In the respiration
strategy, metabolite C has a positive economic potential, so exporting
it (i.e.~converting it into a less valuable metabolite E) would be
uneconomical. Instead, additional enzyme is invested in respiration to
produce more of the valuable A.  Figure \ref{fig:rateversusyield} (c)
shows the conditions for fermentation. Due to lower investments
(e.g.~higher external glucose levels, and thus lower (flux-specific)
transporter demands) C has a negative economic potential.  So a costly
export reaction can satisfy the reaction balance equation.  When this
model was previously analysed by FBA with flux minimisation \co{ref
  schuster}, \co{papers von stefan schuster zitieren: die ergebnisse
  seiner analyse werden bestaetigt!}  the optimal fluxes were found to
depend on the total {\burden} ({\enzymeinvestment} per flux) of
respiration or export flux.  Now we can see, additionally, a
dependence on embodied values, i.e.~at what cost a metabolite has been
produced, given the economic value of external glucose and the cost of
transporter and glycolytic enzymes. The explanation is in line with
previous arguments about rate/yield trade-offs based on thermodynamic
forces and substrate availability (see SI
\ref{sec:SIchoiceMetabolicStrategies}).

\myparagraph{Pure or combined metabolic strategies?} If a cell
``chooses'' between alternative pathways, we can describe this as a
choice between flux modes in which different pathways are active (for
example, elementary flux modes\co{REF} or extreme rays\co{REF}). But
what about mixed metabolic strategies (i.e.~linear superpositions of
such modes)?  For example, if several carbon sources are available,
should a cell consume them one at a time (investing in a single
transporter only), or simultaneously (distributing its investments
over several transporters)?  \co{below we will see, there will usually
  be one best (non-mixed) strategy, and the cell should use only
  this!}  Similar questions concern alternative transporters with
different affinities, isoenzymes, the choice between fermentation or
respiration, \co{REF} or variants of glycolysis with different yield
\cite{fnbl:13}. In all these cases, we can ask: should alternative
reactions or pathways (with different kinetics and costs) be used
separately or in combination?  In VBA, combined {\flow}s can be
economical, but in th eunderlying optimality problems they are usually
not enzyme-optimal unless they are enforced by flux bounds
(\co{proof?}  see SI \ref{sec:SIcombinationOrAlternative}). In kinetic
models (with realistic enzyme parameters), alternative enzymes or
pathways show different cost efficiencies, and the most cost-efficient
flux mode should be preferred (see SI
\ref{sec:SIcombinationOrAlternative}).  FCM makes the same prediction:
since realistic flux cost functions are concave in flux space, optimal
flux modes are vertices of the flux polytope, and mixed flux
strategies will not be optimal \cite{wnfb:18,lieb:18fcm}.  However,
these arguments hold only theoretically. In reality, cells may face
limitations (e.g.~limited membrane space) or a need for preventive
measures that favour \co{allg besseres wort?:} mixed strategies rather
than specialising on one single task. \co{ But: average may also be a
  bet hedging strategy (cite David T's work); FN: Finally, methods
  like geometric FBA assume a linear combination of many EFMs, rather
  than one EFM. In terms of catalytic rates alone, this is probably
  not justified (not knowing which EFM is best does not mean that one
  can justify taking the average).}\footnote{In flux sampling, this is
  a question of practical concern: should we sample ``pure'' flux
  profiles (e.g.~EFMs) or linear combinations of {\flow}s?  If we
  assume that cells realise enzyme-optimal states, we should in fact
  sample pure profiles , i.e.~vertices of the VBA polytope. But if we
  mistrust our optimality assumption or look for economical (but not
  necessarily optimal) strategies, we may sample combined strategies
  from the entire polytope (see Figure \ref{fig:fbacomparison2}).}.

\coout{\begin{figure*}[t!]
  \begin{tabular}{lll}
    \includegraphics[width=5cm]{\psfilesfluxes/knockdown.eps}&
  \end{tabular}
  \caption{\co{Should enzymes after a mutation (which makes an enzyme
      less efficient) be shut off or on?}}
\end{figure*}}

\iftoggle{bookversion}{\section{Conclusions}}{\section{Discussion}}

\coout{Discuss sensible benefit functions.  For pathway: production;
  for cell: biomass production; also low metabolite load!  Discuss
  concentration {{\gain}}; flux benefit; production benefit;
  equivalence between prod and  flux {{\gain}}s; costs}
\coout{The central requirement is the economic principle, the fact
  that the flux, multiplied with the flux {{\myvalue}}, must be
  positive.  In a purely mechanistic kinetic model, this would not
  even be guaranteed for the enzyme catalysing the beneficial
  reaction.}  \coout{Economic analyses may help to predict {\flow}s and
  to understand the selection pressures and constraints under which
  metabolism may have evolved. On the one hand, cells are believed to
  avoid metabolic fluxes if they are not necessary; on the other, flux
  {\fluxpattern}s arise from the thermodynamic force between substrates
  and products and are regulated by various biochemical mechanisms.
  The economic analysis is an attempt to bridge these explain why and
  how this happens But is it true?  Can we phrase such statements
  mathematically terms?} \coout{Trennung: 1 allgemein kinetisch; 2
  flussanalyse; 3 postulat}

\myparagraph{\ \\Economical flux profiles and VBA} \co{WO? say:
  $\partial b / \partial \ln e$ = point benefit = usefulness = value
  production.}  \co{Interpret ``use value as exchange value'' - nach
  CBA labour!}  Optimality problems in  kinetic models lead to a
value structure, a pattern of economic potentials that represent the 
network-wide (indirect) benefit of each metabolite. Understanding this value structure can
help us obtain plausible {\flow}s in FBA: by requiring that {\flow}s
must be compatible with a feasible value structure, we know
that they are realisable by kinetic models in enzyme-optimal states.
{\Flow}s with a positive benefit are called beneficial.  If such a
flux profile satisfies the {\summationcondition} with positive
{\enzymeinvestment}s $y_l$, it is called economical. Such {\flow}s
are in principle able to produce value in every reaction. This
property depends only on flux signs, that is,  active reactions  and flux
directions.  Like thermodynamic constraints, economical flux patterns
can be imposed manually or by evoking general laws.  First, metabolic
value theory provides some useful theoretical concepts: the
{\summationcondition} is an algebraic condition for economical fluxes,
independent of enzyme kinetics \cite{lieb:14a}.  A second, equivalent
criterion states that economical {\flow}s require the existence of
compatible economic potentials, just like thermodynamically feasible
fluxes require the existence of compatible chemical potentials. Third,
just like feasible chemical potentials exclude flux cycles and ensure
thermodynamically feasible fluxes, feasible economic potentials
exclude {\nonbeneficial} flux motifs and ensure economically feasible
fluxes. Non-beneficial flux motifs are flux patters that can never --
under no conditions -- contribute metabolic benefit and that always
imply a waste of resources.

\myparagraph{Local value production as a principle in flux analysis}
To achieve fast (or sustainable) growth, cells need to run their
metabolism efficiently, i.e.~with a high metabolic benefit at a low
enzyme cost.  Metabolic value theory provides concepts to understand this
\cite{lieb:14a}. In the optimal state, value production must match
enzyme investments\footnote{Biomass/enzyme productivity (biomass
  production rate per total metabolic enzyme) can be converted into
  cell growth rates by a simple function \todo{related to} bacterial
  growth laws \cite{wnfb:18}.\co{ref wortel}}, a principle that shapes
metabolic states. First, {\flow}s need to be economical to appear in
optimal states. Whether a flux profile is economical depends only on
flux directions and active reactions, and not the quantitative
fluxes. Second, the economic balance equation quantifies optimal
enzyme investments. It requires economical fluxes, resembles the
thermodynamic flux constraint, and is used in VBA as a constraint for
flux prediction. VBA does not describe the underlying problem directly
and we don't even need to know it precisely -- we just assume that it
exists, and make our FBA \emph{compatible} with it by imposing the
balance equation.  It applies three basic principles -- fluxes must be
stationary, must dissipate Gibbs free energy, and must produce
metabolic value to balance to balance the enzyme
investments\footnote{Ignoring thermodynamic laws may lead to
  paradoxical results \cite{lieb:14a}. Consider a linear pathway: if a
  reaction is fully \co{UEA? FCM, CBA opt} forward-driven (infinite
  thermodynamic force), the downstream enzymes have no flux
  control. If these enzymes are downregulated, their substrate levels
  increase, but the flux does not change. Paradoxically, these enzymes
  provide no point benefit even though they are needed for catalysing
  the flux. \co{STIMMT AUCH NICHT GANZ .. wenn es metab preise gibt,
    dann haben die enzyme einen einfluss auf konz und daher einen
    wert; UEA AENDERN! JA! auch in CBA I!}  In metabolic value theory,
  this paradox can be avoided by prohibiting fully irreversible rate
  laws \cite{liuk:10}. \co{oder requiring upper bounds or costs for
    metabolite concentrations!}}. It provides a physically,
biochemically, and physiologically meaningful theory of metabolic
fluxes, provides a solid definition of futile cycles, and explains
patterns in proteomes.

\myparagraph{Economic potentials and pathway objectives} \co{auch
  geplant fuer CBA lagrange, CBA kin, CBA labour: from global to
  local: internal ec pot as ``abbild'' von ext production
  {{\myvalue}}s; allows for local understanding; extension of models}
In VBA, the economic potentials play a central role. If a small extra
influx $\delta \prodrate_{i}$ of a metabolite (as a ``gift'') allows
for an increase $\delta b$ of the metabolic benefit, the ratio
$\wtoti=\delta b/\delta \prodrate_{i}$ is called the metabolite's
economic potential. Even if the objective scores only a single flux,
this flux still relies on the entire metabolic state. This gives every
metabolite a value: the metabolic potentials, as ``local proxies'',
represent the objective everywhere in the network.  \co{klar von
  gesamtziel (zB prod/enz als proxy for growth) to pathway objective:
  ableitungen des pathway objective als werte / preise auf dem rand;
  deshlab ``stellvertreter fuer alles ausserhalb!}  With their help,
we can describe the local economics of reactions or pathways, even if
benefit is realised elsewhere in the cell. If a pathway objective
represents a cell objective (e.g.~biomass production per total
metabolic enzyme), benefits outside the pathway can be represented by
economic potentials on the pathway boundary, no matter where they
arise in the network, \co{short description of chem / econ pot, also
  as ``local representatives'' of a global quantity to be optimised}
and ``producing a valuable boundary metabolite'' simply describes the
pathway's contribution to global cell benefit. \co{Pathway objectives
  typically attribute benefits to metabolic production and costs to
  metabolite and enzyme levels. Within our pathway or network, this
  objective defines a pattern of economic potentials.}\co{Definition
  of local flux benefit functions via economic potentials of boundary
  metabolites.  modulare modelle und effektive lokale nutzenfunktionen
  ganz kurz erwaehnt?  auf cba local verweisen!} We can then ignore
the surrounding cell and describe the pathway's contribution to cell
fitness by a simple pathway objective: a net production of economic
value on the pathway boundary.  \co{briefly say that the (optimal)
  benefit effects changes of a system (in that case, the environment)
  can be represented by Lagrange multipliers on the boundary (the
  economic potentials); in this way, they can effectively subsume all
  adaptations and fitness effects outside our pathway of interest, to
  ``complete'' our pathway model.}  \co{If we sum the {\benefitshade}s
  from all reactions, we obtain the overall benefit
  $\bvtot\cdot\,\vv$.}  \co{verrueckt, dass man alle arten von
  zielfunktionen (fluss/knz-dependent lokal in produktion (von oek
  potential) umschreiben kann! wo thematisieren?  diskussion?}
\co{noch etwas unklar!  was mit met + enz kosten ist, wie cost und
  benefit (cell) auf cost and benefit (in pathway) aufgeteilt werden
  .. hier noch kurz auf CBA modular verweisen!}  The fact that
economic potentials in a pathway can represent objectives outside the
pathway is important for modelling: it allows us to model a pathway in
isolation (while implicitly accounting for the rest of the cell),
shows how pathway models can be embedded into cell models, and allows
us to construct modular cell models in optinal states.
 
 \textbf{Non-optimality and multi-objective optimality}
 \todo{\co{check order} As noted above, the optimality criteria in VBA
   represent simple, idealised model assumptions.  Optimality-based
   models can tell us what cellular behaviour would result from simple
   resource allocation principles. The optimality problems behind VBA
   asssume that the metabolic state of a cell optimises a simple
   fitness objective, a function of fluxes, metabolite concentrations,
   and enzyme levels (where fitness must decreases with the enzyme
   levels). This objective function determines the direct gains and
   prices that define our VBA problem. In reality, cells may not
   behave optimally -- let alone, optimally with respect to our simple
   criteria! Hence, instead of maximising metabolic efficiency, in
   reality other fitness requirements (e.g., anticipating future
   changes) may be more important than enzyme-saving behaviour.  Cells
   may pursue other objectives, they may express proteins ``just in
   case'' to anticipate future demands, or may simply behave
   non--optimally. Optimal behaviour is not a fact, but a theoretical
   limiting case: even if cells were ``supposed to'' behave optimally,
   gene expression noise and slow adaptation would prevent them from
   reaching optimal states in reality. Alternatively, if we assume
   that cell behaviour is non-optimal, we may try to quantify
   non-optimality: non-optimal behaviour can be described by tension
   in our balance equations, which quantify non-optimality. The
   general form of the balance equations, and how they fit into flux
   modelling, remains unchanged.  To obtain a more realistic
   description of cells, still based on optimality prinicples, we may
   consider multiple objectives and Pareto-optimal states. Instead of
   completely optimising one objective, we obtain a series of
   ``optimal compromise'' states with different flux distributions and
   different linear combinations of the (objective-related) economic
   potentials. This can be handled by VBA.}  \co{mention terms from
   flux bounds (which may represent ``non-optimality'' in the model)}
 \co{Another application of Pareto optimisation concerns cell
   communities. As shown above, flux modelling is useful to describe
   several interacting species, but the fact that each species has its
   own objective and controls only its won reaction fluxes is hard to
   implement in FBA. In VBA this is quite natural.}

 \co{\textbf{General principles}} \co{WO?  uebergang zu interpretation
   of economic variables // wie koennen ergebnisse trotz verschiedener
   theorien allgemeingueltigkeit beanspruchen?  wie kann man die
   theorien verknuepfen?  Bezug zu kinetischen modellen. verknuepfung
   mit kinetik. rechtfertigung und rekonstruktion. (wirklich NOETIG?
   hatten wir schon} \co{wirklich ernstnehmen, dass kosten PRO FLUSS
   das wichtige sind. das gut diskutieren! (kommt auch in CBA labour
   und FCM)} \co{interesting that in the economic potentials (as
   embodied value), substrate and enzyme investments are simply added,
   as if they were the same kind of thing) // possible to replace
   enzyme investment by substrate investment! (in model reformulation)
   (MENTION this also in cba modular - das ist eigentlich der artikel
   tz model reformulation! usw) equivalence between different models!}
 \co{the economic potentials provide a new layer of abstraction
   between kinetic models and flux modelling, hide details of the rate
   laws and metabolic objectives, and make optimality concepts
   comparable between different types of models.}  \co{WO? By linking
   FBA models to kinetic models, VBA sheds new light on some
   epistemological / philosophical questions about model assumptions
   underlying FBA, e.g.: ...  ``Ohm'' principle of higher value
   production - higher flux? Should cells choose between pathways or
   use them in combination? Does it make sense to pick ONE out of many
   (potentially optimal) flux distributions? They are not of practical
   concern for EXISTING flux prediction / optimality problems, but
   concern the very assumptions use to formulate such problems in the
   first place.}

 \myparagraph{New insights about FBA} \co{Absatz: econ FBA and other
   methods (bezug und rechtfertigung); abstraktion (alles dazu in
   einen absatz!)}  Many FBA variants (including FCM, FBA with flux
 minimisation, FBA with molecular crowding, and CAFBA) trade flux
 benefit against enzyme cost. Why is there a need for another method,
 a method that even requires additional variables?  \co{hier sagen,
   dass es kein neues optimierungsproblem ist, sonder eine
   abstraktion, die verschiedene optimierugsprobleme verbindet (kam schon oben, avoid WD)} In
 fact, the abstraction by VBA provides some valuable insights. First,
 it clarifies how different model paradigms, including FCM and kinetic
 models in enzyme-optimal states, are logically related.  \co{In
   contrast to FBA; VBA links economic variables -- economic
   potentials or enzyme investments -- directly to kinetics and
   thermodynamics, which determine variable enzyme efficiencies. For
   example, VBA can be related to FCM with kinetic flux cost
   functions, which in turn is based on kinetic models.}  \co{discuss
   usage in flux modelling; justification of and relation to other
   methods - wie enzymeigenschaften (kcat, kosten) die verwendung von
   enzymen bestimmen -- uebergang zu kinetischen modellen} With VBA as
 an abstraction, we can see that (even in models with metabolite
 concentrations) heuristics like the principle of minimal fluxes can
 be grounded in an enzyme optimisation in kinetic models.  \co{VBA
   resembles FBA with flux cost constraints ({\mfFBA} and FBA with
   molecular crowding), but has a theoretical justification by kinetic
   models, in which fluxes, metabolite concentrations, and enzyme
   levels are explicityl modelled and score by a fitness function
   \cite{lieb:14a}.}  \co{Since, different kinetic models can yield
   different optimal {\flow}s, the value production principle
   \co{``principle of local value production''} does not predetermine
   a single {\flow}. Instead, different high-yield or low-yield
   strategies can be economical.}  Second, frameworks such as \co{uea
   statt methods, auch andere artikel} FCM suppress futile cycles, but
 without explicitly finding them.  Our definition of futile cycles
 allows us to detect and remove such cycles, similar to cycles in
 thermodynamic FBA \cite{prtp:06,nolm:12}.  Third, since VBA, FCM, and
 enzyme-optimised kinetic models predict the same flux modes, we can
 know that FCM solutions, and only those, represent an optimal usage
 of enzymes. This justifies FCM and shows why this method works at
 all. Fourth, economic values (e.g.~shadow values in FCM) can be
 related to enzyme kinetics (e.g.~the cost effects of enzyme
 saturation and thermodynamic forces in kinetic models).  \co{JA!
   datenintegration: enzymkosten besser als constraint on
   fluxes. konkreter, klarer als teil des letzten punktes zeigen}
 Fifth, the comparison between methods shows that cost and benefit
 functions in FBA and FCM do not represent quantities, but
 log-marginal costs and benefits in kinetic models. This clarifies
 some key assumptions in flux analysis. Finally, VBA provides a new
 angle on shadow values in FBA.  Shadow values from FBA can now be
 understood as economic potentials, and by plotting them on the
 network we can verify that fluxes lead towards larger potentials.
 The reaction balance relates potential differences to flux {\burden}s
 which depend on enzyme {\price} and enzyme efficiencies. Therefore,
 estimating these enzyme properties may help us double check FBA
 results. If the economic potentials predicted by FBA are unrealistic
 (e.g.~entailing excessive enzyme investments), we can instead apply
 VBA from the start: by integrating various types of data, we obtain
 fluxes, economic potentials, and enzyme investments that represent a
 plausible metabolic, thermodynamic, and economic state of the system.
 \co{Hence, man kann fcm sowieso verwenden. aber theor rechtfertigung,
   klarer bezug zu kin mod, klarere interpretation .. verbindung
   zwischen theorien (evtl sogar fuer formalismusuebergreifende
   opt-modelle!}

 \co{\textbf{Metabolic strategies} JA! Metabolic values cast light on
   cells' choices between high-yield and low-yield metabolic
   pathways. \co{zusammenfasssung zu usage of pathways: gedankengang:
     Whole network -- zoom in on pathways -- lump pathways and zoom
     out // Genauer: 1. Proteomaps // 2. Boundary, module // 3
     bilanzgleichung von reaktion zu pathway // In particular, \co{The
       link to economic potentials helps us see flux and proteomics
       data as an ``investome''. It helps us predict fluxes and
       proteome, and yields intuitive explanations for flux
       directions, usage of enzymes, and the choices between pathways,
       as well as explanations by local cost-benefit balances.
       Cost-benefit balances also hold along pathways and cycles, in a
       close analogy to thermodynamics} 3a connecting variables in the
     network //4 benutzung von reakt / pathway, ja oder nein?  // 5
     auswahl zwischen wegen? // 6. voroiges {{\investment}} spielt
     eine rolle.}  \co{Discussion: hier kurz: Symmetrieprinzip:
     Abbauen genauso wichtig wie aufbauen; mehr dazu in CBA labour}
   7. wie stark ist der fluss im weg? Each enzyme has an essential
   value, defined simply by its molecular properties.  If the economic
   potential difference in a catlaysed reaction is below its essential
   value, using the enzyme cannot be profitable. As a flux becomes
   more expensive (e.g.~because the substrate concentration decreases
   or the enzyme is inhibited, and the enzyme demand per flux
   increases): at some point, the cost will exceed the flux value, the
   enzyme becomes uneconomical, and the flux may be rerouted to other,
   more cost-efficient reactions.  }

\co{\textbf{Proteome and enzyme investome}
  JA! SCHREIBEN!  \co{FUER runde stunde: d.h. dass man ueber die
    verbindung enzymekosten - proteinmengen (bei linearer
    kostenfunktion) das proteome als ein ``investome'' beschreiben
    kann (wird genauer ausgefuehrt in CBA labour)} \co{say
    ``usefulness''=''investment''.  ``usefulness'' has been claimed in
    proteomaps. now we say: its the POINT BENEFIT of an enzyme ie the
    point objective (eg growth) sensitivity, almost equal to scaled
    growth control coefficient (also called ``usefulness'') - say that
    ``usefulness'' can now be seen as enzyme load of ``value
    production''}} \co{mention exchange value (=labour value) and value
  in use (also in CBA lag, CBA I, sowieso in CBA labour)} 
\co{prominent sagen (auch in anderen artikeln: labour value (=exchange
  value): embodied (``past'') substrate and {\enzymeinvestment}s;
  value in use (``future'' benefit); in optimal state identical! gilt
  fuer gesamtsystem und fuer jedes teilsystem!}

\co{\textbf{Metabolic fluxes and value flows} JA! KURZ. SCHREIBEN!
  (kam schon in CBA lag; text dort oder hier?)  \co{picture here:
    value (called ``cost'') flows in as ``enzyme labour'' and adds to
    the other inflows, giving rise to a flow of ``value in use''}
  Finally, the investome can also be seen as part of a picture of
  value flows. mit ref nach CBA labour \cite{lieb:20a} hier waren es
  immmer fluesse und werte (bzw investmnet + point benefit): but one
  can also interpret th latter directly as value flows, [ substrate
  and enzyme as inflow - accumulation along pathway - benefit as
  outflow] which follow a conservation law and can be directly
  computed and visualised on the network} \co{(and we can see values
  flow and accumulate along metabolic fluxes)}

\section*{Acknowledgements}

I thank Bernd Binder, Mariapaola Gritti, Elad Noor, Tomer Shlomi,
Naama Tepper, and Hermann-Georg Holzh\"utter for thinking with me.
This work was funded by the German Research Foundation (Ll 1676/2-1
and Ll 1676/2-2).

\bibliographystyle{unsrt}
\bibliography{files/biology}


%
%
%
%
%

\end{document}